\documentclass[aps,prc,preprint,superscriptaddress,onecolumn,floatfix]{revtex4-2}
\usepackage{amsmath}
\usepackage{amssymb}
\usepackage{amsfonts}
\usepackage{amscd}
\usepackage{bm}
\usepackage{graphicx}
\usepackage{color}
\usepackage{empheq}
\usepackage{xfrac}
\usepackage{comment}

\def\beq{\begin{equation}}
\def\eeq{\end{equation}}

\DeclarePairedDelimiter\Bra{\langle}{\rvert}
\DeclarePairedDelimiter\Ket{\lvert}{\rangle}
\DeclarePairedDelimiterX\Braket[2]{\langle}{\rangle}{#1 \delimsize\vert #2}

\newcommand{\newlineEq}{\\\times}

\newcommand{\de}[1]{\mathrm{d}#1}
\newcommand{\bra}[1]{\Bra{#1}}
\newcommand{\ket}[1]{\Ket{#1}}
\newcommand{\braket}[2]{\Braket{#1}{#2}}
\newcommand{\matrixelement}[3]{\bra{#1}\,#2\,\ket{#3}}

\newcommand{\ve}[1]{\mathbf{#1}}
\newcommand{\versor}[1]{\ve{\hat{#1}}}
\newcommand{\Dmatrix}[3]{D^{#1}_{#2}\left[#3\right]}

\newcommand{\CG}[6]{\left(#1,#2,#3,#4\vert#5,#6\right)}

\newcommand{\Y}[2]{Y_{#1}(#2)}
\newcommand{\Yconj}[2]{Y^\ast_{#1}(#2)}

\newcommand{\energy}[3]{E_#1({#2#3})}
\newcommand{\energyel}[3]{E_#1#3}

\newcommand{\oneHalf}{\sfrac{1}{2}}
\newcommand{\mt}{\tau}
\newcommand{\deutL}[1]{\phi_{D,#1}}

\begin{document}

\title{Electron and neutrino scattering off the deuteron in a relativistic framework}

\author{A. Grassi}
\affiliation{M. Smoluchowski Institute of Physics, Jagiellonian University, PL-30348 Krak\'ow, Poland}
\author{J. Golak}
\affiliation{M. Smoluchowski Institute of Physics, Jagiellonian University, PL-30348 Krak\'ow, Poland}
\author{W. N. Polyzou}
\affiliation{Department of Physics and Astronomy, The University of Iowa, Iowa City, Iowa 52242, USA}
\author{R. Skibi{\'n}ski}
\affiliation{M. Smoluchowski Institute of Physics, Jagiellonian University, PL-30348 Krak\'ow, Poland}
\author{H. Wita{\l}a}
\affiliation{M. Smoluchowski Institute of Physics, Jagiellonian University, PL-30348 Krak\'ow, Poland}
\author{H. Kamada}
\affiliation{Department of Physics, Faculty of Engineering,
Kyushu Institute of Technology, Kitakyushu 804-8550, Japan}

\date{\today}

\begin{abstract}
We build a relativistic model to perform calculations of exclusive,
semi-exclusive and inclusive unpolarized cross sections and various
polarization observables in electron and neutrino
scattering experiments with deuteron targets.
The strong interaction dynamics is defined by an explicit dynamical unitary
representation of the Poincar\'e group, where
representations of space translations and
rotations in the interacting and non-interacting
representations are the same.
The Argonne V18 potential is used to construct a relativistic
nucleon-nucleon interaction reproducing
the experimental deuteron binding energy and
nucleon-nucleon scattering observables.
Our formalism does not include the pion production channel
and neglects two-body contributions in the electromagnetic as well as in the weak
nuclear current operator. We show that it is applicable
to processes at kinematics, where the internal two-nucleon energy
remains below the pion production threshold but the magnitude of the three-momentum
transfer extends at least to several GeV.
\end{abstract}

\maketitle

\section{Introduction} 

The deuteron is the simplest bound system of nucleons.  Because of its
simplicity it is an ideal system for detailed investigations of strong
interaction dynamics.  The deuteron can be modeled in terms of
experimentally observable particle degrees of freedom or in terms of
sub-nucleon degrees of freedom.  As long as the energy scale is
limited, both representations can in principle be used to calculate the
same experimentally observable scattering matrix.  Understanding the
relation between these two representations is a central question in
nuclear physics applications.
One of the most useful ways to study the dynamics and structure of the
deuteron is by scattering deuterons with photons, electrons or
neutrinos.  This is because the scattering reaction can be accurately
approximated in the lowest non-trivial order in the electroweak
interaction.  In this approximation the scattering operator is linear
in the hadronic matrix elements of hadronic current operators.  The
hadronic current operators encode the density and motion of the
strongly interacting charged constituent particles.  The
representation of the hadronic current is largely determined by the
representation of the interaction.  In the coordinate representation a
locally gauge invariant Hamiltonian can be constructed by replacing
momentum operators in the Hamiltonian by gauge covariant derivatives.
The current is the coefficient of the part of the gauge invariant
Hamiltonian that is linear in the vector potential.  In general it has
a cluster expansion which is a sum of interaction-independent
one-body and interaction-dependent many-body operators. 

In a scattering experiment involving electroweak probes the initial
and final hadronic states are in different frames related by the
momentum transferred to the target deuteron by the electron, photon or
neutrino.  Probes with sufficient resolution to be sensitive to
sub-nucleon degrees of freedom must have high momentum transfers
which require a relativistic description of the strong interaction
dynamics.

The most detailed information about the hadronic current is contained
in exclusive spin-dependent matrix elements of the current.
Calculations using realistic interactions with controlled
approximations are possible for spin-dependent elastic scattering and breakup reaction observables.  The available phase space that can be
explored is large and comparison of detailed computations with
experiment can put strong constraints on the model interactions, which
have implications for larger nuclei.  The deuteron is also a special
system because quasi-elastic scattering off deuteron targets supplies
important information about electroweak scattering from neutrons.

The purpose of this work is to develop tools to perform consistent
relativistic calculations of spin-dependent observables in electroweak
scattering experiments on deuteron targets. In this work the strong
interaction dynamics is defined by an explicit dynamical unitary
representation of the Poincar\'e group \cite{Wigner1939}.  The dynamical
Poincar\'e generators are constructed using a relativistic
re-interpretation \cite{KAMADA2007119} of the Argonne V18 interaction \cite{Wiringa:1994wb} that is designed to reproduce
the experimentally observable deuteron binding energy and
nucleon-nucleon scattering observables.  The dynamical
representation is chosen so representations of space translations and
rotations are identical in the interacting and non-interacting
representations \cite{Bakamjian:1953kh}.  The focus in this work is on the
spin-dependent observables in elastic and inelastic scattering from
deuterons.

The scope of this work is limited in two ways.  First, a pion
production channel is not included.  The second limitation is that the
dynamical two-body contributions to the current that arise from local
gauge invariance are not taken into account, although current covariance and
current conservation can be satisfied by using the Wigner-Eckart theorem
for the Poincar\'e group.  The model can be extended to overcome both
of these limitations.  A realistic treatment of production reactions
would require developing new nucleon-nucleon and production
interactions that are consistent with nucleon-nucleon scattering data both
above and below the pion production threshold.  While a consistent
derivation of two-body currents is possible; ignoring them can be used
to determine which reactions are sensitive to the dynamical parts of
the current.  These limitations, which require additional development,
can be addressed in subsequent investigations.  The present
investigation should provide some clarity on which observables are
sensitive to both two-body currents and/or production channels.

The next section discusses the structure of the theory, including the
construction of the dynamical unitary representation of the Poincar\'e
group.  It also discusses the assumptions that are needed to justify
the approximations used in the subsequent sections.  Section~III
describes relativistic and non-relativistic kinematic
relations that are used in the calculations.
Results of numerical calculations for unpolarized cross sections and 
spin dependent observables are shown in Sec.~IV.  
The summary and conclusions follow in Sec.~V.
Appendix A discusses the construction of nuclear current matrix
elements in this formalism.

\section{Theory}

In this work a relativistically invariant quantum mechanical model is
defined by a unitary representation, $U(\Lambda,a)$, of the Poincar\'e
group acting on the Hilbert space of the theory.  This ensures that
quantum observables (probabilities, expectation values and ensemble
averages) are independent of the choice of inertial reference frame~\cite{Wigner1939}.

The dynamics is solved by simultaneously diagonalizing the mass and
spin Casimir operators.  This decomposes $U(\Lambda,a)$ into a direct
integral of irreducible representations.  This is the relativistic
analog of diagonalizing the non-relativistic center of mass
Hamiltonian.

The model Hilbert space is a multi-particle space determined by the
particle content of the reaction.  In this work the particles of the
model are nucleons and leptons. 
Single-particle Hilbert spaces
${\cal H}_i$ are represented by square integrable functions of the particle's
linear momentum and magnetic quantum numbers:
\beq
\langle \psi \vert \psi \rangle := 
\sum_{\mu=-j}^j\int \vert \langle (m,s) \mathbf{p},\mu \vert \psi \rangle \vert^2
d\mathbf{p} =1 .
\label{t1}
\eeq
Unless otherwise mentioned the non-covariant normalization above is assumed.

There are single-particle unitary representations of the Poincar\'e group,
$U_k(\Lambda,a)$, that act on each ${\cal H}_k$.  The representation
determines the interpretation of the
magnetic quantum numbers.
For a particle of mass $m$ and spin $j$
a single-particle unitary representation of the Poincar\'e group
is defined on the single particle basis by
\[
U(\Lambda ,a) \vert (m,j)\mathbf{p}, \mu \rangle =
\]
\beq
e^{-ip'\cdot a} \sum_{\nu=-j}^j
\vert (m,j)\mathbf{p}', \nu \rangle
\sqrt{\frac{E(\mathbf{p}')}{E(\mathbf{p})}} \,
D^{j}_{\mu\nu}[R_w(\Lambda,p)],
\label{t2}
\eeq
where
\beq
E(\mathbf{p}) = \sqrt{m^2 + \mathbf{p}^2}  
\label{t3}
\eeq
is the particle's energy,
\beq
p^{\prime \mu} := \Lambda^{\mu}{}_{\nu}p^{\nu}
\eeq
is the transformed four momentum,
\beq
R_w(\Lambda,p) := B^{-1}(\mathbf{p'}/m)\Lambda B(\mathbf{p}/m)  
\eeq
is a $SU(2)$ Wigner rotation
and $B(\mathbf{p}/m)^{\mu}{}_{\nu}$ is a rotationless Lorentz
transformation that maps $(m,0,0,0)$ to $(E(\mathbf{p}),\mathbf{p})$:
\beq
B(\mathbf{p}/m)^0{}_0 = E(\mathbf{p})/m ,
\qquad
B(\mathbf{p}/m)^i{}_0 =B(\mathbf{p}/m)^0{}_i = {p}^i/m ,
\label{eq6}
\eeq
\beq
B(\mathbf{p}/m)^i{}_j = \delta^{ij} + p^ip^j/(m (m+E(\mathbf{p}))) .
\label{eq7}
\eeq
In representation (\ref{t2}) the spin observable in an arbitrary frame
is defined as the spin that would be measured in the particle's rest
frame if it was boosted to the rest frame using
$B^{-1}(\mathbf{p}/m)$. The $SL(2,C)$ version of $B(\mathbf{p}/m)$,
which appears in the Wigner $D$-function, is $e^{\pmb{\rho}\cdot
  \pmb{\sigma}/2}$, where $\pmb{\rho}$ is the rapidity of the boost
and $\pmb{\sigma}$ are the Pauli matrices.

The multi-particle Hilbert space is the tensor product of suitably symmetrized
single-particle Hilbert spaces:
\beq
{\cal H}= \otimes_i {\cal H}_i .
\label{t6}
\eeq
The free dynamics on ${\cal H}$ is given by the tensor product of the
single particle unitary representations, $U_i(\Lambda ,a)$, of the Poincar\'e group
\beq
U_0(\Lambda ,a) = \otimes_i U_i(\Lambda ,a).
\label{t7}
\eeq

The infinitesimal generators of $U_0(\Lambda ,a)$ are the free four momentum
$P_0^{\mu} := \sum_i P_{0i}^{\mu}$
and the free Lorentz generators
$J_0^{\mu\nu} := \sum_i J_{0i}^{\mu\nu}$.
The free mass Casimir operator and canonical spin operators are
functions of the
infinitesimal generators of $U_0(\Lambda ,a)$:
\beq
M_0 = \sqrt{ g_{\mu\nu} P_0^{\mu} P_0^{\nu}} 
\label{t8}
\eeq
and
\beq
\mathbf{j}_0^k = 
\frac12 \epsilon^{ijk}
B^{-1}(\mathbf{P_0}/M_0)^{i}{}_{\mu}B^{-1}(\mathbf{P_0}/M_0)^{j} {}_{\nu}J_0^{\mu\nu}
\label{t9}
\eeq
where in (\ref{t9}) 
$B(\mathbf{P_0}/M_0)^{\mu}{}_{\nu}$ 
is a canonical boost,(\ref{eq6})-(\ref{eq7}), valued matrix of the operators $P^{\mu}_0/M_0$.
The inverse is obtained by replacing $\mathbf{P_0}\to -\mathbf{P_0}$
in the expression above.  The free spin $\mathbf{j}_0$ is Hermitian,
the components satisfy $SU(2)$ commutation relations and commute with
$P^{\mu}_0$.  It represents the angular momentum of the system of
particles in the rest frame of the non-interacting system assuming
that the system was transformed to the rest frame with a
non-interacting rotationless (canonical) Lorentz boost.

To construct the dynamical representation of the Poincar\'e group
the first step is to construct simultaneous eigenstates of the
commuting observables
$M_0,\mathbf{P_0},\mathbf{j}_0^2,\mathbf{j}_0\cdot \hat{\mathbf{z}}$,
\beq
\vert (m_0,j_0)\mathbf{p},\mu,d \rangle
\label{t10}
\eeq
for the non-interacting system, where $d$ represents kinematically
invariant degeneracy parameters. 
These eigenstates can be expressed as linear combinations of
single particle states using Clebsch-Gordan
coefficients of the Poincar\'e group, see Eq.~(\ref{b.4}) below.
The next step is to add 
interactions, $V$, that commute with and are independent of $\mathbf{P_0}$
and commute with all three components of $\mathbf{j}_0$ to the
non-interacting mass Casimir operator.  This defines a dynamical mass
operator
\beq
M=M_0+V
\label{t11}
\eeq
where matrix elements of the interaction in the basis (\ref{t10}) have the form
\beq
\langle (m',j')\mathbf{p}', \mu' ,d' \vert V \vert
(m,j)\mathbf{p}, \mu ,d \rangle =
\delta (\mathbf{p}'-\mathbf{p}) \delta_{j'j}\delta_{\mu\mu'}
\langle m' \, d' \Vert V^j \Vert m \, d \rangle .
\label{t12}
\eeq
In this expression $d$ represents kinematically invariant degeneracy
quantum numbers:
\beq
U_0 (\Lambda,a) d U^{\dagger}_0 (\Lambda,a)=d .
\label{t13}
\eeq
The degeneracy parameters depend on the model Hilbert space, but
typically involve quantities like invariant masses and squares of
angular momenta of subsystems.  The kernel, 
$\langle m' , d' \Vert V^j \Vert m , d \rangle$, of the reduced potential is the analog of a partial
wave potential.  Since $V$ is not diagonal in the degeneracy
observables they will no longer be invariant in the dynamical
representation.

For $V$ satisfying (\ref{t12}) the operators $M,\mathbf{P_0},\mathbf{j}_0^2,\hat{\mathbf{z}}
\cdot \mathbf{j}_0$ are mutually commuting self-adjoint operators.
The dynamical mass operator $M$ can be diagonalized in the basis
(\ref{t10})
resulting in simultaneous eigenstates of 
$M,\mathbf{P_0},\mathbf{j}_0^2,\hat{\mathbf{z}}
\cdot \mathbf{j}_0$:
\beq
\vert (m,j) \mathbf{p},\mu, \tilde{d} \rangle
\label{t14}
\eeq
where $\tilde{d}$ represent new dynamically
invariant degeneracy parameters. 
These eigenstates transform like (\ref{t2}) with the single-particle mass
replaced by the eigenvalues, $m_I$, of the dynamical mass operator $M$:
\[
U(\Lambda ,a) \vert (m_I,j)\mathbf{p}, \mu, \tilde{d} \, \rangle =
\]
\beq
e^{-ip'\cdot a} \sum_{\nu=-j}^j
\vert (m_I,j)\mathbf{p}', \nu, \tilde{d} \, \rangle \,
\sqrt{\frac{E(\mathbf{p}')}{E(\mathbf{p})}} \,
D^{j}_{\mu\nu}[R_w(\Lambda,p)] \, ,
\label{t15}
\eeq
where
\beq
E(\mathbf{p}) = \sqrt{m_I^2 + \mathbf{p}^2}  \, .
\label{t16}
\eeq
This construction is due to Bakamjian and
Thomas~\cite{Bakamjian:1953kh}.  It results in an explicit unitary
representation of the Poincar\'e group.  It has the property that the
Lorentz boosts are interaction dependent.  The interaction dependence
appears in the dependence of the right side of (\ref{t15}) on the
dynamical mass eigenvalues, so dynamical boosts can be computed once
the mass operator (\ref{t11}) is diagonalized.

For the applications in this paper the construction
discussed above will be used to model the strong interactions
while the weak and electromagnetic interactions will be treated
using the one-boson exchange approximation.

In the one-boson exchange approximation $U(\Lambda,a)$ factors into a tensor
product of unitary representations of the Poincar\'e group
for the strongly interacting baryons (B) and the leptons (L):
\beq
U(\Lambda,a) \approx U_B(\Lambda,a)\otimes U_L(\Lambda,a).
\label{t18}
\eeq
The coupling is through a current that couples to the exchanged boson.
The current is a sum of a weak and
strong current plus an interaction current
\beq
J^{\mu}(x) = J_B^{\mu}(x)+  J_L^{\mu}(x) + J_I^{\mu}(x) \to
J_B^{\mu}(x)+  J_L^{\mu}(x) ,
\label{t19}
\eeq
where the interaction term includes the parts of the current that do
not contribute to the one-boson exchange approximation
and they will be ignored in what follows. The strong and
weak currents transform covariantly with respect to the baryonic,
$U_B(\Lambda,a)$, and
leptonic, $U_L(\Lambda,a)$,  representations of the Poincar\'e group
\beq
U_A(\Lambda,a) J_A^{\mu}(x) U_A^\dagger(\Lambda,a)=
(\Lambda^{-1})^{\mu}{}_{\nu}J_A^{\nu}(\Lambda x+ a) \qquad \mbox{where} \qquad
A=B,L. 
\label{t20}
\eeq 

Both $J_B^{\mu}(x)$ and $J_L^{\mu}(x)$
have cluster expansions as sums of one-body, two-body, $\cdots$
operators
\beq
J_A^{\mu}(x) = \sum_i J_{Ai}^{\mu}(x) + \frac12 \sum_{i\not=j} 
J_{Aij}^{\mu}(x) + \cdots
\label{t21}
\eeq
While the leptons can be approximately treated at tree level,
where current covariance holds up to higher order corrections, 
the $2 \cdots N$-body parts of the baryon currents must be non-zero in order
to satisfy the covariance and current conservation.

This can be seen from the commutation relations of the current operator with
the dynamical generators of the Poincar\'e group.  The cluster expansions
for the current and rotationless boost generators, $\mathbf{K}$, have the form
\beq
J_B^{\mu}(0) = J_{B0}^{\mu}(0) + J_{BI}^{\mu}(0) \, ,
\qquad 
J_{B0}^{\mu}(0)
= \sum_i J_{Bi}^{\mu}(0)
\eeq
\beq
\mathbf{K} = \mathbf{K_0} + \mathbf{K_I} \, ,
\qquad 
\mathbf{K_0}
= \sum_i \mathbf{K}_i
\eeq
Since the current $J^{\mu}(0)$ transforms like $P^{\mu}$ under Lorentz transformations
it has the same commutation
relations with the boost generators as $P^{\mu}$:  
\beq
[K_0^i+K_I^i, J^i(0)] = i J^0(0) \, , 
\qquad
[K_0^i+K_I^i, J^0(0)] = i J^i(0)
\label{t22}
\eeq
Cluster properties mean that (\ref{t22}) holds when the particles are
asymptotically separated where $K_I^i\to K_0^i$ and $J_I^{\mu}(0) \to J_0^{\mu}(0)$.
Canceling the one-body terms 
means that the interacting parts of the current $J_I^{\mu}(0)$ must satisfy
\[
[K_0^i, J_I^i(0)] +  [K_I^i, J_I^i(0)] -i J_I^0(0) =  [ J_0^i(0),K_I^i] ,
\]
\beq
[K_0^i, J_I^0(0)] + [K_I^i, J_I^0(0)] - i J_I^0(0) =
[J_0^0(0),K_I^i].
\label{t23}
\eeq
If the right side of either equation is non zero then the current must have
many-body parts in order to satisfy current covariance.  Similar conditions
follow if the current is conserved.

In general, these many-body contributions to the current have to be supplemented by the many-body
currents that arise from physical processes such as exchange of charged mesons.
They are not uniquely determined from current covariance.  

One way to ensure covariance is to use current matrix elements.  Since
current matrix elements transform covariantly, all current matrix
elements can be generated from any independent set of matrix elements
using covariance.  Any model of the current can be used to compute an
independent set of current matrix elements, while the remaining
elements can be computed by requiring covariance.  While this
implicitly generates covariant current matrix elements, the current
matrix elements will depend on the choice of independent current
matrix elements.  If the current operator used to calculate the
independent matrix elements was exactly covariant the results would be
independent of the choice of independent current matrix elements.
In this work, since boosts are dynamical, impulse approximations
in one frame are not equivalent to impulse approximations 
in another frame.
Violations of current covariance at the operator level can be investigated by
comparing calculations performed in different frames or based on different choices of independent
current matrix elements.

A model of the strong interaction dynamics is defined by the
interacting mass operator $M$.  For the two nucleon-system it should
have a discrete one-body eigenstate with the mass of a deuteron and
should produce measured scattering observables.  The scattering
operator is a unitary operator that can be expressed in the form
$S=e^{2i\delta}$ where $\delta$ is the phase shift operator.  Since
$S$ is relativistically invariant, so are the phase shifts.
Phenomenological non-relativistic interactions are constructed by (1)
using data from experimentally measured cross sections (2) using
correct relativistic kinematics to transform the laboratory cross
sections to the center of momentum (3) adjusting potential parameters so
the wave functions obtained by solving the Schr\"odinger equation give
the correct phase shifts as a function of the center of momentum momenta,
$\mathbf{k}$, of one of the particles.  {\it An important observation
  is that even though the non-relativistic Schr\"odinger equation is
  used to extract the phase shifts, the data knows nothing about
  non-relativistic limits and the transformation of the data to the
  two-body rest frame is performed using the correct relativistic
  kinematics.}  The phase shifts are experimental quantities that
parameterize the relativistic scattering operator.  The only place
where the difference between the relativistic and non-relativistic
treatment appears is whether the experimental phase shifts are
identified as functions of energy, $\delta (E_{NR}) =\delta (E_R)$ 
for $E_{NR} = E_{R}$
or
as functions of center of momentum momenta $\delta
(\mathbf{k}_{NR}^2)= \delta (\mathbf{k}_{R}^2)$
for $|\mathbf{k}_{NR}| = |\mathbf{k}_{R}|$.
This depends on how
the potential is constructed; for the V18 potential used in this work
the phase shifts are fit as functions $\mathbf{k}^2$.  This means that
if the dynamical mass operator is a function of the non-relativistic
Schr\"odinger Hamiltonian, the wave functions and phase shifts will be
identical to the non-relativistic wave functions and phase shifts as a
function of the relative momentum.  These phase shifts are the
``experimental'`` phase shifts that define the physical (relativistic)
scattering operator.

For 
\[
k^{\mu} := B^{-1}({(\mathbf{p}_n +\mathbf{p}_p)/ m_{np}})^{\mu}{}_{\nu}
\frac12 
(p^{\nu}_p-p^{\nu}_n)
\]
the free invariant mass (for equal mass nucleons)
has the form
\beq
M_0 = 2 \sqrt{m^2 +\mathbf{k}^2} .
\label{t24}
\eeq
The operator
\beq
M:=2 \sqrt{m^2+ 2 \mu 
\left( \frac{\mathbf{k}^2}{2\mu}+V_{NR} \right)},
\label{t25}
\eeq
where $\mu$ is the reduced mass of the two-body system and
$V_{NR}$ a realistic non-relativistic two-body interaction,
is a function of the non-relativistic Hamiltonian that becomes the
two-body invariant mass in the limit that the interaction vanishes.
A relativistic interaction is  defined as the difference
$V_R=M-M_0$:
\beq
V_R:= 2 \sqrt{m^2+ 2 \mu  \left( \frac{\mathbf{k}^2}{2\mu}+V_{NR} \right)} -
2 \sqrt{m^2+  2 \mu \left( \frac{\mathbf{k}^2}{2\mu} \right)} .
\label{t26}
\eeq
There are a
number of ways to diagonalize $M$,  however
the wave functions and phase shifts are identical to the ones obtained
by solving the nonrelativistic Schr\"odinger equation.  In this work $V_R$ is calculated
directly using the method outlined in~\cite{KAMADA2007119}, which involves solving a
non-linear equation for $V_R$.  In this work $V_{NR}$
is taken as the Argonne V18 potential~\cite{Wiringa:1994wb}.

This model is formally applicable to calculations at energies below
the threshold for pion production.  The model can be extended to
include pion degrees of freedom, but that extension is not considered
in this work.

The general expression for the differential cross section is
\beq
d\sigma
= \frac{(2 \pi)^4}{ v_r } \,
\vert \langle
\mathbf{p}_1, \mu_1, \cdots, \mathbf{p}_N, \mu_N
\Vert T (E+i \epsilon) \Vert \mathbf{p}_B, \mu_B, \mathbf{p}_T, \mu_T
 \rangle \vert ^2 \,
\Pi_i d \mathbf{p}_i  \, \delta^4 (P_f - P_i) \, ,
\label{t27}
\eeq
where $v_r$ is the relative speed between the projectile and target, and
$P^{\mu}_f$ and $P^{\mu}_i$ are the total final and initial four momentum of
the system, and $T(z)$ is the transition operator with the total
momentum conserving delta function removed:
\beq
\langle \mathbf{P}_t' \cdots' \vert T \vert \mathbf{P}_t \cdots \rangle = \delta(\mathbf{P}'_t-\mathbf{P}_t)
\langle \cdots' \Vert T \Vert \cdots \rangle .
\eeq

The differential cross section (\ref{t27}) can be expressed as a product of three invariant quantities
\[
d\sigma
= \frac{(2 \pi)^4}{ \sqrt{(p_B\cdot p_T)^2 - m_T^2 m_B^2}}
\times
\]
\[
\Pi_i E_i(\mathbf{p}^{\prime 2}_i)
\vert \langle \mathbf{p}_1', \mu_1', \cdots \mathbf{p}_N', \mu_N'  \Vert T (E_T+i \epsilon) \Vert \mathbf{p}_B, \mu_B, \mathbf{p}_T, \mu_T 
\rangle \vert ^2  E_B(\mathbf{p}^2_B)E_T(\mathbf{p}^2_T) \times
\]
\beq
 \delta^4 (P_f - P_i)
\label{t28}
\eeq
where the subscripts $B$ and $T$ stand for beam and target respectively.
This form can be utilized in any frame.  Specific cross sections are derived from (\ref{t28}) by integrating over the unmeasured kinematical quantities, including
those fixed by the four-momentum conserving delta function.

In the one boson-exchange approximation, 
the transition matrix elements for
a beam of electrons or neutrinos is
\[
\langle \mathbf{p}'_L, \mu'_L, \mathbf{p}'_1, \mu'_1, \cdots \mathbf{p}'_N \mu'_N \Vert T (E_T+i \epsilon) \Vert \mathbf{p}_L, \mu_L, \mathbf{p}_T, \mu_T 
\rangle =
\]
\beq
- (2 \pi)^3 g^2
\langle \mathbf{p}_L', \mu_L' \vert J_L^{\mu}(0) \vert \mathbf{p}_L, \mu_L
\rangle G_{\mu\nu} (p_l-p_l') 
\langle (\mathbf{p}_1', \mu_1' \cdots \mathbf{p}_N', \mu_N')^- \vert 
J_B^{\nu}(0) \vert \mathbf{p}_B', \mu_B' \rangle \, ,
\label{t29}
\eeq
where
\beq
\vert  (\mathbf{p}_1', \mu_1' \cdots \mathbf{p}_N', \mu_N')^- \rangle
\eeq
can be a bound or scattering eigenstate, 
\beq
G_{\mu\nu} (k) = i \int \langle 0 \vert T(V_{\mu} (x) V_{\nu}(0)) \vert 0
\rangle e^{ik\cdot x} d^4x \, ,
\label{t30}
\eeq
where $V^{\mu} (x)$ is the field of the exchanged boson and $g$ is the coupling
constant of 
the interaction of the current with the exchanged boson
\beq
V_c = g \int d\mathbf{x} J^{\mu}(\mathbf{x},t)V_{\mu}(\mathbf{x},t) 
\label{t31}
\eeq
for fixed $t$.

\section{Kinematics and matrix elements}
	
\subsection{Kinematics and matrix elements for electron or neutrino elastic 
scattering off the deuteron}
	
	
	\noindent
	The four-momentum conservation 
	for the $ e + d \rightarrow e' + d'  $ reaction 
	in a general frame, where the total energy is $E_t$ 
	and the total momentum is denoted $\ve{P}_t$, reads
	\begin{eqnarray}
		E_t  \equiv   E_e + E_D           =  \energyel{e}{\ve{p}_e}{'} + {E_D'}\, , \nonumber \\
           \ve{P}_t  \equiv  \ve{p}_e + \ve{p}_D  =  \ve{p}_e' + \ve{p}_D' \, ,
	\label{elas_kin}
	\end{eqnarray}
	where 
	$\ve{p}_e$ and $\ve{p}_D$ ($\ve{p}_e'$ and $\ve{p}_D'$)
	are the initial (final) electron and deuteron momenta, 
$E_e= \sqrt{ m_e^2 + |\ve{p}_e|^2 }$ and 
$E_D= \sqrt{ m_D^2 + |\ve{p}_D|^2 }$
($E_e'= \sqrt{ m_e^2 + |\ve{p}_e'|^2 }$ and 
$E_D'= \sqrt{ m_D^2 + |\ve{p}_D'|^2 }$)
	are the corresponding total energies
	with $m_e$ and $m_D$ being the electron and deuteron masses. 
	The system of equations (\ref{elas_kin}) can be solved analytically 
	to yield $|\ve{p}_e'|$ for a given
	electron scattering angle $\theta_e$, which is the angle 
	between $\ve{P}_t$ and $\ve{p}_e'$. For $\ve{P}_t = 0$ the electron scattering angle is taken between 
	the initial $\ve{p}_e$ and final $\ve{p}_e'$ electron momentum.
	There is no restriction for $\theta_e$: $ 0 \le \theta_e \le 180^\circ$.
	We obtain 
	\begin{eqnarray}
	| \ve{p}_e' | & = & 
		\frac{{E_t} \sqrt{ H + \left( 2 {m_e} {|\ve{P}_t|} {\cos\theta_e} \right)^2}}
	{2 ({E_t}-{\cos\theta_e} {|\ve{P}_t|}) ({E_t}+{\cos\theta_e} {|\ve{P}_t|})} \nonumber \\
	& + &
	\frac{|\ve{P}_t| \left({E_t}^2-{m_D}^2+{m_e}^2-{|\ve{P}_t|}^2\right)\, {\cos\theta_e} }
	{2 ({E_t}-{\cos\theta_e} {|\ve{P}_t|}) ({E_t}+{\cos\theta_e} {|\ve{P}_t|})}
		\label{pep}
	\end{eqnarray}
	with
	$ H \equiv E_t^4 + \left( m_D^2 - m_e^2 + {|\ve{P}_t|}^2 \, \right)^2 - 2 E_t^2 (m_D^2 + m_e^2 + {|\ve{P}_t|}^2)$.
In the laboratory frame, where $\ve{P}_t = \ve{p}_e$, $E_t = m_D + \energyel{e}{\ve{p}_e}{} $, 
and neglecting the electron mass
equation~(\ref{pep}) reduces to the simple result well known from the Compton scattering: 
	\begin{eqnarray}
	| \ve{p}_e' | = 
	\frac{ m_D |\ve{p}_e| }
	{ m_D + {|\ve{p}_e|} \left( 1  - {\cos\theta_e} \right) } \, ,
	\label{pepme0}
	\end{eqnarray}
	which is sufficient for all our calculations performed in this frame.

	The transition matrix elements for this reaction in the one-photon-exchange approximation are 
	given as contractions of the electron 
	\[
	\langle \ve{p}_e', \mu_e'  \vert J_e^{\nu} (0) \vert \ve{p}_e,\mu_e \rangle
	\]
	and nuclear (here deuteron "D")
	\[
	\langle \ve{p}_D', \mu_D' ,D \vert J_{nuc,EM}^{\mu} (0) \vert \ve{p}_D,\mu_D ,D \rangle  
	\]
	matrix elements:
	\begin{eqnarray}
		\langle \ve{p}_D',\mu_D',D, \ve{p}_e', \mu_e' \Vert T_{eD} \Vert 
		\ve{p}_D, \mu_D,D, \ve{p}_e ,\mu_e \rangle = \nonumber \\
	- e^2 (2 \pi)^3
	\langle \ve{p}_D', \mu_D' ,D \vert J_{nuc,EM}^{\mu} (0) \vert \ve{p}_D,\mu_D ,D \rangle  
	\,
	\frac{g_{\mu \nu}}{ (p_e'-p_e)^2 +i \epsilon} \,
	\langle \ve{p}_e', \mu_e'  \vert J_e^{\nu} (0) \vert \ve{p}_e,\mu_e \rangle \, ,
	\label{qft17}
	\end{eqnarray} 
	with $ e^2 = 4 \pi \alpha $, where $\alpha \approx \frac1{137}$ is the fine structure constant.
	The differential cross section in terms of (\ref{qft17}) becomes 
	\begin{eqnarray}
	\de\sigma =
	{(2 \pi)^4 {E_D} \energyel{e}{\ve{p}_e}{}\over  \sqrt{(p_D\cdot p_e)^2- m_D^2m_e^2}} 
\vert \langle \ve{p}_D',\mu_D',D, \ve{p}_e', \mu_e' \Vert T_{eD} \Vert \ve{p}_D, \mu_D,D, \ve{p}_e ,\mu_e \rangle \vert^2 
		\nonumber \\
	\times \, {E_D'}\energyel{e}{\ve{p}_e}{'}
	\delta^4 (p_D+p_e - p_D'-p_e') 
	{\de\ve{p}_D' \over {E_D'}}
	{\de\ve{p}_e' \over \energyel{e}{\ve{p}_e}{'}} 
	\label{qft18}
	\end{eqnarray}
	and is a product of the following three invariant factors
	\begin{eqnarray}
	{(2 \pi)^4 \over  \sqrt{(p_D\cdot p_e)^2- m_D^2m_e^2}} \, ,
	\label{qft19}
	\end{eqnarray}
	\begin{eqnarray}
		{E_D'} \energyel{e}{\ve{p}_e}{'} \,
		\vert \langle \ve{p}_D',\mu_D',D, \ve{p}_e', \mu_e' \Vert T_{eD} \Vert \ve{p}_D, \mu_D,D, \ve{p}_e ,\mu_e \rangle \vert^2
\,	{E_D}\energyel{e}{\ve{p}_e}{} \, ,
	\label{qft20}
	\end{eqnarray}
	and
	\begin{eqnarray}
	\delta^4 (p_D+p_e - p_D'-p_e') {\de\ve{p}'_D \over {E_D'}}
	{\de \ve{p}'_e \over \energyel{e}{\ve{p}_e}{'}} \, .
	\label{qft21}
	\end{eqnarray}
	(Note the momentum eigenstates have a delta function normalization 
	$ \langle \ve{p}' \vert \ve{p} \rangle = \delta (\ve{p}'-\ve{p}) $.)
	
	In terms of Dirac spinors 
\[
	u(\ve{p},\mu)=
		\sqrt{ \frac{ \sqrt{ m^2 + |\ve{p}|^2 \, } + m}{2 m }\, } \,
	\left(
	\begin{matrix}
	\chi_{\mu}\\
	\frac{\ve{p} \cdot \bm{\sigma}}
		{ \sqrt{ m^2 + |\ve{p}|^2 \, } + m}
		\chi_{\mu}
	\end{matrix} \right)
	\]
	with the Bjorken-Drell \cite{BD} conventions and normalization $\bar{u}(\ve{p},\mu) u( \ve{p},\mu) = 1$,
	the electron current matrix elements can be expressed as
	\begin{eqnarray}
	\langle \ve{p}_e', \mu_e'  \vert J_e^{\nu} (0) \vert \ve{p}_e,\mu_e \rangle 
	& = &  {1 \over (2\pi)^3}\sqrt{m_e^2 \over \energyel{e}{\ve{p}_e}{} \energyel{e}{\ve{p}_e}{'}}
	\bar{u}_e(\ve{p}_e',\mu_e')\gamma^{\nu} u_e(\ve{p}_e,\mu_e) \label{qft22-1} \\
	& \equiv &  {1 \over (2\pi)^3}\sqrt{1 \over 4  \energyel{e}{\ve{p}_e}{} \energyel{e}{\ve{p}_e}{'}} \, 
	L_e^{\nu} \left( \ve{p}_e',\mu_e', \ve{p}_e,\mu_e \right) \, ,
	\label{qft22}
	\end{eqnarray}
	where 
	in (\ref{qft22}) the mass factor $\frac1{2 m_e}$ is extracted
from $\bar{u}_e(\ve{p}_e',\mu_e')\gamma^{\nu} u_e(\ve{p}_e,\mu_e) $. The latter form 
	can be used also in the reactions with (approximately massless) neutrinos.
	
	In order to calculate the deuteron current matrix element,
	\begin{eqnarray}
	\langle \ve{p}_D', \mu_D' ,D \vert J_{nuc,EM}^{\mu} (0)  \vert \ve{p}_D,\mu_D ,D \rangle \, \equiv \,
{1 \over (2\pi)^3}\, 
		N_{eD}^{\mu} \left( \ve{p}_D', \mu_D' , \ve{p}_D , \mu_D \, \right) \, ,
	\label{qft23}
	\end{eqnarray}
	we have to recall our choice of noninteracting irreducible
	states and the resulting Poincar\'e Clebsch-Gordan coefficients
	\cite{moussa,relform1,kei91}
	\begin{eqnarray}
	\langle  \ve{p}_1, \mu_1,
	\ve{p}_2, \mu_2  \vert (j,k ) \ve{p} ,\mu ; l ,s \rangle
	\nonumber \\
	=\sum_{\mu_l \mu_s \mu_1'\mu_2'}
	\delta (\ve{p} - \ve{p}_1 - \ve{p}_2)
	\frac {\delta (k -k (\ve{p}_1, \ve{p}_2)) }{k^2} \,
	{\cal N}^{-1}(\ve{p}_1, \ve{p}_2)  \,  
		Y_{l \mu_l }(\hat{\ve{k}}(\ve{p}_1,\ve{p}_2) )
	\nonumber \\
        \times \, 
	D^{\oneHalf}_{\mu_1 \mu_1'}[R_w(B(\ve{p}/m_{120}), \ve{k}_1)] 
	\,
	D^{\oneHalf}_{\mu_2 \mu_2'}[R_w(B(\ve{p}/m_{120}), \ve{k}_2)]
	\nonumber \\
	\times \,  ( l, \mu_l, s, \mu_s \vert  j, \mu )
	( \oneHalf, \mu_1', \oneHalf, \mu_2' \vert  s, \mu_s )
	\nonumber \\
	= \int \de \hat{\ve{k}}\sum_{\mu_l \mu_s \mu_1'\mu_2'}
	\delta (\ve{p}_1 - \ve{p}_1(\ve{p},\ve{k}))
	\delta (\ve{p}_2 - \ve{p}_2(\ve{p},\ve{k})) \,
		{\cal N} (\ve{p}_1, \ve{p}_2) \, 
		Y_{l \mu_l }(\hat{\ve{k}})
	\nonumber \\
	\times \, 
	D^{\oneHalf}_{\mu_1 \mu_1'}[R_w(B(\ve{p}/m_{120}), \ve{k}_1)] 
	\,
	D^{\oneHalf}_{\mu_2 \mu_2'}[R_w(B(\ve{p}/m_{120}),\ve{k}_2)] 
	\nonumber \\
        \times \, 
	( l, \mu_l, s, \mu_s \vert  j, \mu )
	( \oneHalf, \mu_1', \oneHalf, \mu_2' \vert  s, \mu_s ) \, .
	\label{b.4}
	\end{eqnarray}
 	In these expressions $m_{120} = \sqrt{m^2 + \ve{p}_1^2} + \sqrt{m^2 + \ve{p}_2^2}$
	is replaced by $\sqrt{\ve{k}^2}$, where $m$ is the nucleon mass, 
	$(j_1, \mu_1, j_2, \mu_2 \vert j_3 , \mu_3 )$ are $SU(2)$ Clebsch-Gordan coefficients, 
	$D^{\oneHalf}_{\mu' \mu}[R]$ is the Wigner D-matrix for spin $\oneHalf$.
	The $\ve{k}_i$ are the three-vector components of $k_i = B^{-1}(\ve{p}/m_{120}) p_i$.
	Arguments of the latter, $R_w(B(\ve{p}/m_{120} ), \ve{k}_i )$, are Wigner rotations resulting 
	from a product of three rotationless Lorentz transformations
	\begin{equation}
        R_w(B(\ve{p}/m_{120}), \ve{k}_i)= B^{-1}(\ve{p}_i/m)B(\ve{p}/m_{120})
        B(\ve{k}_i/m) \, ,
	\end{equation}
	where $B(\ve{k}_i/m)$ takes a particle of mass $m$ at rest to momentum $\ve{k}_i$,
	$B(\ve{p}/m_{120})$ takes a system of two particles with the same
        mass, $m$, and 
	momenta $\ve{k}=\ve{k}_1$ and $-\ve{k}=\ve{k}_2$, respectively, 
	to the total two-particle momentum $\ve{p}$,
	by which the momentum  $\ve{k}_i$ is changed to $\ve{p}_i$.
	Finally, $B^{-1}(\ve{p}_i/m)$ brings the particle with the momentum $\ve{p}_i$
	to its rest frame.  Here $B(\mathbf{p}/m)^{\mu}{}_{\nu}$ is the
        rotationless Lorentz transformation that takes $(m,0,0,0)$ to
	$(E(\mathbf{p}),\mathbf{p})$ (see Eqs. (\ref{eq6})--(\ref{eq7})).
The normalization coefficients
	\begin{eqnarray}
		{\cal N}^{-2}(\ve{p}_1, \ve{p}_2) & = & 
	\frac {\energy{{}}{\ve{k}}{} \energy{{}}{\ve{k}}{} (\energy{{}}{\ve{p}_1}{} 
		+ \energy{{}}{\ve{p}_2}{}) }{ \energy{{}}{\ve{p}_1}{} \energy{{}}{\ve{p}_2}{} (\energy{{}}{\ve{k}}{} + \energy{{}}{\ve{k}}{})} \, ,
	\label{b.6}
	\end{eqnarray}
        ensure unitarity of the Clebsch Gordon coefficients for basis states
        with delta function normalizations.

        The Bjorken and Drell spinors are also representations of the
        canonical boost;  the spins undergo the same Wigner rotations
        under Lorentz transformations:
        \beq
        \sum\limits_{b}
         S(\Lambda)_{ab}u_b(\mathbf{p},\mu) =
         \sum_\nu u_a({\Lambda}{p},\nu )D^{\oneHalf}_{\nu \mu}[R_w(\Lambda, \mathbf{p})] ,
          \eeq
          where $S(\Lambda)$ is the $4\times 4$ Dirac spinor representation of the Lorentz group.
The relativistic counterpart of the center of mass
        relative momentum is
        obtained by replacing a Galilean boost applied to
        half of the relative momentum to the zero momentum frame
        by a canonical boost to the zero momentum frame
        of the non-interacting two body system:
	\begin{eqnarray}
	  \ve{k} &\equiv& \ve{k}(\ve{p}_1, \ve{p}_2)=
		\ve{B}^{-1}(\mathbf{p}/m_{120}) \left( \frac{1}{2} \left({p}_1 - {p}_2 \right) \right) =
         \nonumber \\
&=& \frac{1}{2} \left(\ve{p}_1 - \ve{p}_2 - 
{({\energy{{}}{\ve{p}_1}{}  - \energy{{}}{\ve{p}_2}{}) \, (\ve{p}_1 + \ve{p}_2 )  } \over{\energy{{}}{\ve{p}_1}{} + \energy{{}}{\ve{p}_2}{} 
		+ \sqrt{(\energy{{}}{\ve{p}_1}{}  + \energy{{}}{\ve{p}_2}{})^2 - (\ve{p}_1 + \ve{p}_2   )^{2}}  } } \right) \, .
	\label{b.21}
	\end{eqnarray}
	Conversely, the individual momenta $\ve{p}_1$ and $\ve{p}_2$ 
	can be calculated from $\ve{p}$ and $\ve{k}$ in the following way:
	\begin{eqnarray}
	\ve{p}_1  & \equiv&   \ve{p}_1 (  \ve{p} ,  \ve{k} )  
	=  \ve{k} + \frac12 \ve{p} + \frac{ ( \ve{p} \cdot \ve{k} ) \, \ve{p} }
	{2 \energy{{}}{\ve{k}}{}  ( E_{12}( \ve{p} ,  \ve{k}  )  + 2 \energy{{}}{\ve{k}}{} ) } \, , \nonumber \\
	\ve{p}_2  &\equiv&   \ve{p}_2 (  \ve{p} ,  \ve{k} )   =  
	-\ve{k} + \frac12 \ve{p} - \frac{ ( \ve{p} \cdot \ve{k} ) \, \ve{p} }
	{2 \energy{{}}{\ve{k}}{}  ( E_{12}( \ve{p} ,  \ve{k}  )  + 2 \energy{{}}{\ve{k}}{} ) } \, 
	\label{p1p2fromkp}
	\end{eqnarray}
where $E_{12} ( \ve{p} ,  \ve{k}  )  = \sqrt{ \left( 2 E ( \ve{k} )   \right)^2 + \ve{p}^2 }$.
	Note that $\ve{k}$ does not transform like the space component of a four vector; 
	instead it undergoes Wigner rotation 
	$\ve{k} \rightarrow \ve{k}' = R_w(\Lambda,\mathbf{p})\, \ve{k} $
        for $ p' = \Lambda p $.  This means $k^2:=\mathbf{k}^2$ is kinematically invariant. 
        The quantum numbers $l$ and $s$ are also kinematically invariant
        degeneracy parameters
	that distinguish representations with the same mass
	($k$) and spin ($j$). For a two-nucleon system they have the same spectrum
	as the orbital and spin angular momentum operators in a partial
	wave representation of the nonrelativistic basis~\cite{moussa,relform1,kei91}.

	This information is used to express the deuteron state in terms of the 
	``relativistic partial waves'':
	\begin{eqnarray}
	\langle (j,k ) \ve{P} ,\mu ; l ,s ; t , \mt \vert 
	\ve{p}_D,\mu_D ,D \rangle \, 
	= \, 
	\delta\left(  \ve{P} - \ve{p}_D \, \right) \, 
	\delta_{ j 1} \, 
	\delta_{ \mu \mu_D} \, 
	\delta_{ s 1 } \, \delta_{t 0} \, 
	\delta_{\mt 0 } \,
	\deutL{l} (k) \, ,
	\label{D_PWD}
	\end{eqnarray}
	where $\deutL{l}(k)$ are the s ($l=0$) and d ($l=2$) components of the deuteron 
	wave function.
	Note that we added here isospin quantum numbers: $t$ is the total 
	two-nucleon isospin and $\mt$ is the value of its $z$-component.
	The formal structure of the current matrix element (\ref{qft23}) is 
	\begin{eqnarray}
	\langle \ve{p}_D', \mu_D' ,D \vert J_{nuc,EM}^{\mu} (0)  \vert \ve{p}_D,\mu_D ,D \rangle \nonumber \\
	= \, \int \de  \ve{p}' \, \sum\limits_{l'=0,2} \int \de k' {k'}^2 \, 
	\int \de \ve{p}_1' \, \sum\limits_{ \mu_1', \tau_1'} \, \int \de \ve{p}_2' \, \sum\limits_{\mu_2' , \tau_2'} \nonumber \\ 
	\int \de \ve{p}_1 \, \sum\limits_{ \mu_1, \tau_1} \, \int \de \ve{p}_2 \, \sum\limits_{\mu_2 , \tau_2} \,
	\int \de  \ve{p} \, \sum\limits_{l=0,2} \int \de k {k}^2 \nonumber \\ 
	\langle \ve{p}_D', \mu_D' ,D \vert (1,k' ) \ve{p}' ,\mu' ; l' ,1;  0 , 0 \rangle \nonumber \\
	\langle (1,k' ) \ve{p}' ,\mu' ; l' ,1 ; 0 , 0 \vert \ve{p}_1', \mu_1', \tau_1' , \ve{p}_2', \mu_2' , \tau_2'  \rangle    \nonumber \\
	\langle \ve{p}_1', \mu_1', \tau_1' , \ve{p}_2', \mu_2' , \tau_2' \vert J_{nuc,EM}^{\mu} (0)  \vert 
	\ve{p}_1, \mu_1, \tau_1 , \ve{p}_2, \mu_2 , \tau_2 \rangle \nonumber \\
	\langle \ve{p}_1, \mu_1, \tau_1 , \ve{p}_2, \mu_2 , \tau_2 \vert (1,k ) \ve{p} ,\mu ; l ,1 ; 0 , 0  \rangle \nonumber \\
	\langle (1,k ) \ve{p} ,\mu ; l ,1 ; 0 , 0 \vert \ve{p}_D,\mu_D ,D \rangle \, ,  
	\label{structureA}
	\end{eqnarray}
	where $\tau_i$ and $\tau_i'$ denote the isospin projections in the single-nucleon states.
	Matrix elements (\ref{structureA}) comprise contributions from single-nucleon current 
	and two-nucleon current operators:
	\begin{eqnarray}
	& & \langle \ve{p}_1', \mu_1', \tau_1' , \ve{p}_2', \mu_2' , \tau_2' \vert J_{nuc,EM}^{\mu} (0)  \vert
	\ve{p}_1, \mu_1, \tau_1 , \ve{p}_2, \mu_2 , \tau_2 \rangle \nonumber \\
	& = &
	\delta (\ve{p}_1'-\ve{p}_1) \, \delta_{ \mu_1'  \mu_1} \, \delta_{ \tau_1' \tau_1 } \,
		\langle \ve{p}_2' , \mu_2', \tau_2'  \vert J_{2,EM}^{\mu}(0) \vert \ve{p}_2 , \mu_2 , \tau_2 \rangle \nonumber \\
	& + & \delta (\ve{p}_2'-\ve{p}_2) \, \delta_{ \mu_2'  \mu_2} \, \delta_{ \tau_2' \tau_2 } \,
		\langle \ve{p}_1' ,  \mu_1', \tau_1'  \vert J_{1,EM}^{\mu}(0) \vert \ve{p}_1 , \mu_1 , \tau_1 \rangle \nonumber \\
		& + & \langle \ve{p}_1', \mu_1', \tau_1' , \ve{p}_2', \mu_2' , \tau_2' \vert J_{\{1,2\},EM}^{\mu} (0)  \vert
	\ve{p}_1, \mu_1, \tau_1 , \ve{p}_2, \mu_2 , \tau_2 \rangle  \, .
	\label{1N+2N}
	\end{eqnarray}
	In this paper we neglect two-nucleon current contribution and discuss consequences of such an approximation in Sec.~IV. Since the deuteron state and the two-nucleon 
	scattering states are antisymmetric with respect to the exchange of nucleons 1 and 2, it is sufficient
	to consider the contribution from $J_{1,EM}^{\mu}(0)$ and multiply the result by $2$ in order 
	to account for the $J_{2,EM}^{\mu}(0)$ part. The electromagnetic single-nucleon current operator
	has a well known form 
	\begin{eqnarray}
		\langle \ve{p}', \mu' , \mt'  \vert J_{k,EM}^{\mu} (0) \vert \ve{p}, \mu , \mt \rangle
	= \delta_{ \mt' \mt} \,
	{1 \over (2\pi)^3}\sqrt{m^2 \over \energy{{}}{\ve{p}}{'}\energy{{}}{\ve{p}}{}} 
	\nonumber \\
	\times \, \bar{u}(\ve{p}',\mu') 
		\left (\gamma^{\mu} F_{1, \tau} (Q^2) + i {(p'_{\alpha} - p_{\alpha})  \sigma^{\mu \alpha} \over 2m} 
	F_{2, \tau} (Q^2) \right )  u(\ve{p},\mu) \, ,
	\label{qft36}
	\end{eqnarray}
	where $F_{1, \tau}$ and $F_{2, \tau}$ are the Dirac and Pauli proton (for $\mt= \oneHalf$) 
	or neutron (for $\mt = -\oneHalf$) electromagnetic form factors,
	which depend on the square of the four momentum transferred to the nucleon, $Q^2 \equiv - (p' - p)_{\alpha} (p' - p)^{\alpha} $. 
	(Note that we neglect the small difference
	between the proton and neutron mass, so $m$ is the average nucleon mass.)
	For selected observables we compare, in Sec.~IV, results based on a few recent models 
	of the electromagnetic nucleon form factors \cite{BBA03,Kelly04,Lomon02,PRC86.035503}.

	Formulas for nuclear current matrix elements not only in elastic reactions 
	but also in deuteron breakup processes follow from Eq.~(\ref{1N+2N}),
	the Poincar\'e Clebsch-Gordan coefficients (\ref{b.4}) and the deuteron (\ref{D_PWD})
	or the two-nucleon scattering wave function. More details can be found in the appendix.

	We now return to Eq.~(\ref{qft18}) and calculate the cross section, 
	first for elastic electron-deuteron scattering in the laboratory frame.
	We define the energy transfer $\omega \equiv E_e - E_e'$ and set the three-momentum transfer 
	$  \ve{q} \equiv  \ve{p}_e - \ve{p}_e' $
	to be parallel to the $z$-axis. In this frame thus
	\begin{eqnarray}
	{p_e}_x & = & | \ve{p}_e|  | \ve{p}_e'| \sin\theta_e / | \ve{q} | \, , \nonumber \\
	{p_e}_y & = & 0 \, , \nonumber \\
	{p_e}_z & = & | \ve{p}_e| \left( | \ve{p}_e| - | \ve{p}_e'| \cos\theta_e  \,  \right) /  | \ve{q} | \, , \nonumber \\
	{p_e'}_x & = &  {p_e}_x  \, , \nonumber \\
	{p_e'}_y & = & 0 \, , \nonumber \\
	{p_e'}_z & = & | \ve{p}_e'| \left( -| \ve{p}_e'| + | \ve{p}_e| \cos\theta_e \,  \right) /  | \ve{q} | 
	\, , \nonumber \\
	\ve{p}_D & = & 0 \, , \nonumber \\
	\ve{p}_D' & = & \ve{q} \, .
	\label{labkinematics}
	\end{eqnarray}
	Further steps are standard.
	The electron mass $m_e$ is neglected, 
	nuclear current conservation
	\begin{eqnarray}
		\omega \, N_{eD}^0 = \ve{q} \cdot \ve{N}_{eD}
	\end{eqnarray}
	is used to express $N_{eD,z} \equiv N_{eD}^3$ in terms of $N_{eD}^0$,
	\begin{eqnarray}
	N_{eD,z} = \frac{ \omega  }{ | \ve{q} |} \, N_{eD}^0 \, ,
	\end{eqnarray}
	some factors are used to build the Mott cross section
	\begin{eqnarray}
	\sigma_{Mott} = \frac { \alpha^2 \cos^2\theta_e} {4 | \ve{p}_e |^2 \, \sin^4 \frac{\theta_e}{2}} \, ,
	\end{eqnarray}
	the phase space factor $\rho_{elas}$ yields
	\begin{eqnarray}
	\rho_{elas} \equiv \int \de  \ve{p}_e' \, \int \de \ve{p}_D' \, \delta^4 (p_D+p_e - p_D'-p_e') 
	= \int  {\de \mathbf{\hat p}_e' }  \,
		\frac{ {E_D'}| \ve{p}_e' |^3 }{ m_D  |\ve{p}_e| } \, 
	\end{eqnarray}
	and finally, the contraction of the electron and nuclear matrix elements 
	\begin{eqnarray}
	\left(L_{e,\alpha} \left( \ve{p}_e',\mu_e', \ve{p}_e,\mu_e \right) \, \right)^\ast \, 
	L_{e,\beta} \left( \ve{p}_e',\mu_e', \ve{p}_e,\mu_e \right)
		\, \left( N_{eD}^{\alpha} \right)^\ast N_{eD}^\beta \,
	\end{eqnarray}
	is evaluated~\cite{DR1986}. 
	The latter can be done under various assumptions. In the simplest case 
	the initial electron is unpolarized and polarization of the final electron is 
	not measured. Then we average over $\mu_e$ and sum over $\mu_e'$.
	Alternatively, the initial 
	electron can have a definite helicity $h= 1$ or $h=-1$, 
	while we still sum over $\mu_e'$. 
In such a case the two-component spinor $\chi_{{\mu}_e}$ in (\ref{qft22-1}) and (\ref{qft22}) 
is chosen to fulfill
\[
	{\ve{p}_e \cdot \bm{\sigma}} \chi_{{\mu}_e} = h |\ve{p}_e|  \chi_{{\mu}_e} \, .
\]
	The deuteron spin quantum numbers 
	(not shown here for the sake of brevity) can be still chosen at will.
	However, the unpolarized laboratory frame differential cross section is studied most often (see for example \cite{Coester75,Huang09,Filin21}).
	It contains the structure functions $A(Q^2)$ and $B(Q^2)$, which experimentally are obtained through 
	a Rosenbluth separation~\cite{Ros}:
	\begin{eqnarray}
	\frac{ \de\sigma}{\de \mathbf{\hat p}_e' } ( Q^2, \theta_e )  
	= \sigma_{Mott} \, \left( A(Q^2) + B(Q^2) \tan^2 \left( \frac{\theta_e}{2} \right) \, \right) \,
	\frac{ | \ve{p}_e' | }{ |\ve{p}_e| } \, ,
	\label{sigmaunpo}
	\end{eqnarray}
	where $Q^2$ is now the square of the four momentum transferred to the deuteron.
	The frequently considered polarization observable is the deuteron tensor 
	analyzing power $T_{20} (Q^2 , \theta_e)$ at $\theta_e$= 70$^\circ$~\cite{Haftel80},
	which is obtained from the cross sections for the unpolarized initial and final electrons
	but for the initial deuterons with canonical spin
	polarizations $\mu_D$= 1 and $\mu_D$= 0:
	\begin{eqnarray}
	T_{20} (Q^2 , \theta_e) =
	\frac{\sqrt{2} \, \left( 	\frac{ \de\sigma}{\de \mathbf{\hat p}_e' } ( Q^2, \theta_e ; \mu_D=1 )  
		- \frac{ \de\sigma}{\de \mathbf{\hat p}_e' } ( Q^2, \theta_e ; \mu_D=0 ) \right) }
	{
		\frac{ \de\sigma}{\de \mathbf{\hat p}_e' } ( Q^2, \theta_e ) } \, .
	\label{T20}
	\end{eqnarray}
	
\vspace{1cm}

	The kinematics of the neutral-current (NC) driven $\nu + d \rightarrow \nu' + d'$ reaction
	is the same as in elastic electron-deuteron scattering, provided the electron mass is neglected.
	We will describe this reaction using the approximate ``current-current'' theory, which allows us
	to employ the same formalism as before for electron scattering.
	
	The transition matrix element for this process is
	\begin{eqnarray}
		\langle \ve{p}_D',\mu_D',D, \ve{p}_\nu', \mu_\nu' \Vert T_{\rm WNC} \Vert 
		\ve{p}_D, \mu_D,D, \ve{p}_\nu ,\mu_\nu \rangle = \nonumber \\
	- \frac{G_F}{\sqrt{2}} (2 \pi)^3
		\langle \ve{p}_D', \mu_D' ,D \vert J_{\rm WNC}^{\alpha} (0) \vert \Phi_{D} \vert
	\ve{p}_D,\mu_D ,D \rangle  
	{g_{\alpha \beta}}
	\langle \ve{p}_\nu', \mu_\nu'  \vert J_\nu^{\beta} (0) \vert \ve{p}_\nu,\mu_\nu \rangle \, ,
	\label{Ted_elastic}
	\end{eqnarray} 
	where $G_F$ is the Fermi constant.
	The neutrino current matrix element is written as
	\begin{eqnarray}
	\langle \ve{p}_\nu', \mu_\nu'  \vert J_\nu^{\beta} (0) \vert \ve{p}_\nu,\mu_\beta \rangle 
	\equiv {1 \over (2\pi)^3} \frac1{\sqrt{4 |\ve{p}_\nu| |\ve{p}_\nu'|}} \, 
	L_\nu^{\beta} \left( \ve{p}_\nu',\mu_\nu', \ve{p}_\nu,\mu_\nu \right) \, ,
	\end{eqnarray}
	with 
	\begin{equation}
	L_\nu^{\beta} \left( \ve{p}_\nu',\mu_\nu', \ve{p}_\nu,\mu_\nu \right)
	=\bar{u}_\nu(\ve{p}_\nu',\mu_\nu')\gamma^{\beta}\left(1-\gamma_5\right) u_\nu(\ve{p}_\nu,\mu_\nu)
	\end{equation}
	and the Dirac spinors for massless neutrinos defined as
	\begin{equation}
	u_\nu(\ve{p}_\nu,\mu_\nu)=
	\sqrt{|\ve{p}_\nu|}
	\left(
	\begin{matrix}
	\chi_{\mu_\nu}\\
	\frac{\ve{p}_\nu \cdot \bm{\sigma}}{|\ve{p}_\nu|}\chi_{\mu_\nu}
	\end{matrix}
	\right) \, .
	\end{equation}
	Also for weak reactions we include in the nuclear matrix elements
	\[
		\langle \ve{p}_D', \mu_D' ,D \vert J_{\rm WNC}^{\alpha} (0)  \vert \ve{p}_D,\mu_D ,D \rangle \, \equiv \,
	{1 \over (2\pi)^3}\, 
	N_{\nu D}^{\alpha} \left(
	\ve{p}_D', \mu_D' , \ve{p}_D , \mu_D \,
	\right) \, ,
	\]
	only the single-nucleon contributions:
	\begin{eqnarray}
		\matrixelement{\ve{p}',\mu',\tau'}{J_{k,{\rm WNC}}^\mu(0)}{\ve{p},\mu , \tau} \nonumber \\
	= \delta_{ \tau' \tau } \,
	\bar u(\ve{p}',\mu')
	\left(
	F_{1, \tau}^N(Q^2) \gamma^\mu
	+ \frac{i}{2m}\sigma^{\mu\nu}q_\nu F_{2, \tau}^N(Q^2) \right. \nonumber \\
	\left. 
	+F_{A, \tau}^N(Q^2) \gamma^\mu \gamma_5 
	+  \frac{q^\mu}{m}\gamma_5F_{P , \tau}^N(Q^2)
	\right)
	u(\ve{p},\mu)\, ,
	\label{j1wnc}
	\end{eqnarray}
	where $q^\mu=p'^\mu-p^\mu$ and the weak neutral-current nucleon form factors $F_{i,\tau}^N$ depend on the nucleon isospin.
	For these quantities we use the parametrizations from Refs.~\cite{PRC86.035503,BBA03}.
	The part with $F_{P, \tau}^N$ gives no contribution in Eq.~(\ref{j1wnc}) in the case of massless neutrinos
	but we keep it, since the single nucleon charged current has the same functional form, with a different 
	isospin dependence.
	
	The steps leading to the cross section are standard.
	We denote in particular 
	\begin{equation}
	\widetilde{L}_{\alpha \beta} \left( \ve{p}_\nu' , \ve{p}_\nu \right) \equiv
	\sum_{\mu_\nu \mu_\nu'}
	\left(	L_{\nu, \alpha} \left( \ve{p}_\nu',\mu_\nu', \ve{p}_\nu,\mu_\nu \right)\right)^\ast \,
	L_{\nu, \beta} \left( \ve{p}_\nu',\mu_\nu', \ve{p}_\nu,\mu_\nu \right)
	\end{equation}
	and obtain an intermediate result for the unpolarized case
	\begin{eqnarray}
	\de\sigma =
	\frac{G_F^2}{32\pi^2} \frac{{E_D}}{|\ve{p}_\nu'|}
	\frac1{|p_D\cdot p_\nu| }
	\widetilde{L}_{\alpha \beta} \left( \ve{p}_\nu' , \ve{p}_\nu \right) 
	\nonumber \\
		\times \, 
	\frac13 \sum_{ \mu_D' \mu_D }
	\left( N^{\alpha} \left( \ve{p}_D', \mu_D' , \ve{p}_D , \mu_D \, \right) \right)^\ast 
	N^{\beta} \left( \ve{p}_D', \mu_D' , \ve{p}_D , \mu_D \, \right)
	\nonumber \\
		\times \, 
	\delta^4 (p_D+p_\nu - p_D'-p_\nu') \de\ve{p}_D' 
	\de \ve{p}_\nu' \, .
	\label{dsigma_inter}
	\end{eqnarray}
	This result can be used together with the laboratory frame kinematics (\ref{labkinematics})
	or to calculate the differential cross section in the total momentum zero frame (``c.m.''), where
	\begin{eqnarray}
	{p_\nu}_x & = & w \cos\left( {\theta_{c.m.}}/2 \right) \, , \nonumber \\
	{p_\nu}_y & = & 0 \, , \nonumber \\
	{p_\nu}_z & = & w \sin\left( {\theta_{c.m.}}/2 \right) \, , \nonumber \\
	{p_\nu'}_x & = &  {p_\nu}_x  \, , \nonumber \\
	{p_\nu'}_y & = & 0 \, , \nonumber \\
	{p_\nu'}_z & = & -{p_\nu}_z  \, , \nonumber \\
	\ve{p}_D & = & -\ve{p}_\nu \, , \nonumber \\
	\ve{p}_D' & = &  -\ve{p}_\nu' 
	\label{cmkinematics}
	\end{eqnarray}
	and $w = |\ve{p}_D| = |\ve{p}_\nu| = |\ve{p}_D'| = |\ve{p}_\nu'|$.
	In the ``c.m.'' frame the differential cross section can be written as
	\begin{equation}
	\left. \frac{\de{\sigma}}{\de{\versor{p}_\nu'}} \right|_{c.m.}
	=
	\frac{G_F^2}{32\pi^2} \frac{\energy{D^2}{\ve{p}_D}{}}{s}
		\widetilde{L}_{\alpha\beta} \, \frac13 \sum_{ \mu_D' \mu_D } N_{\nu D}^{*\alpha} N_{\nu D}^\beta\,,
	\label{eq:nuDelasticCMcross}
	\end{equation}
	with $s \equiv (p_\nu + p_D)^2 = m_D \left( m_D + 2 |\ve{p}_{\nu, {lab}}| \, \right) $
	and $w = \frac{s - m_D^2}{ 2 \sqrt{s}} $.
	The corresponding expression in the laboratory frame reads
	\begin{equation}
	\left. \frac{\de{\sigma}}{\de{\versor{p}_\nu'}} \right|_{lab}
	=
	\frac{G_F^2}{32\pi^2}
	\frac{|\ve{p}_\nu'|^2}{|\ve{p}_\nu|^2}
	\frac{{E_D'}}{m_D}
		\widetilde{L}_{\alpha\beta} \, \frac13 \sum_{ \mu_D' \mu_D } N_{\nu D}^{*\alpha} N_{\nu D}^\beta\,.
	\label{eq:nuDelasticLABcross}
	\end{equation}
	In both frames we can use the analytical result for $\widetilde{L}_{\alpha\beta}$, 
	\begin{eqnarray}
	\widetilde{L}_{\alpha\beta}  = 
	8 \left(
	{p_\nu}_\alpha {p_\nu'}_\beta  +  
	{p_\nu'}_\alpha {p_\nu}_\beta - 
	( p_\nu \cdot p_\nu' ) g_{\alpha \beta} + 
	i \epsilon_{\alpha \beta \rho \sigma } {p_\nu}^\rho {p_\nu'}^\sigma \, 
	\right) \, ,
	\end{eqnarray}
	to evaluate the contraction $ \widetilde{L}_{\alpha\beta} \, \frac13 \sum_{ \mu_D' \mu_D } N_{\nu D}^{*\alpha} N_{\nu D}^\beta$.
	Here $ \epsilon_{\alpha \beta \rho \sigma } $ is the totally antisymmetric Levi-Civita symbol 
	with $\epsilon_{0123} = 1$.
	Note that for the choices of kinematics (\ref{labkinematics}) or (\ref{cmkinematics}) some of the terms 
	in this sum are identically zero. 
	
	Finally, we evaluate the total elastic cross section $\sigma^{el}$
	\begin{equation}
	\sigma^{el} = \int \de{\versor{p}_\nu'} \frac{\de{\sigma}}{\de{\ve{p}_\nu'}} \equiv 
	\int_0^{2\pi} \de \phi' \int_0^{\pi} \de\theta' \sin\theta' \, \frac{\de{\sigma}}{\de{\ve{p}_\nu'}}=
	2\pi \int_0^{\pi} \de\theta' \sin\theta' \, \left. \frac{\de{\sigma}}{\de{\ve{p}_\nu'}} \right|_{\phi' = 0} \,,
	\end{equation}
	where $\theta'$ and $\phi'$ are the polar and azimuthal angles corresponding to $\ve{p}_\nu'$.
	A comparison of results for $\sigma^{el}$ obtained in the ``c.m.'' and laboratory frames 
	can be used to test the relativistic character of the calculations. While in principle they should be the same, 
	current covariance requires two-body currents which are not treated in these calculations. 
	The difference in the cross section calculations in different frames provides a measure of the impact 
	of the two-body currents.

        In our nonrelativistic calculations we use nonrelativistic kinematics and a nonrelativistic 
	form of the current operator, which can be derived from Eq.~(\ref{j1wnc}). In addition to the strict nonrelativistic
	limit one can retain $(p/m)^2$ corrections 
	stemming from a $p/m$ expansion of the relativistic current operator. The density part is then 
	\begin{multline}
		\matrixelement{\ve{p}',\mu',\tau'}{J_{k, {\rm NR-WNC}}^{0}(0)}{\ve{p},\mu , \tau} \\=
	\delta_{\tau'\tau}
	\chi^\dagger_{\mu'}\bigg[F_{1, \tau}^N
	-(F_{1, \tau}^N+2F_{2, \tau}^N)\frac{(\ve{p}'-\ve{p})^2}{8m^2}+
	(F_{1, \tau}^N+2F_{2, \tau}^N)i\,\frac{(\ve{p}'\times\ve{p})\cdot\bm{\sigma}}{4m^2}\\
	+F_{A, \tau}^N\frac{\bm{\sigma}\cdot(\ve{p}+\ve{p}')}{2m}
	+F_{P, \tau}^N\frac{(\ve{p}'^2-\ve{p}^2)}{4m^2} \frac{\bm{\sigma}\cdot(\ve{p}'-\ve{p})}{m}
	\bigg]\chi_\mu\,.\qquad
	\end{multline}
	and the vector part becomes
	\begin{multline}
		\matrixelement{\ve{p}',\mu',\tau'}{\ve{J}_{k, {\rm NR-WNC}}(0)}{\ve{p},\mu , \tau}=\\
	\delta_{\tau'\tau}
	\chi^\dagger_{\mu'}\bigg\{
	F_{1, \tau}^N\frac{\ve{p}'+\ve{p}}{2m}
	-\frac{1}{2m}(F_{1, \tau}^N+F_{2, \tau}^N)i\,\bm{\sigma} \times (\ve{p}'-\ve{p})\\
	+ F_{A, \tau}^N \left[1-\frac{(\ve{p}'+\ve{p})^2}{8m^2}\right] \bm{\sigma}
	+\frac{F_{A, \tau}^N}{4m^2}\bigg[(\ve{p}\cdot \bm{\sigma})\ve{p}'
	+(\ve{p}'\cdot \bm{\sigma})\ve{p}\\
	+i(\ve{p}\times\ve{p}')
	\bigg]
	+F_{P, \tau}^N\frac{(\ve{p}'-\ve{p})}{m}\frac{\bm{\sigma}\cdot(\ve{p}-\ve{p}')}{2m}
	\bigg\}\chi_\mu\,.\qquad
	\end{multline}
	Note that the above formula can be also used (with the proper isospin operators and nucleon 
	form factors) for electron scattering processes as well as for the charged-current induced reactions 
	addressed in the next subsection.
	
\subsection{Kinematics, matrix elements and observables for electron or neutrino induced deuteron breakup}
	
\begin{figure}[hb!]
		\includegraphics[width=9cm]{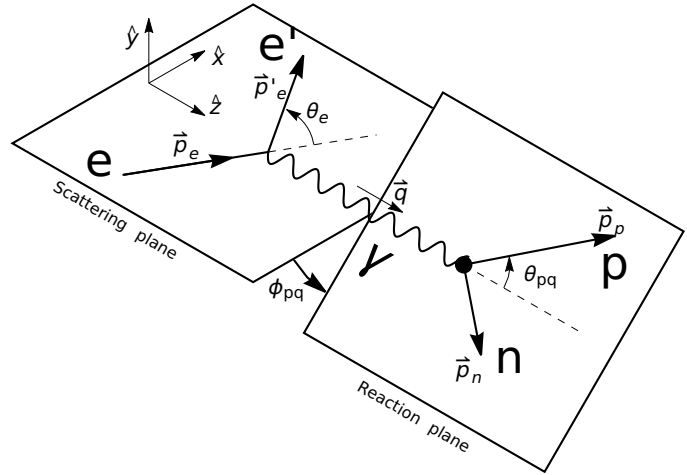}
		\caption{
			Kinematics of the $(e,e^{\prime},p)$ reaction with the definitions 
			of the kinematic variables~\cite{asymmetry}.
		}
		\label{fig:kinematics_diagram}
	\end{figure}

	
	\noindent
	The four-momentum conservation 
	for the $ e + d \rightarrow e' + p + n $ reaction 
	in the laboratory frame reads 
	\begin{eqnarray}
	\energyel{e}{\ve{p}_e}{} + m_D & = & 
	\energyel{e}{\ve{p}_e}{'} + \energy{p}{\ve{p}_p}{} +
	\energy{n}{\ve{p}_n}{} \nonumber \\
	\ve{p}_e  =  
		\ve{p}_e' + \ve{p}_p + \ve{p}_n & \equiv &
	\ve{p}_e' + \ve{p}_1' + \ve{p}_2' 
		\equiv 
	\ve{p}_e' + \ve{q} \, ,
	\end{eqnarray}
	where $ \energy{i}{\ve{p}}{} \equiv \sqrt{ m_i^2 + \ve{p}^2 \, } $.
	A diagram representing kinematics of this process is shown in Fig.~\ref{fig:kinematics_diagram}.
	
	The phase space for the breakup reaction can be naturally 
	described with the following variables:
	\begin{enumerate}
		\item 
		electron scattering angle $\theta_e$
		\item
		energy of the outgoing electron, which for given $\theta_e$ lies in the 
		interval $ m_e \le \energyel{e}{\ve{p}_e}{'} \le 
		{E_e'}^{\rm max}$.
		The maximal outgoing electron energy ${E_e'}^{\rm max}$
		can be obtained from the condition that the total energy 
		of the $(p,n)$ system in the total-momentum-zero frame
		is not less than $m_p+m_n$:
		\begin{eqnarray}
		\left( \energy{p}{\ve{p}_p}{} + \energy{n}{\ve{p}_n}{} \, \right)^2 - 
		\left( \ve{p}_p + \ve{p}_n \, \right)^2 \ge 
		\left( m_p + m_n \, \right)^2 \, ,
		\end{eqnarray}
		which means 
		\begin{eqnarray}
		\left( \energyel{e}{\ve{p}_e}{} + m_D - \energyel{e}{\ve{p}_e}{'} \, \right)^2 - \left( \ve{p}_e - \ve{p}_e' \, \right)^2 \ge
		\left( m_p + m_n \, \right)^2 \, .
		\label{max1}
		\end{eqnarray}
		Inequality (\ref{max1}) can be analytically solved with the result 
		\begin{eqnarray}
		{| \ve{p}_e' |}^{\rm max} = 
			\frac{ 
			G {| \ve{p}_e | } \cos\theta_e }
			{2 \left( \energyel{e}{\ve{p}_e}{} + m_D \right)^2 -
			2 {| \ve{p}_e | }^2 \cos^2 \theta_e } 
\nonumber \\
	+  \frac{ \left( \energyel{e}{\ve{p}_e}{} + m_D  \right) \, \sqrt{4 m_e^2 
				{| \ve{p}_e | }^2
				\cos^2 \theta_e  + G^2 - 4 \left( \energyel{e}{\ve{p}_e}{} + m_D \right)^2 m_e^2} }
			{2 \left( \energyel{e}{\ve{p}_e}{} + m_D \right)^2 -
			2 {| \ve{p}_e | }^2 \cos^2 \theta_e } 
		\label{pepmax}
		\end{eqnarray}
		and
		\begin{eqnarray}
	{E_e'}^{\rm max} =
		\sqrt{ m_e^2 + \left({| \ve{p}_e' |}^{\rm max} \right)^{\, 2} \, } \, ,
		\end{eqnarray}
		where 
		\[
		G= 2 m_e^2 + m_D^2 - (m_p + m_n)^2 + 2 m_D \, \energyel{e}{\ve{p}_e}{} \, .
		\]
		The three quantities ${| \ve{p}_e |}$, $\theta_e$ and 
		${| \ve{p}_e' |}$ define the total total energy and total three-momentum
		of the $(p,n)$ system and thus its internal energy 
		\[
E_{c.m.} = \sqrt{ \left( \omega + m_D \, \right)^2  \, - \, {\ve  q}^{ 2} \, } - 2 m \, ,
\]
		so we can calculate also the magnitude of the 
		relative momentum ${| \ve{k} |}$ in the two-nucleon total-momentum-zero frame. 
		Two additional variables, which fully determine the exclusive kinematics might be 
		\item
		polar ($0 \le \theta_k \equiv \theta_{pq} \le \pi$) and
		\item 
		azimuthal ($0 \le \phi_k \equiv \phi_{pq} \le 2 \pi$) 
		angle of $\ve{k}$ measured with respect to $\ve{q}$.
	\end{enumerate}
	Obviously all the four choices can be made for any 
	azimuthal angle of the outgoing electron momentum
	($0 \le \phi_e \le 2 \pi$).
	
	\vspace{1cm}
	
	\noindent
	For fixed ${| \ve{p}_e |}$, $\theta_e$,
	${| \ve{p}_e' |}$ and 
	$ \ve{k} \equiv |\ve{k} | \, \left( \sin\theta_k \, \cos \phi_k, 
	\sin\theta_k \, \sin \phi_k, \cos\theta_k \, \right) $
	we get
	\begin{eqnarray}
	\ve{p}_p & = & \ve{k} + 
	\left( \frac {  \energy{p}{\ve{k}}{} }{M_0} + 
	\frac1{M_0 ( M_0 +E_0) } \ve{k} \cdot \ve{q} \, \right)\, \ve{q} \, ,
	\nonumber \\
	\ve{p}_n & = & -\ve{k} + 
	\left( \frac {  \energy{n}{\ve{k}}{} }{M_0} - 
	\frac1{M_0 ( M_0 +E_0) } \ve{k} \cdot \ve{q} \, \right)\, \ve{q}
	\, ,
	\label{pppn}
	\end{eqnarray}
	where $ M_0 = \energy{p}{\ve{k}}{} + \energy{n}{\ve{k}}{} $
	and $E_0 = \sqrt{ M_0^2  + | \ve{q} |^2 }$~\cite{PRC77.034004}.
	
	\noindent
	Very often variables of interest are 
	\begin{eqnarray}
	Q^2 = | \ve{q} |^2 - 
	\left( \energyel{e}{\ve{p}_e}{} - \energyel{e}{\ve{p}_e}{'} \, \right)^2 
	\end{eqnarray}
	and
	\begin{eqnarray}
	p_{\rm miss} = \sqrt { \left (  \energy{n}{\ve{p}_n}{} \right)^2 - m_n^2 }  \, .
	\end{eqnarray}
	In the practical calculations we will use approximations: we will neglect not
	only the neutrino but also the electron mass $m_e$ and the difference
	between the proton and neutron masses, $m_p = m_n = m$.
	Then Eqs.~(\ref{pppn}) coincide with Eqs.~(\ref{p1p2fromkp}).
	
	The transition matrix elements for the breakup reaction are 
	\begin{eqnarray}
		\langle \ve{p}_1',\mu_1', \tau_1' , \ve{p}_2',\mu_2', \tau_2' , \ve{p}_e', \mu_e' \Vert T_{\rm epn}
	\Vert \ve{p}_D, \mu_D,D, \ve{p}_e ,\mu_e \rangle \nonumber \\
	= -e^2 (2 \pi)^3\,
	{{}^{(-)}\langle} \ve{p}_1', \mu_1' , \tau_1' , \ve{p}_2', \mu_2' , \tau_2' \vert J_{nuc,EM}^{\mu} (0) \vert \ve{p}_D,\mu_D ,D \rangle \nonumber \\
		\times \,
	{g_{\mu \nu}  \over (p_e'-p_e)^2 +i \epsilon}
	\langle \ve{p}_e', \mu_e'  \vert J_e^{\nu} (0) \vert \ve{p}_e,\mu_e \rangle \nonumber \\
	\equiv  -e^2 (2 \pi)^3\,
		{1 \over (2\pi)^3} \, N_{\rm epn}^{\mu} \left(  \ve{p}_1', \mu_1' , \tau_1' , \ve{p}_2', \mu_2' , \tau_2' , \ve{p}_D , \mu_D \, \right) \nonumber \\
		\times \,
	{g_{\mu \nu}  \over (p_e'-p_e)^2 +i \epsilon}
	{1 \over (2\pi)^3} \, \frac1{\sqrt{4 \, \energyel{e}{\ve{p}_e}{} \, \energyel{e}{\ve{p}_e}{'}}} 
	L_e^{\nu} \left( \ve{p}_e',\mu_e', \ve{p}_e,\mu_e \right) \, ,
	\label{qft172N}
	\end{eqnarray}
	where the final two-nucleon bound state $ \vert \ve{p}_D' , \mu_D' , D \rangle$
	is replaced by the corresponding scattering state 
	$ \vert  \ve{p}_1', \mu_1' , \tau_1' , \ve{p}_2', \mu_2' , \tau_2' {\rangle}^{(-)} $,
	which is calculated as 
	\begin{eqnarray}
	{{}^{(-)}\langle} \ve{p}_1', \mu_1' , \tau_1' , \ve{p}_2', \mu_2' , \tau_2' \vert J_{nuc,EM}^{\mu} (0) \vert \ve{p}_D,\mu_D ,D \rangle \nonumber \\
	=
	\langle \ve{p}_1', \mu_1' , \tau_1' , \ve{p}_2', \mu_2' , \tau_2' \vert J_{nuc,EM}^{\mu} (0) \vert \ve{p}_D,\mu_D ,D \rangle \label{planewave}
	\\
	+ 
	\langle \ve{p}_1', \mu_1' , \tau_1' , \ve{p}_2', \mu_2' , \tau_2' \vert 
	t( E + i \epsilon) 
	G_0 ( E + i \epsilon) 
	J_{nuc,EM}^{\mu} (0) \vert \ve{p}_D,\mu_D ,D \rangle \label{rescatt} \, ,
	\end{eqnarray}
	where $E = \energyel{e}{\ve{p}_e}{} + m_D - \energyel{e}{\ve{p}_e}{'}$,
	$G_0 (E)$ is the relativistic free two-nucleon propagator 
	and $t(E)$ is the ``boosted'' $t$-matrix, which obeys 
\begin{eqnarray}
t(E;|\ve{q}|) = v(|\ve{q}|) + t(E;|\ve{q}|) \, G_0 ( E + i \epsilon) v(|\ve{q}|) \, ,
\end{eqnarray}
where 
$ v(|\ve{q}|) = \sqrt{ 4 \left( |\ve{k}|^2 + m^2 + 2 m V  \right) + |\ve{q}|^2 } 
   -  \sqrt{ 4 \left( |\ve{k}|^2 + m^2 \right) + |\ve{q}|^2 } $ 
is the ``boosted'' potential \cite{KAMADA2007119}.
	For the single-nucleon contribution in the nuclear matrix element 
	the plane-wave part (\ref{planewave}) is given in Eq.~(\ref{termB}).
	Various types of the rescattering parts of the nuclear matrix elements are evaluated in the appendix.

	The generic formulas for any type of the cross section in terms of (\ref{qft172N}) is 
	\begin{eqnarray}
	\de\sigma =
	{(2 \pi)^4 {E_D} \energyel{e}{\ve{p}_e}{}\over  \sqrt{(p_D\cdot p_e)^2- m_D^2m_e^2}} 
\vert \langle \ve{p}_1', \mu_1', \tau_1', \ve{p}_2', \mu_2', \tau_2', \ve{p}_e', \mu_e' \Vert T_{\rm epn} \Vert \ve{p}_D, \mu_D,D, \ve{p}_e ,\mu_e \rangle \vert^2 \nonumber \\
	\times \, \energy{{}}{\ve{p}_1}{'}
	\energy{{}}{\ve{p}_2}{'}
	\energyel{e}{\ve{p}_e}{'} \,
	\delta^4 (p_D+p_e - p_1'-p_2'-p_e') 
	{\de\ve{p}_1' \over \energy{{}}{\ve{p}_1}{'}}
	{\de\ve{p}_2' \over \energy{{}}{\ve{p}_2}{'}}
	{\de \ve{p}_e' \over \energyel{e}{\ve{p}_e}{'}} \, .
	\label{qft182N}
	\end{eqnarray}
	We take essentially the same steps as for elastic electron-deuteron scattering cross section
	to obtain in the laboratory frame \cite{DR1986,Rep05}
	\begin{eqnarray}
	\frac{ \de^5\sigma}{\de \mathbf{\hat p}_e' \de|\ve{p}_e'| \de{\versor{p}}_1' } 
	= \sigma_{Mott} \, 
	\sum_{{\rm physical}\ |\ve{p}_1'|} \, 
	\frac{|\ve{p}_1'|^2 }
	{\left| \frac{|\ve{p}_1'|}{\energy{{}}{\ve{p}_1}{'}} + \frac{|\ve{p}_1'| - |\ve{q}|\cos\theta_1}{\energy{{}}{\ve{p}_2}{'}} \right|} 
	\nonumber \\
	\times \, \Big(
	v_L R_L + v_T R_T + v_{TT} R_{TT} + v_{TL} R_{TL}
	\, + \, 
	h \left( v_{T '} R_{T '} + v_{TL '} R_{TL '} 
	\, \right) \, \Big) \, ,
	\label{sigmaEMBR}
	\end{eqnarray}
	where we assume that polarization of the final electron is not measured 
	and that the spin of the initial electron is parallel ($h=1$) 
	or antiparallel ($h=-1$) to its momentum.
	Note that $\theta_1$ is the angle between $\ve{p}_1'$ and $\ve{q}$. 
	The $v_i$ functions are given in terms of
	the four momentum transfer squared $Q^2$ (positive),
	three momentum transfer squared $|\ve{q}|^2$,
	and the electron scattering angle $\theta_e$
	\begin{eqnarray}
	v_L = \frac{(Q^2)^2}{|{\mathbf q}|^4} \, , \nonumber \\
	v_T = \frac12 \frac{Q^2}{|{\mathbf q}|^2} + \tan^2 \frac{\theta_e}2  \, ,\nonumber \\
	v_{TT} = -\frac12 \frac{Q^2}{|{\mathbf q}|^2}  \, ,\nonumber \\
	v_{TL} = -\frac1{\sqrt{2}}\, \frac{Q^2}{{|\mathbf q|}^{2}} \,
	\sqrt{\frac{Q^2}{|{\mathbf q}|^2} + \tan^2 \frac{\theta_e}2 }  \, ,\nonumber \\
	v_{T '} = \sqrt{ \frac{Q^2}{|{\mathbf q}|^2} + \tan^2 \frac{\theta_e}2 } \, \tan \frac{\theta_e}2 \, ,
	\nonumber \\
	v_{TL '} = -\frac1{\sqrt{2}} \frac{Q^2}{|{\mathbf q}|^2} \, \tan \frac{\theta_e}2  .
	\label{obse15}
	\end{eqnarray}
	The nuclear response functions $R_i$ are
	\begin{eqnarray}
		R_L = \mid N_{\rm epn}^0 \mid ^2  \, , \nonumber \\
		R_T = \mid N_{{\rm epn},+1} \mid ^2  + \mid N_{{\rm epn},-1} \mid ^2  \, , \nonumber \\
		R_{TT} = 2 \Re \left( N_{{\rm epn},+1} N_{{\rm epn},-1}^{\, *} \right) \, , \nonumber \\
		R_{TL} = -2 \Re \left( N_{\rm epn}^0 ( N_{{\rm epn},+1} - N_{{\rm epn},-1} )^{\, *}  \right) \, , \nonumber \\
		R_{T '} =  \mid N_{{\rm epn},+1} \mid ^2  - \mid N_{{\rm epn},-1} \mid ^2  \, , \nonumber \\
		R_{TL '} = -2 \Re \left( N_{\rm epn}^0 ( N_{{\rm epn},+1} + N_{{\rm epn},-1} )^{\, *}  \right) \, ,
	\label{obse16}
	\end{eqnarray}
	where 
\begin{eqnarray*}
	N_{{\rm epn},+1}  =  -\frac1{\sqrt{2}} \left( N_{\rm epn}^1 + i N_{\rm epn}^2 \right) 
 	   \equiv  -\frac1{\sqrt{2}} \left( N_{{\rm epn},x} + i N_{{\rm epn},y} \right) \, ,  \\ 
 	N_{{\rm epn},-1}  =  \frac1{\sqrt{2}} \left( N_{\rm epn}^1 - i N_{\rm epn}^2 \right) 
 	   \equiv   \frac1{\sqrt{2}} \left( N_{{\rm epn},x} - i N_{{\rm epn},y} \right) 
\end{eqnarray*}
are spherical components of $\ve{N}_{\rm epn}$.
	The two-nucleon scattering states 
	in $N_{\rm epn}^0$, $N_{{\rm epn},+1}$ and $N_{{\rm epn},-1}$ 
	are antisymmetrized and polarizations
	of the hadronic states in Eq.~(\ref{sigmaEMBR}) can be still chosen at will. 
	Equation (\ref{sigmaEMBR}) is a starting point for 
	defining target-spin independent or target-spin dependent helicity asymmetries
	as well as the various target analyzing powers.
	
	The choice of individual momentum in Eq.~(\ref{sigmaEMBR}) is not always
	convenient and often (see for example \cite{Arenhovel05}) another 
	form of the phase space factor for fixed $\ve{p}_e'$ is used:
	\begin{eqnarray}
	\rho_2 \equiv 
	\de\ve{p}_1' \, 
	\de\ve{p}_2' \,
	\delta^4 (p_D+p_e - p_e' - p_1'-p_2') 
	\nonumber \\
	= 
	\de\ve{p}' \, 
	\de\ve{k}' \,
	\left|
	\frac{\partial \left( \ve{p}_1' , \ve{p}_2' \,  \right) }
	{\partial \left( \ve{p}' , \ve{k}' \,  \right) }
	\right| \,
	\delta^4 (p_D+p_e - p_e' - p_1'-p_2') 
	\nonumber \\
	=
	\de\ve{p}' \, 
		\de\ve{k}' \, {\cal N}^2(\ve{p}_1' , \ve{p}_2' ) \, 
	\delta^4 (p_D+p_e - p_e' - p_1'-p_2') \nonumber \\
	= \de\mathbf{\hat k}' \, \frac14 \, E \, |\ve{k}'|_{\rm physical} \,  
	{\cal N}^2( \ve{p}_1' , \ve{p}_2' ) \, , 
	\end{eqnarray}
	which leads to to 
	\begin{eqnarray}
	\frac{ \de^5\sigma}{\de \mathbf{\hat p}_e' \de|\ve{p}_e'| \de\mathbf{\hat k}' } 
	= \sigma_{Mott} \, 
		\frac14 \, E \, |\ve{k}'|_{\rm physical} \,  {\cal N}^2( \ve{p}_1' , \ve{p}_2' ) \, 
	\nonumber \\
	\times \, \Big(
	v_L R_L + v_T R_T + v_{TT} R_{TT} + v_{TL} R_{TL}
	\, + \, 
	h \left( v_{T '} R_{T '} + v_{TL '} R_{TL '} 
	\, \right) \, \Big) \, .
	\label{sigmaEMBR2}
	\end{eqnarray}
	Further, we calculate semi-exclusive cross sections 
	\begin{eqnarray}
	\frac{ \de^3\sigma}{\de \mathbf{\hat p}_e' \de|\ve{p}_e'| } 
	= \int \de{\versor{k}'} \, 
	\frac{ \de^5\sigma}{\de \mathbf{\hat p}_e' \de|\ve{p}_e'| \de\mathbf{\hat k}' } \nonumber \\
	= \int_0^{2\pi} \de \phi_{k'} \int_0^{\pi} \de\theta_{k'} \sin\theta_{k'} \, 
	\frac{ \de^5\sigma}{\de \mathbf{\hat p}_e' \de|\ve{p}_e'| \de\mathbf{\hat k}' } \, .
	\end{eqnarray}

\section{Numerical results}

For the reactions on the deuteron we show predictions for 
different processes and kinematics. For electron scattering we start 
with results for the elastic scattering $e +d \rightarrow e' + d'$ observables 
and discuss predictions for the structure functions $A(Q^2)$ and $B(Q^2)$, 
which constitute the differential cross section
$\frac{ \de\sigma^{\rm el}}{\de \versor{p}_e' } ( Q^2, \theta_e )$. 
We consider also the standard polarization observable, the deuteron tensor analyzing power $T_{20}$.
For the breakup reaction, 
$e +d \rightarrow e' + p + n$,
we can calculate not only the corresponding 
$\frac{ \de\sigma^{\rm br}}{\de \versor{p}_e' } ( Q^2, \theta_e )$
cross sections but also 
fully exclusive ones,
$\frac{ \de^5\sigma}{\de \versor{p}_e' \de|\mathbf{p}_e'| \de\versor{p}_1' }$,
where ${\hat p}_1'$ denotes the direction of the outgoing nucleon momentum
determined on top of the final electron momentum $\mathbf{p}_e'$.
It is also very frequent that other kinematical variables, for example $Q^2$
or the so-called missing momentum $p_{miss}$, are used to define parts of the studied phase space.

In the case of neutrino-induced reactions we focus mainly on the total elastic and breakup 
cross sections but show also examples of the angular distributions of the cross sections.

\subsection{Elastic electron-deuteron scattering}

\noindent

Elastic electron-deuteron scattering has been studied by many authors;
see for example~\cite{PhysRevC.75.014001,Huang09,Epelbaum2014,Walzl2021}.
Standard observables are the structure functions $A(Q^2)$ and $B(Q^2)$ as well as
the tensor analyzing power $T_{20}(Q^2,\theta)$ at $\theta_{\rm lab}$= 70$^\circ$~\cite{Huang09}.

\begin{figure}
\begin{center}
\includegraphics[width=0.45\textwidth,clip=true]{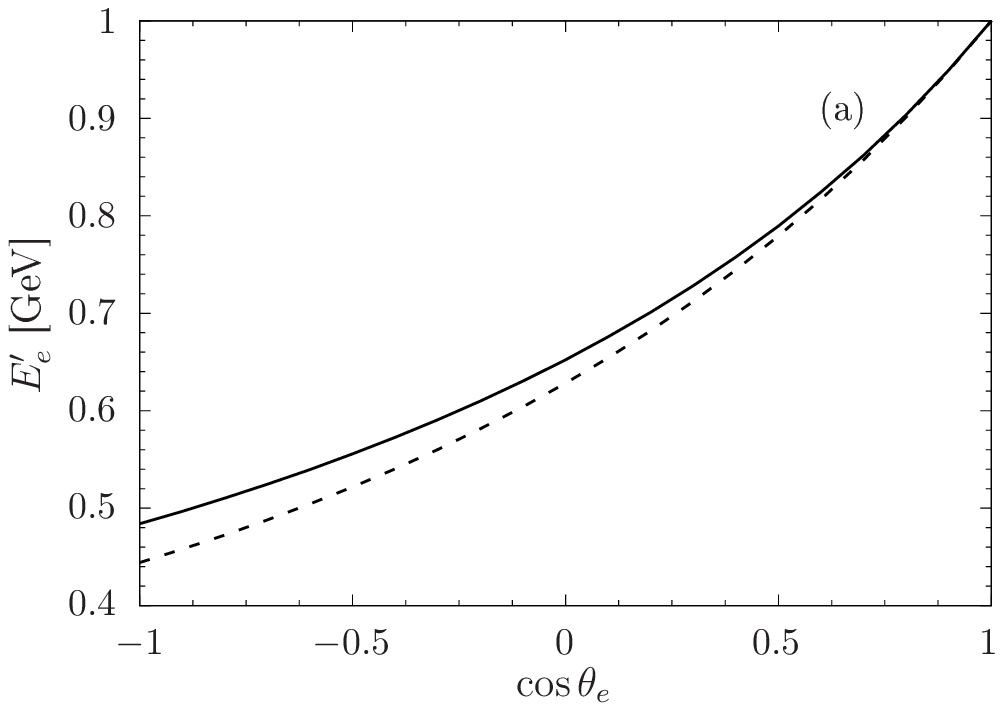}
\includegraphics[width=0.45\textwidth,clip=true]{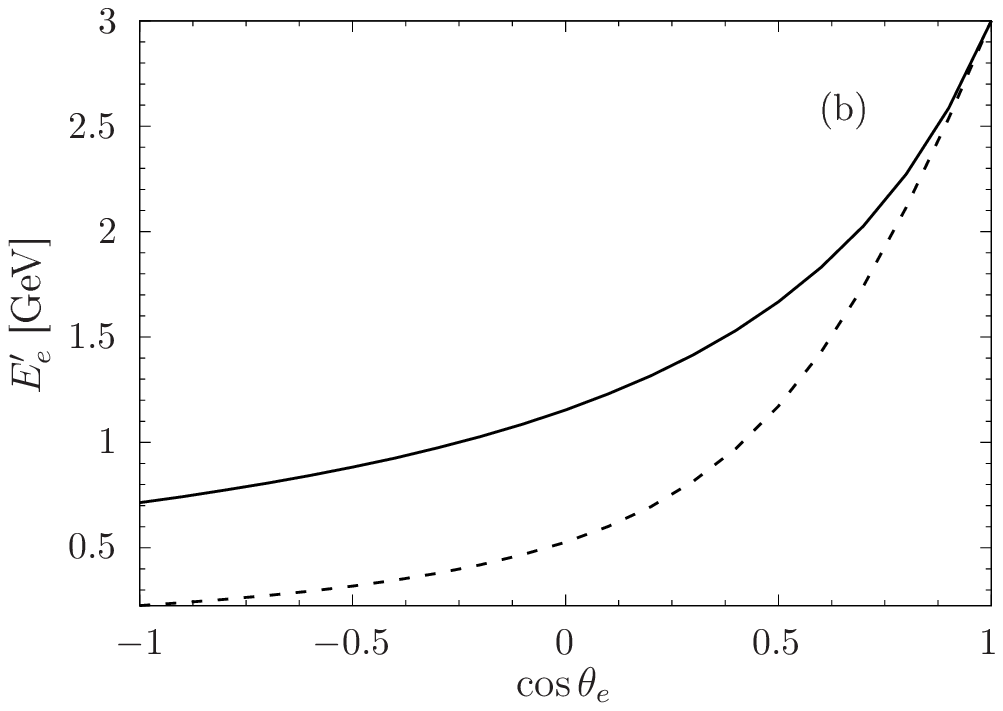}
	\caption{The final electron energy $\energyel{e}{\mathbf{p}_e}{'}$ calculated relativistically (solid line)
	and nonrelativistically (dashed line) as a function of 
	the cosine of the electron scattering angle $\cos \theta_e $ 
	for the initial electron energy $\energyel{e}{\mathbf{p}_e}{}$= 1~GeV (a) and 3~GeV (b).
}
\label{Felastic_eD_Eprime}
\end{center}
\end{figure}

\begin{figure}
\begin{center}
\includegraphics[width=0.45\textwidth,clip=true]{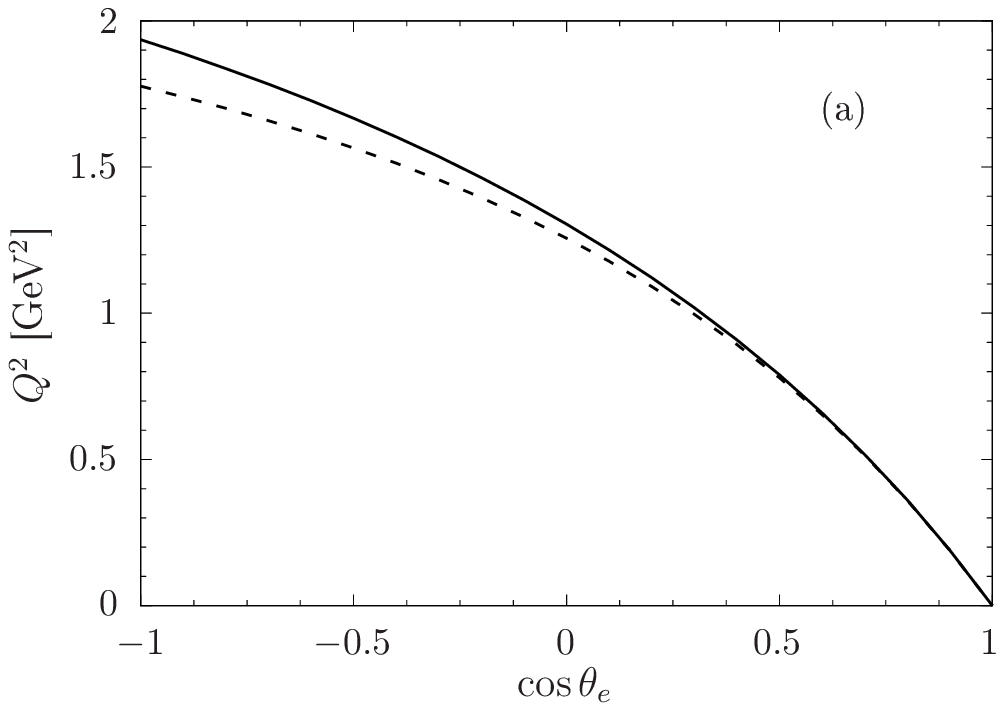}
\includegraphics[width=0.45\textwidth,clip=true]{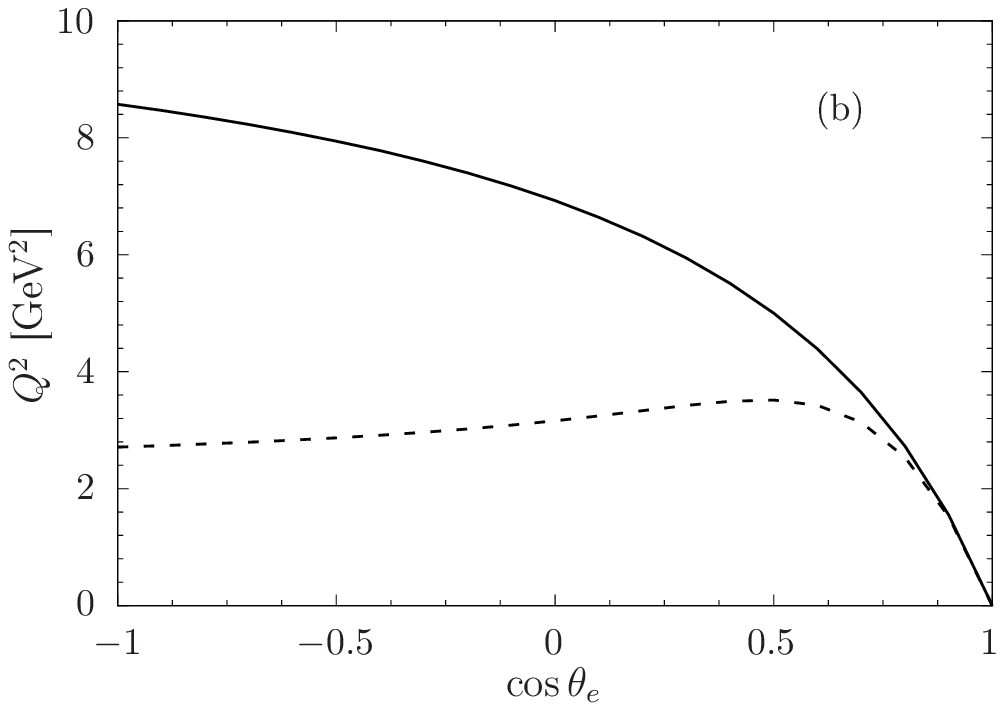}
	\caption{Same as in Fig.~\ref{Felastic_eD_Eprime} but for
	the four-momentum transfer squared, $ Q^2$.
}
\label{Felastic_eD_qq}
\end{center}
\end{figure}

The observables can be investigated for various ranges of the four-momentum transfer squared.
In order to reach higher $Q^2$-values one has to consider a sufficiently high 
initial electron energy. This raises difficulties, if we want to compare
results of strictly nonrelativistic calculations with the relativistic 
ones, since already the nonrelativistic kinematics yields results very different 
from the relativistic kinematics. This is shown for two basic
quantities, the final electron energy and the four-momentum transfer squared
in Figs.~\ref{Felastic_eD_Eprime}-\ref{Felastic_eD_qq} at the initial electron
energy 1 and 3 GeV. That is why our ``nonrelativistic'' predictions 
mean that in this case the nonrelativistic single-nucleon current operator 
and wave functions are combined with the {\em relativistic} kinematics.

\begin{figure}
\begin{center}
\includegraphics[width=0.45\textwidth,clip=true]{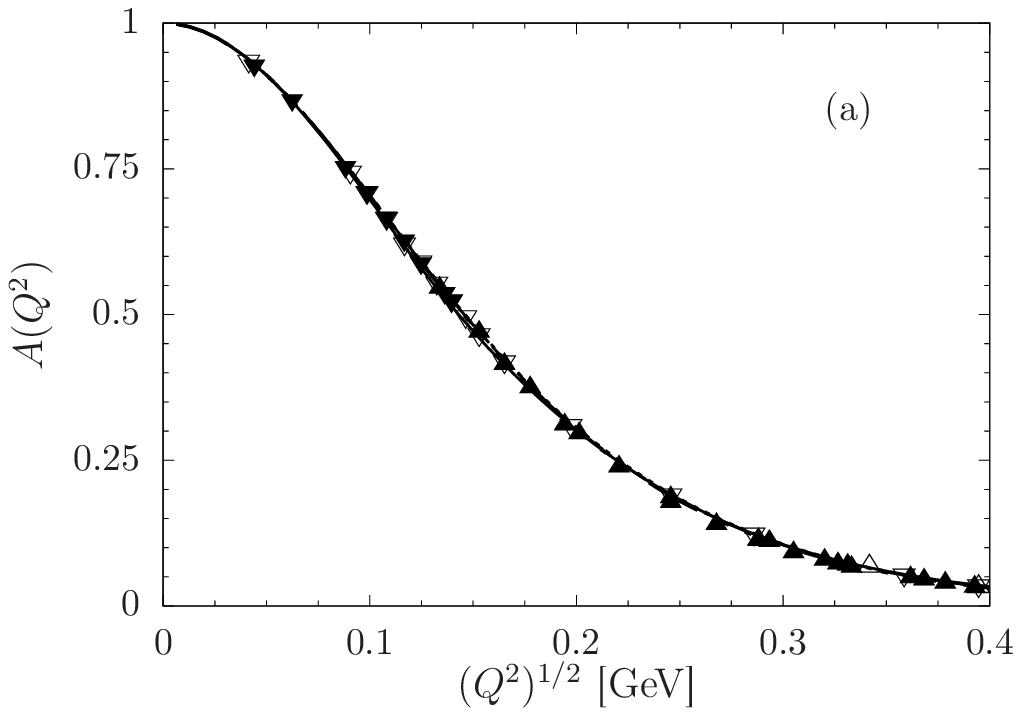}
\includegraphics[width=0.45\textwidth,clip=true]{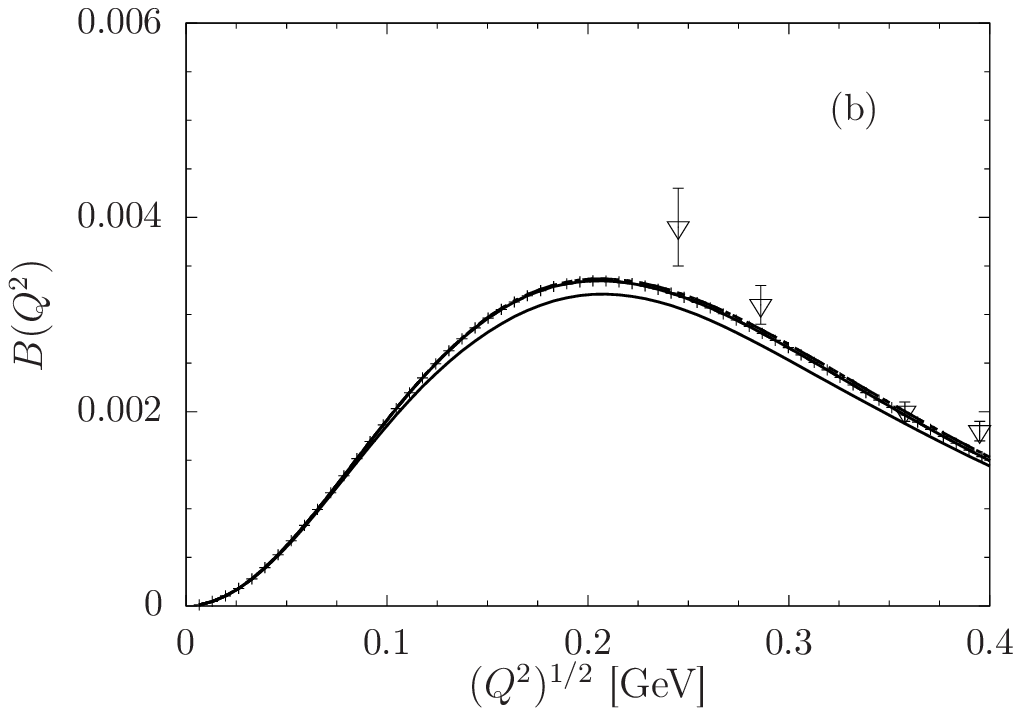}
	\caption{The structure function $A(Q^2)$ (left panel) and $B(Q^2)$ (right panel) 
	as a function of the square root of the absolute value 
	of the four-momentum transfer squared, $ Q^2$, for low $ Q^2$-values. 
	Various nonrelativistic predictions are compared with the relativistic result.
	Experimental data for (a) are from Refs.~\cite{Orsay} (white triangles), \cite{Monterey} (black down triangles) and \cite{SaclayALS.2} (black triangles).
	Experimental data for (b) are from \cite{Mainz} (white down triangles).
	To be compared with Figs.~1 and~2 in Ref.~\cite{Walzl2021}.
	The dashed line (solid line with crosses) is the structure function evaluated for non-relativistic kinematics with the non-relativistic (non-relativistic plus relativistic corrections) current operator, while the dotted  (dash-dotted) line is the same quantity with relativistic  kinematics and non-relativistic (non-relativistic plus relativistic corrections) current operator.
	Finally the solid line is the structure function evaluated fully relativistically.
	}
\label{FAB1}
\end{center}
\end{figure}

Figure~\ref{FAB1} shows for $ Q^2 \le $ 0.4 GeV$^2$ that
the relativistic corrections to the nonrelativistic single-nucleon current are negligible for the 
two deuteron form factors. In this $ Q^2$ range it is also 
possible to employ the nonrelativistic kinematics for the deuteron, making thus
the nuclear part of the calculation more consistent. 

Data for the observables were collected in Ref.~\cite{Huang09}
and comprise many sets. For $A(Q^2)$ they include those from
Refs.~\cite{StanfordMarkIII,CEA,Orsay,SLACE101,SaclayALS,DESY,Bonn,JLabHallC,JLabHallA,Monterey}.
Data for $B(Q^2)$ come from Refs.~\cite{SLACNPSANE4,Martin,Bonn,SaclayALS.2,Mainz,StanfordMarkIII}.

\begin{figure}
\begin{center}
\includegraphics[width=0.45\textwidth,clip=true]{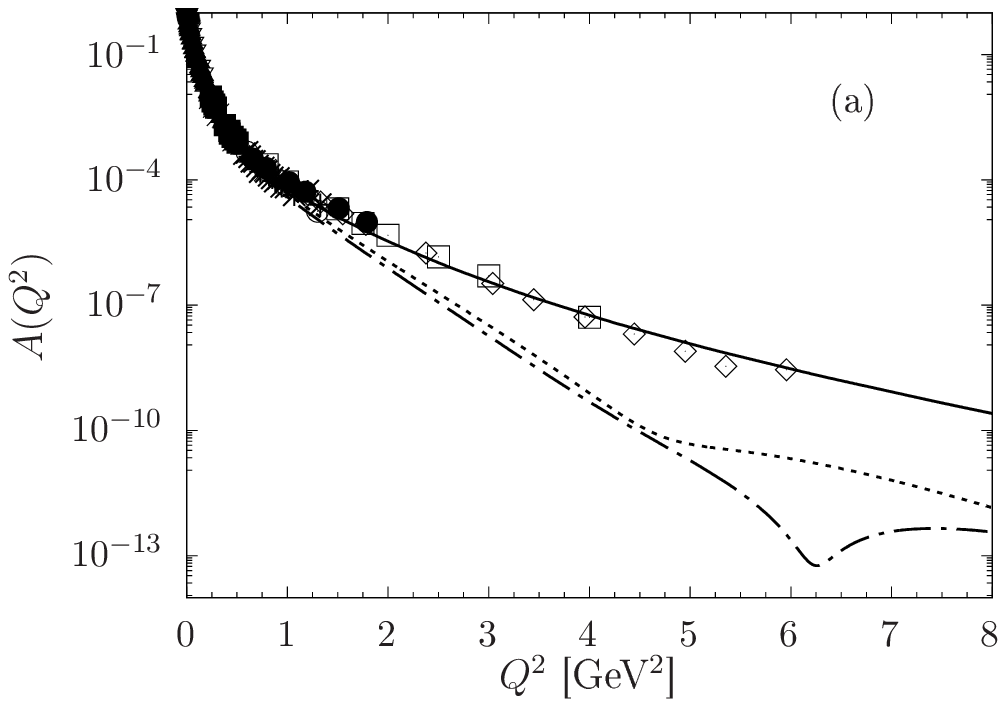}
\includegraphics[width=0.45\textwidth,clip=true]{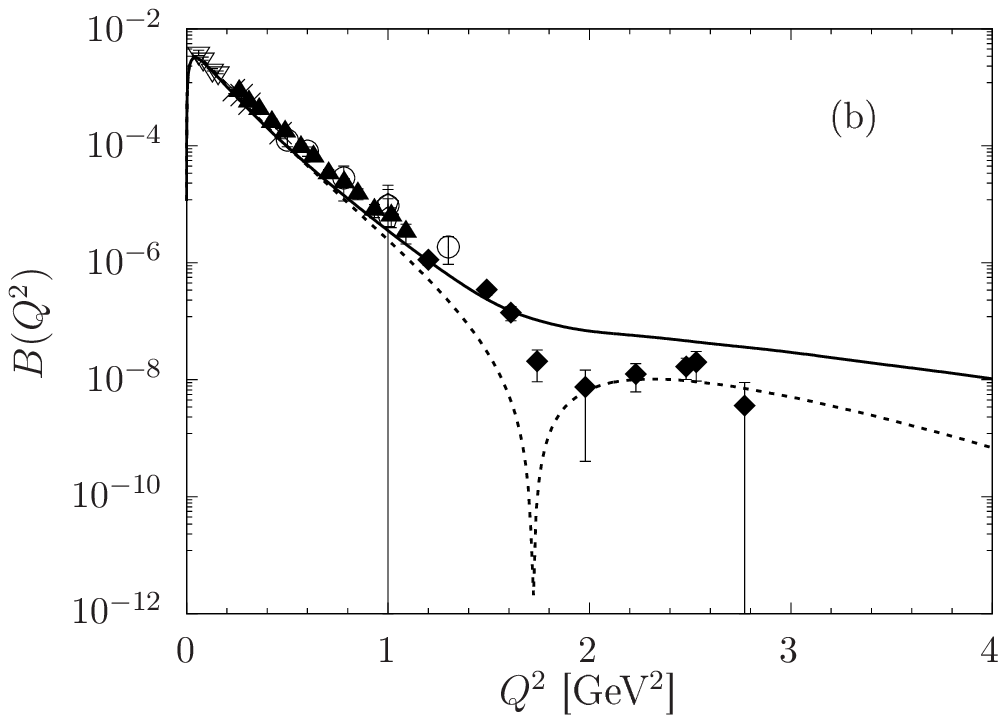}
	\caption{The structure function $A(Q^2)$ (left panel) and $B(Q^2)$ (right panel) 
	as a function of the absolute value 
	of the four-momentum transfer squared, $ Q^2$,
	shown on a logarithmic scale for a broader range of $ Q^2$-values. 
	To be compared with Figs.~9 and~10 in Ref.~\cite{Huang09}.
	The used lines are the same as in Fig.~\ref{FAB1}. Experimental data are from Ref.~\cite{StanfordMarkIII} (x),
	\cite{CEA} (*),
	\cite{SLACE101} (white squares),
	\cite{DESY} (black squares),
	\cite{Bonn} (white circles),
	\cite{JLabHallC} (black circles),
	\cite{JLabHallA} (white diamonds),
	\cite{SLACNPSANE4} (black diamonds),
	\cite{Martin} (white pentagon). The rest of the symbols for the data points has been described in Fig.~\ref{FAB1}.
}
\label{FAB2}
\end{center}
\end{figure}

In Fig.~\ref{FAB2} we display the structure functions $A(Q^2)$ and $B(Q^2)$
on a logarithmic scale for a broader range of $ Q^2$-values.
Since the nonrelativistic kinematics is no longer applicable,
the nonrelativistic results mean in this case only the nonrelativistic
single-nucleon current operator and the deuteron wave function.
We show clear effects of the relativistic corrections in the single-nucleon 
current operator emerging at higher $ Q^2$ values. Including these corrections 
actually makes the difference between the relativistic and nonrelativistic results larger.
This behavior is visible for the structure function $A(Q^2)$, since the corrections affect only the 
charge density operator.

Without 2N contributions in the nuclear current operator we are not able to describe 
the data properly~\cite{Huang09,Epelbaum2014,Walzl2021}.
The gap between our relativistic (and thus ``best'') results and the data
is clearly visible for the structure function $B(Q^2)$ already 
for lower $ Q^2$-values.

\begin{figure}
\begin{center}
\includegraphics[width=0.45\textwidth,clip=true]{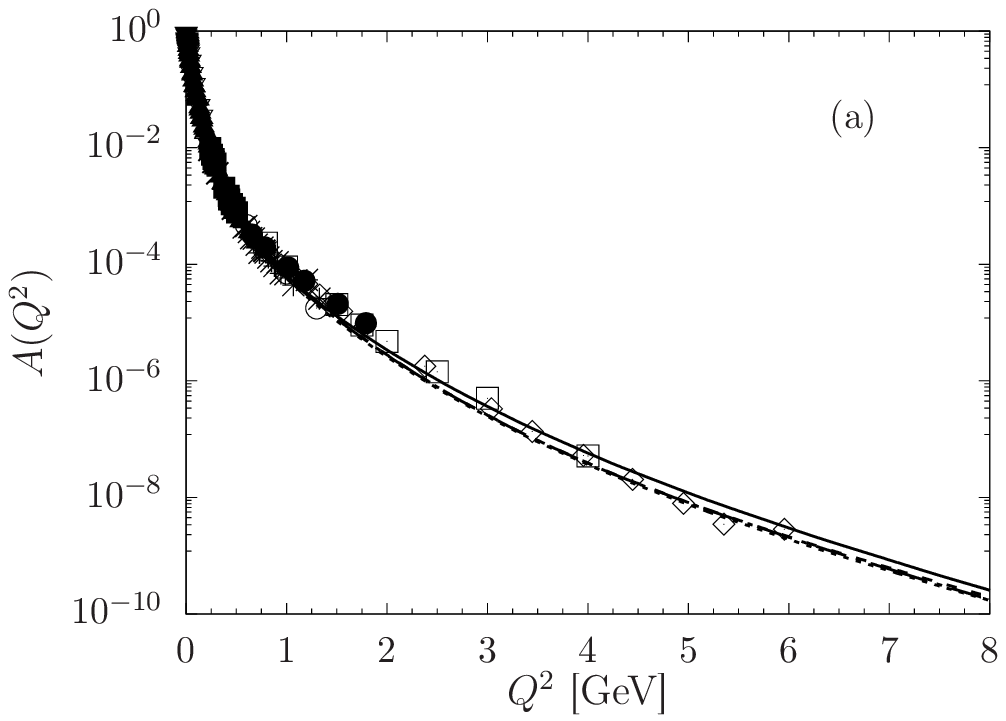}
\includegraphics[width=0.45\textwidth,clip=true]{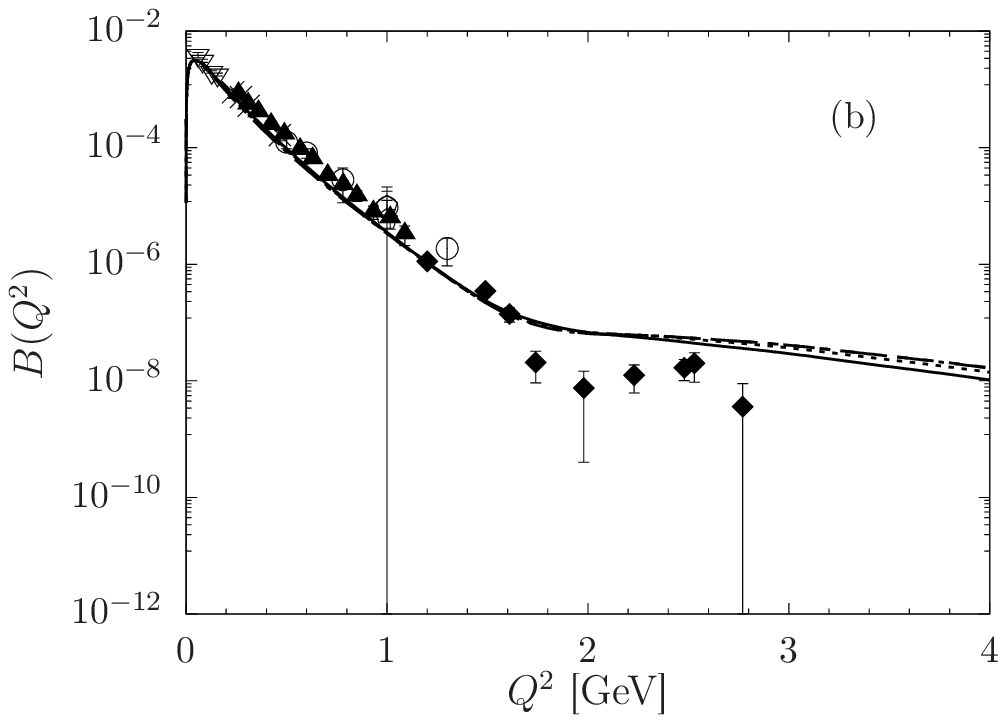}
	\caption{The structure function $A(Q^2)$ (left panel) and $B(Q^2)$ (right panel) 
	as a function of the absolute value 
	of the four-momentum transfer squared, $ {Q^2}$,
	calculated with four different parametrizations of the electromagnetic
	nucleon form factors. The solid line is the dipole parametrization from Ref.~\cite{PRC86.035503}, the dashed  line is from Ref.~\cite{Kelly04}, the dotted line is from Ref.~\cite{Lomon02} and the dash-dotted line is from Ref.~\cite{BBA03}. 
	The symbols for the data points are the same as in Fig.~\ref{FAB2}.
}
\label{FAB3}
\end{center}
\end{figure}

The bulk of our results is obtained with the simple dipole 
parametrization of the nucleon electromagnetic (and also weak) 
form factors~\cite{PRC86.035503}.
For the present investigation the difference between various parametrizations
is not important but we checked also results obtained with the recent 
parametrizations from Budd, Bodek, and Arrington~\cite{BBA03}; 
Kelly~\cite{Kelly04}; and Lomon~\cite{Lomon02}.
For higher $ Q^2$-values some spread between results based 
on various form factor parametrizations develops and is demonstrated 
in Fig.~\ref{FAB3}. This spread is essentially due to the difference
between the predictions using the dipole parametrization~\cite{PRC86.035503}
and the three others~\cite{BBA03,Kelly04,Lomon02}.

We checked also the effects of the Wigner spin rotations for 
these two deuteron structure functions. They are very small and predictions
calculated with the Wigner $D$-functions replaced by the $ 2 \times 2$ identity matrix (not shown here)
are extremely close to the complete results. 

Figures~\ref{FAB1}--\ref{FAB3} were generated choosing the initial electron energy $\energyel{e}{\mathbf{p}_e}{}$= 3 GeV
and taking a fixed step (${\frac18}^\circ$) in the electron scattering angle. 
In this way we could cover the selected range of the $ Q^2$-values. 

The third observable usually studied for elastic electron-deuteron scattering 
is the deuteron tensor analyzing power $T_{20}$. It is investigated 
at fixed laboratory electron scattering angle $\theta_e$= 70$^\circ$
as a function of $ Q^2$. Thus it is sufficient to change the initial electron
energy to generate this $ Q^2$-dependence. We took a 2-MeV step up to 2 GeV.

\begin{figure}
\begin{center}
\includegraphics[width=0.45\textwidth,clip=true]{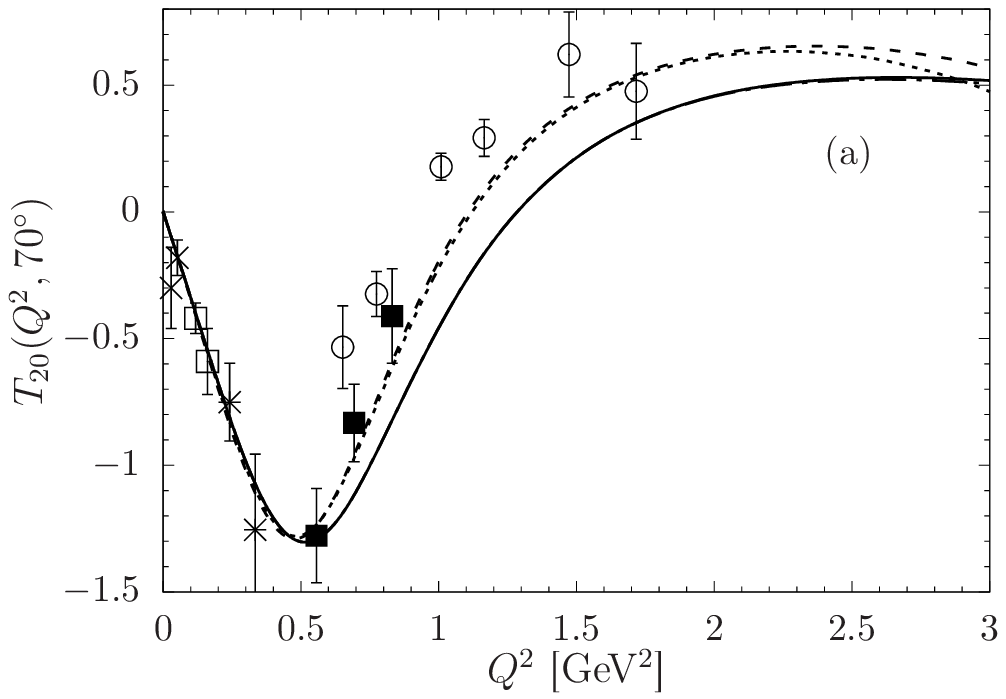}
\includegraphics[width=0.45\textwidth,clip=true]{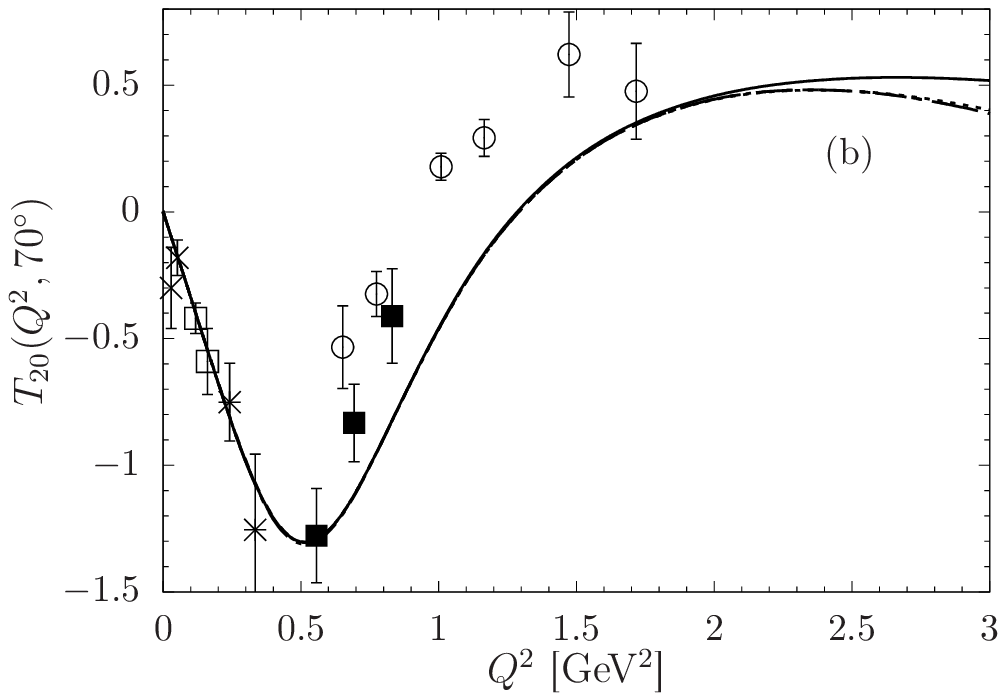}
	\caption{The deuteron tensor analyzing power $T_{20}$ 
	for the electron laboratory scattering angle $\theta_e$= 70$^\circ$
	as a function of the absolute value 
	of the four-momentum transfer squared, $ {Q^2}$.
	In the left panel two different nonrelativistic predictions 
	are compared with the relativistic results. 
The dashed (dotted) line represents predictions obtained with the non-relativistic current operator without (with) $(p/m)^2$ corrections, while the relativistic results (relativistic results ignoring the Wigner spin rotations) are displayed with the solid (dash-dotted) line. 
	A very small effect of the Wigner spin rotations is hardly visible even at the largest shown 
	$ {Q^2}$-values since the solid and dash-dotted lines overlap.
	In the right panel this polarization observable is shown for
	four different parametrizations of the electromagnetic
	nucleon form factors \cite{PRC86.035503} (solid line), \cite{BBA03} (dash-dotted line), 
	\cite{Kelly04} (dashed line), and \cite{Lomon02} (dotted line). 
	Note that the dash-dotted, dashed and dotted lines overlap. 
	All the calculations are based on the relativistic kinematics. Experimental data are from
       Refs.~\cite{Novosibirsk-85-1,Novosibirsk-85-2} (x), \cite{Novosibirsk-90} (*), \cite{Bates-84} (white squares), \cite{Bates-91} (black squares) and \cite{JLabHallC.2} (white circles).
}
\label{FT20}
\end{center}
\end{figure}

In the left panel of Fig.~\ref{FT20} it is shown that up to approximately 0.4~GeV$^2$
all the predictions coincide but for the higher $ {Q^2}$-values the relativistic
results are significantly below the two nonrelativistic calculations. 
Relativistic $(p/m)^2$ corrections to the single-nucleon current operator become important
at $ {Q^2}$= 2~GeV$^2$. Still the effect of the Wigner spin rotations
is hardly visible, even at $ {Q^2}$= 3~GeV$^2$. 
In the right panel $T_{20}$ is shown for the 
	four parametrizations of the electromagnetic
	nucleon form factors \cite{PRC86.035503,BBA03,Kelly04,Lomon02}.  Up to approximately $ {Q^2}$= 1.7~GeV$^2$ all the four curves overlap, but for the higher $Q^2$ values the prediction obtained with the simple dipole parametrization from 
	Ref.~\cite{PRC86.035503} differs from the other three, which remain essentially indistinguishable. 
	
While consistent relativistic calculations are possible by computing 
independent current 
matrix elements and generating the rest using covariance, the comparison of the calculations 
with data points to a need for two-body currents.
Note however, that also in this case we employ the relativistic kinematics 
in the nonrelativistic calculations.

\subsection{Exclusive deuteron electrodisintegration}

\noindent

\begin{figure}
\begin{center}
\includegraphics[width=0.45\textwidth,clip=true]{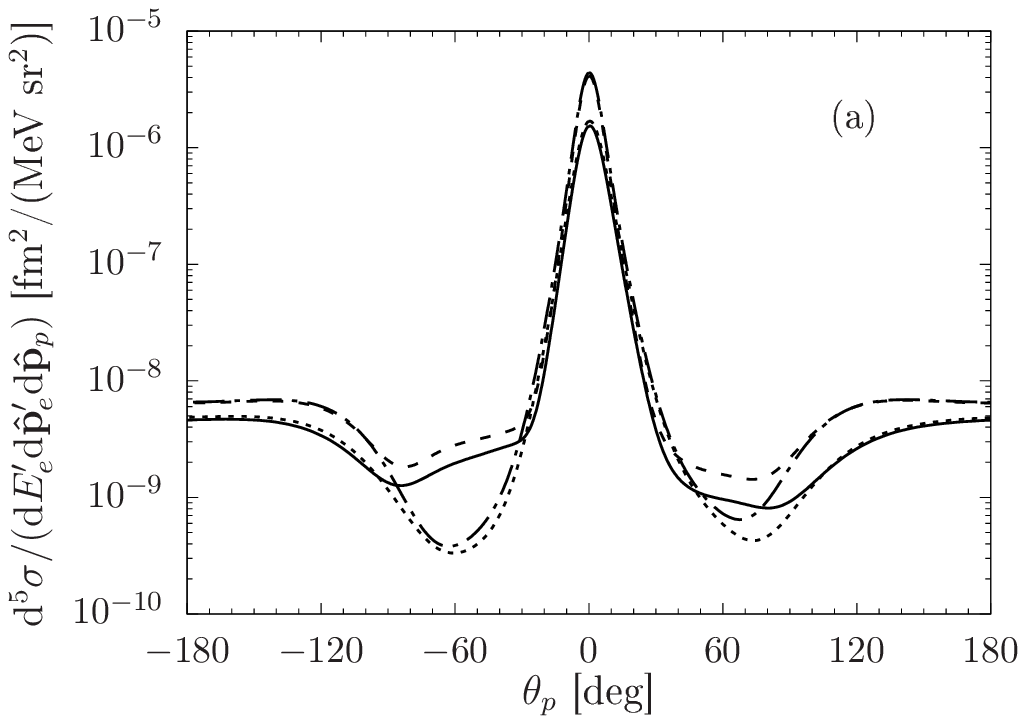}
\includegraphics[width=0.45\textwidth,clip=true]{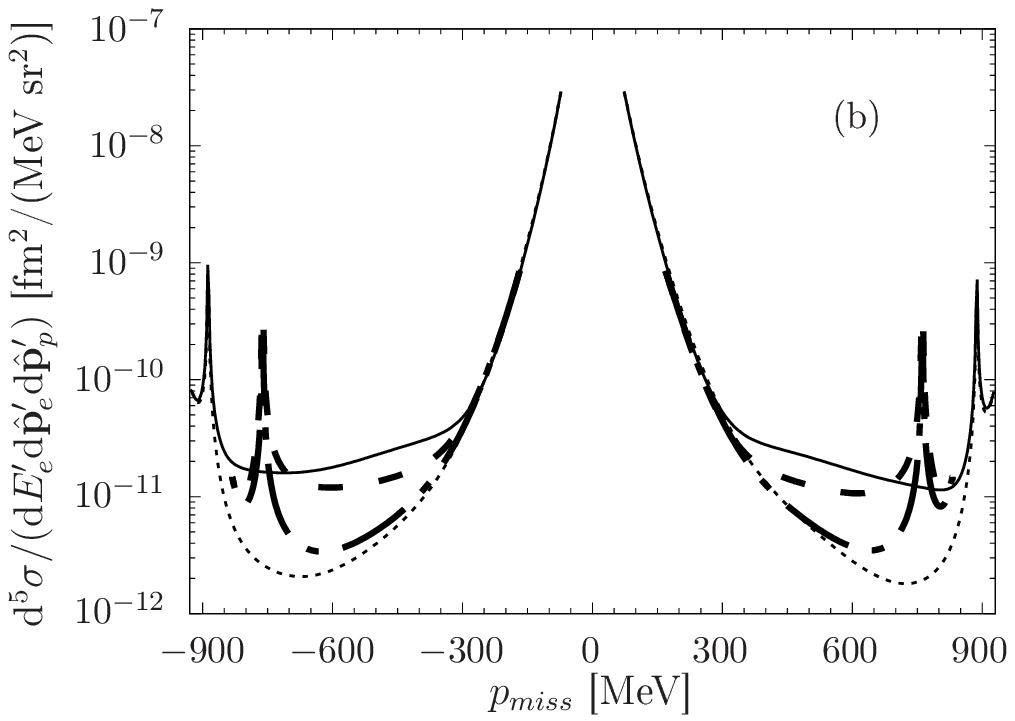}
	\caption{Predictions for the five fold
		differential cross section ${\de^5\sigma}/({\de\versor{p}_e' \de E_e' \de\versor{p}_p})$
		for the exclusive $^2{\rm H}(e,e'p)n $ reaction at
		fixed outgoing electron parameters 
		((a) $E_e$= 800~MeV, 
		$\theta_e$= 38.7$^\circ$,
		$E_e'$= 635.3~MeV, (b) 
		($E_e$= 800~MeV, 
		$\theta_e$= 106.5$^\circ$, 
		$E_e'$= 414.4~MeV)) 
		with (a) $E_{c.m.}$= 100 MeV and $|\ve{q}|$= 500~MeV,
		     (b) $E_{c.m.}$= 150 MeV and $|\ve{q}|$= 1000~MeV,
		shown as a function of
		(a) the polar angle of the outgoing proton momentum,
		(b) the magnitude of the momentum of the undetected (here neutron) nucleon (``missing'' momentum).
		Negative values of $\theta_p$ 
                or $p_{miss}$ correspond to $\phi_{pq}$= 0
		and their positive values to $\phi_{pq}$= 180$^\circ$. 
		The dotted (solid) line represents the relativistic results without (including) 
		the rescattering contribution. The corresponding nonrelativistic predictions 
		without (with) the rescattering part are shown with the dash-dotted (dashed) line.
}
\label{EX1-2}
\end{center}
\end{figure}

The $e + d \rightarrow e' + p + n$ reaction offers more possibilities than the corresponding elastic scattering process
due to the richer phase space. On top of the electron
parameters ($E_e$, $\theta_e$, $E_e'$)  additional quantities are needed to fix the
exclusive kinematics. Due to very small values of the cross sections in the so-called 
point-like geometry, they are very hard to measure in a realistic experiment but we decided to 
demonstrate essential features of the exclusive cross sections.
In Fig.~\ref{EX1-2}(a) 
we choose (arbitrary) initial electron energy $E_e$= 800 MeV and consider two electron kinematics.
In the first one
         ($\theta_e$= 38.7$^\circ$
         $E_e'$= 635.3~MeV),
         where $E_{c.m.}$= 100 MeV and $|\ve{q}|$= 500 MeV,
there is no restriction on the angle $\theta_p$
between the outgoing proton momentum $\ve{p}_p$ and $\ve{q}$,
so this angle can be chosen to label the exclusive kinematics.
We restrict ourselves to the case, where either $\phi_{pq}$= 0
(negative $\theta_p$) or $\phi_{pq}$= 180$^\circ$ (positive $\theta_p$).
There are regions, where the rescattering effects given by the difference between
the solid and dotted lines (relativistic calculations)
or the dashed and dash-dotted lines (nonrelativistic results) are very strong.
In particular in the very proton knockout peak rescattering effects reduce
the values of the {\em plane wave} (obtained without rescattering contribution) cross sections.
Also here we see clear differences between the relativistic and nonrelativistic predictions.

In the second kinematics
 considered in Fig.~\ref{EX1-2}(b) 
($\theta_e$= 106.5$^\circ$, $E'_e$= 414.4 MeV) the internal two-nucleon energy
$E_{c.m.}$= 150 MeV is much smaller than
the magnitude of the three-momentum transfer $|\ve{q}|$= 1000 MeV.
For such electron parameters $0 \le \theta_p < 90^\circ$ and for each $\theta_p$
there are two physical solutions. In this case the so-called ``missing'' momentum, $p_{miss}$,
the magnitude of the momentum
of the undetected neutron, is convenient to label the exclusive kinematics.
The physical ranges of $\theta_p$ and $p_{miss}$ are quite different
in the nonrelativistic and relativistic calculations,
which makes the comparison more difficult. The sharp peaks correspond to the maximal $\theta_p$
values, where the phase space factor becomes singular. Small $p_{miss}$ values
coincide with the proton knockout peak and here the rescattering effects are very small.

\subsection{The cross section in the ${^2{\rm H}}(e,e'p)$ process at low energy transfer and close to threshold}

\noindent

In this subsection we consider the cross section for the ${^2{\rm H}}(e,e'p)$
reaction at a low initial electron energy and momentum transfer 
for the measurement reported in Ref.~\cite{PRL88}.
As in many other papers by H. Arenh\"ovel (see for example \cite{Arenhovel05,asymmetry}) the data are shown 
in a ``mixed" representation: the electron energies and the electron scattering angle
are defined in the laboratory frame but the proton angles are given in the two-nucleon c.m. frame.
It is then natural to start from the cross section 
$ {\de^5\sigma} /{\left( \de \energyel{e}{\mathbf{p}_e}{'} \de\versor{p}_e' \de\versor{k} \right) } $ defined in Eq.~(\ref{sigmaEMBR2}). 

\begin{figure}[hb!]
\includegraphics[width=7cm]{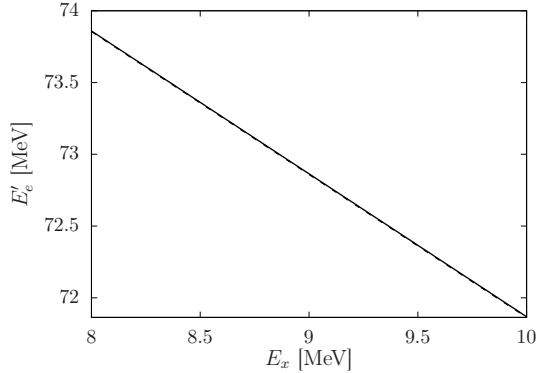}
\caption{
The final electron energy $\energyel{e}{\mathbf{p}_e}{'}$ as a function of the excitation energy $E_x$
in the ${^2{\rm H}}(e,e^{\prime},p)$ reaction calculated relativistically 
and nonrelativistically for the initial electron energy $E$= 85 MeV and the 
electron scattering angle $\theta_e$= 40$^\circ$. The two curves fully overlap.
}
\label{fig:Ex}
\end{figure}

\begin{figure}[hb!]
\includegraphics[width=7cm]{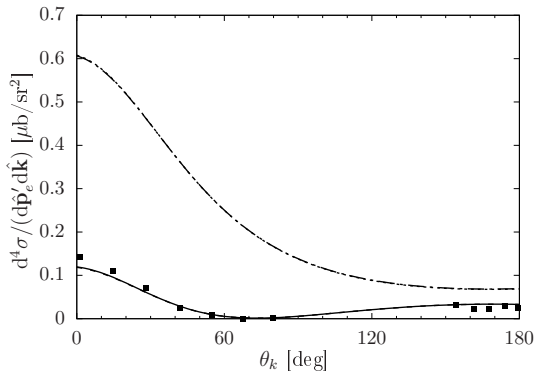}
\caption{
The four fold differential cross section 
	${\de^4\sigma}/ \left( { \de\versor{p}_e' \de\versor{k} } \right) $ 
	of the ${^2{\rm H}}(e,e'p)$
reaction at $E$= 85 MeV for an excitation energy bin
$ 8\, {\rm MeV} \le E_x \le 10\, {\rm MeV}$ of the breakup spectra 
as a function of the polar proton emission angle $\theta_k$
defined with respect to the three-momentum transfer in the two-nucleon
	c.m. frame (the same as $\theta_{pq}$ in Fig.~\ref{fig:kinematics_diagram}). 
	The solid (dashed) line represents the calculation for the cross section with all ({\em plane wave}) contributions for the fully relativistic calculations.
	The dotted (dash-dotted) line represents the calculation for the cross section with all ({\em plane wave}) contributions for the non-relativistic calculations.
	Note that the plane wave results (upper curves) of the nonrelativistic and relativistic calculations
	overlap, as expected. The same is true for the full predictions (lower curves).
        The data come from Ref.~\cite{PRL88}.
}
\label{fig:PRL88}
\end{figure}

The excitation energy used in \cite{PRL88} is the kinetic energy in the two-nucleon c.m. frame 
and is uniquely related to the final electron energy. In the relativistic case 
the connection reads
\[
\energyel{e}{\mathbf{p}_e}{'}= \frac{{E_x}^2+4 {E_x} {m}-2 {\energyel{e}{\mathbf{p}_e}{}}
   {m_D}-{m_D}^2+4 {m}^2}{2 ({\energyel{e}{\mathbf{p}_e}{}} \cos \theta_e
   -{\energyel{e}{\mathbf{p}_e}{}}-{m_D})}
\] 
and is in fact more complicated for the nonrelativistic kinematics
\begin{eqnarray*}
	\energyel{e}{\ve{p}_e}{'} & = &  \frac{1}{\sqrt{2}}\bigg(
8
{\energyel{e}{\ve{p}_e}{}} {m}-8 {m}
({E_x}-{m_D}+{m})-{\energyel{e}{\ve{p}_e}{}}^2+\
{\energyel{e}{\ve{p}_e}{}}
({\energyel{e}{\ve{p}_e}{}} \cos 2 \theta_e 
-8 {m} \cos \theta_e )
\bigg)^{1/2}  \nonumber \\
	& + & {\energyel{e}{\ve{p}_e}{}} \cos \theta_e -2 {m} \, .
\end{eqnarray*}
For this kinematics we do not seek any visible relativistic effects, since already 
the connection between the excitation energy $E_x$ and the outgoing electron energy $E_e'$ shown in
Fig.~\ref{fig:Ex} suggests that nonrelativistic framework should be fully adequate.
We simply check that at low energies and momenta our relativistic framework is consistent with the 
calculations performed for example in Refs.~\cite{PhysRevC.98.015501,PhysRevC.100.064003}.

To compare our predictions with the cross section measured in the finite 
excitation energy bin~\cite{PRL88} we calculated numerically the following integral
\[
\frac{\de^4\sigma}{ \de\versor{p}_e' \de\versor{k} } 
= \int\limits_{E_e'^{\rm min}}^{E_e'^{\rm max}} \, \de \energyel{e}{\mathbf{p}_e}{'} \,
\frac{\de^5\sigma}{ \de \energyel{e}{\mathbf{p}_e}{'} \de\versor{p}_e' \de\versor{k} } \, .
\]

Figure~\ref{fig:PRL88} shows our predictions calculated 
with the single-nucleon current. The agreement with the data is reasonable, 
which means that this low-energy observable is not sensitive to the details
of the current operator - the cross section is dominated by the charge density
part. However, as noted in Ref.~\cite{PRL88}, the separation of various 
parts in the cross section would reveal true drawbacks of the theoretical
framework. Measurements at low energies, providing more detailed 
observables, can be really used to test 
important dynamical ingredients.

\subsection{The semi-exclusive cross sections in the ${^2{\rm H}}(e,e')$ process}

\noindent

The semi-exclusive process, where only the final electron is detected, still
allows one to vary independently the energy and the magnitude of the three-momentum transfer
for fixed initial electron energy. The values of the cross sections are much higher
than in the exclusive case and many measurements (see for example 
\cite{21,22,23}) focused especially on broad maxima, which appear for the
quasi-elastic scattering domain in the ($\theta_e$, $E_e'$) plane, where
the magnitude of the three-momentum transfer $|\ve{q}|$ and the energy 
transfer $\omega $ are related by $|\ve{q}|^2 \approx 2 m \omega $ with $m$ being 
the nucleon mass.
The cross section in this region is dominated by the single-nucleon 
current operator, which means that our framework should give reliable predictions 
for such kinematics.

\begin{figure}
\begin{center}
\includegraphics[width=.45\textwidth,clip=true]{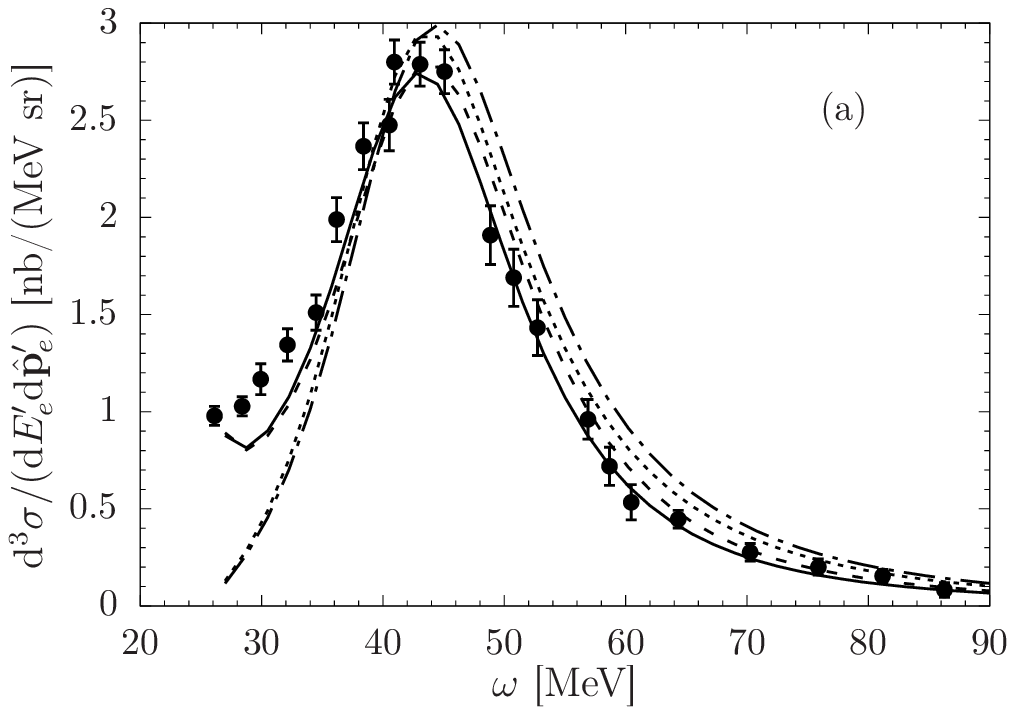}
\includegraphics[width=.45\textwidth,clip=true]{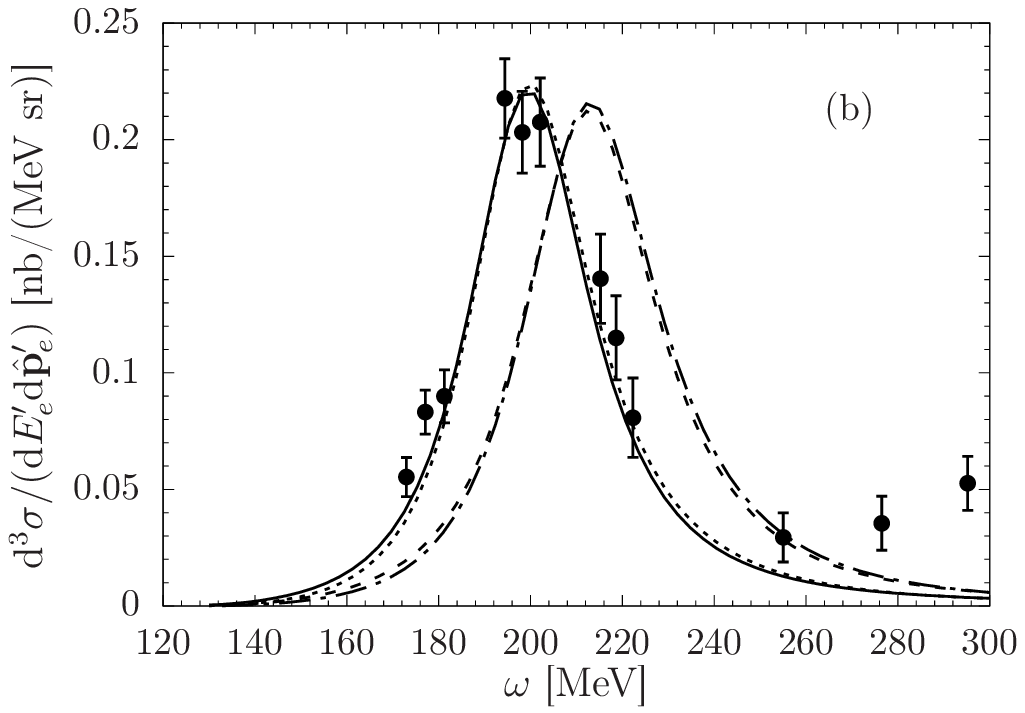}
\includegraphics[width=.45\textwidth,clip=true]{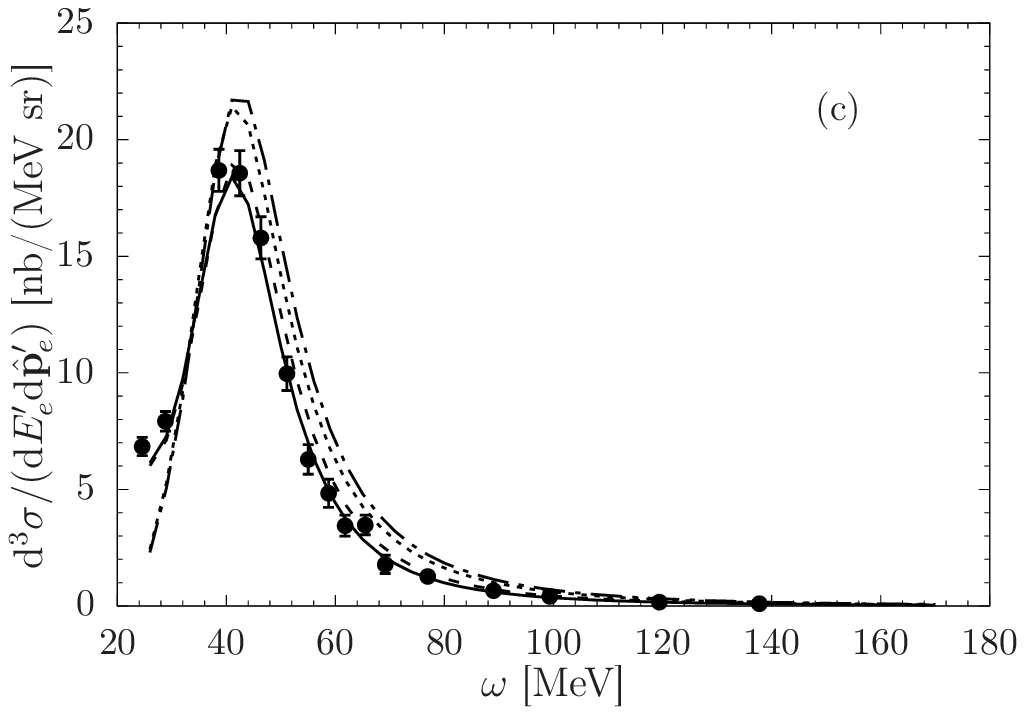}
\includegraphics[width=.45\textwidth,clip=true]{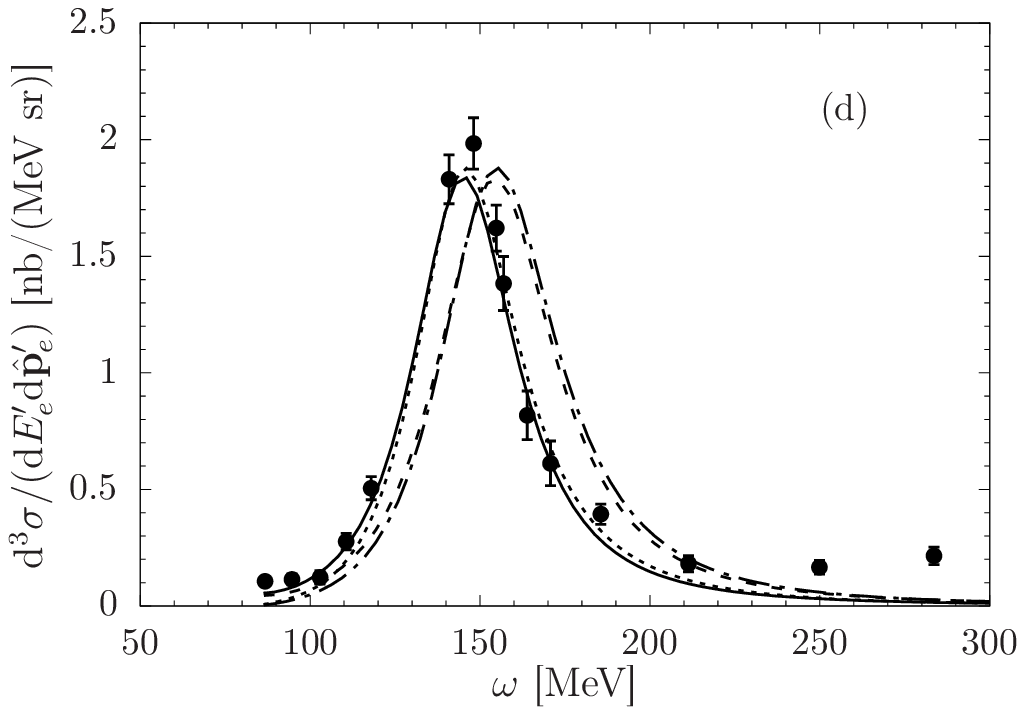}
	\caption{Results of the nonrelativistic {\em plane wave} (dash-dotted line)
	and {\em full} (dashed line) calculations as well as the relativistic {\em plane wave} (dotted line) 
	and {\em full} predictions (solid line) 
	for the semi-exclusive cross section ${\de^3 \sigma}/\left( {\de \energyel{e}{\mathbf{p}_e}{'} \de{\versor{p}_e'}} \right)$
	are shown as a function of the energy transfer $\omega$.
The sequence of subfigures follows Fig.~8 in Ref.~\cite{PRC37_1609}. 
}
\label{PRC37_1609_FIG08}
\end{center}
\end{figure}

\begin{figure}
\begin{center}
\includegraphics[width=.45\textwidth,clip=true]{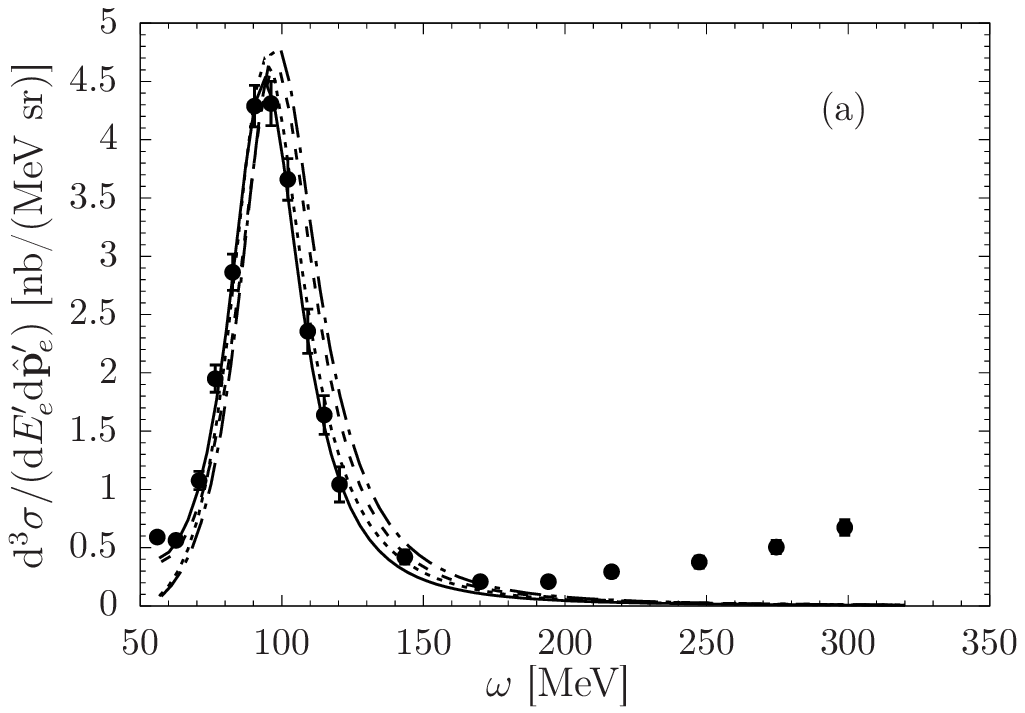}
\includegraphics[width=.45\textwidth,clip=true]{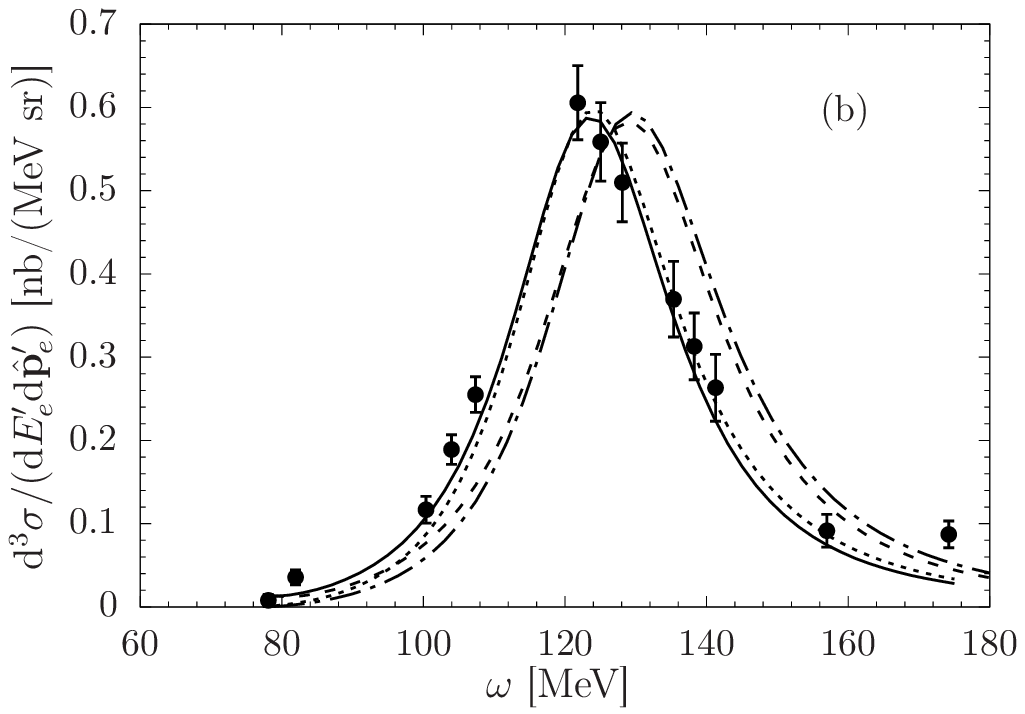}
\includegraphics[width=.45\textwidth,clip=true]{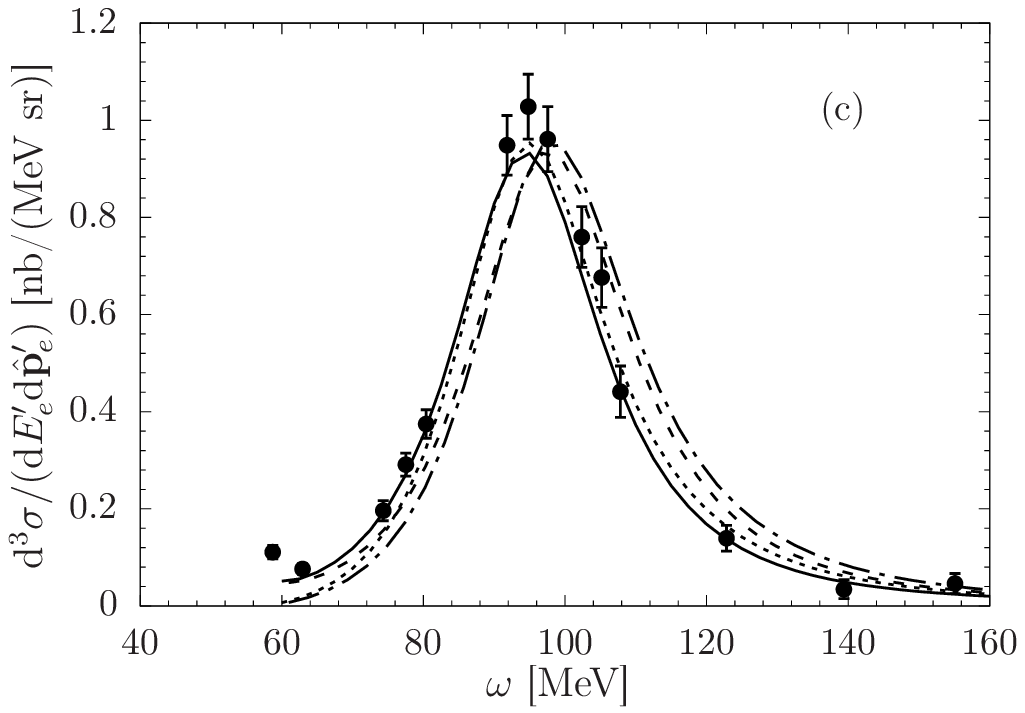}
\caption{The same as in Fig.~\ref{PRC37_1609_FIG08} 
but for the configurations from Fig.~9 in Ref.~\cite{PRC37_1609}. 
}
\label{PRC37_1609_FIG09}
\end{center}
\end{figure}

\begin{figure}
\begin{center}
\includegraphics[width=.45\textwidth,clip=true]{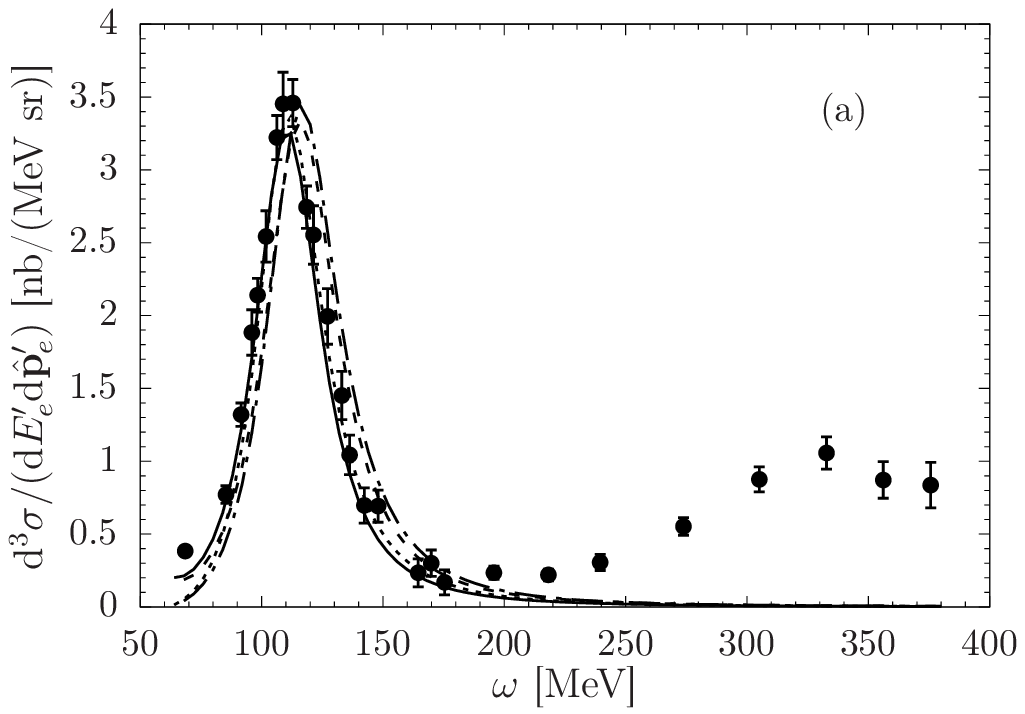}
\includegraphics[width=.45\textwidth,clip=true]{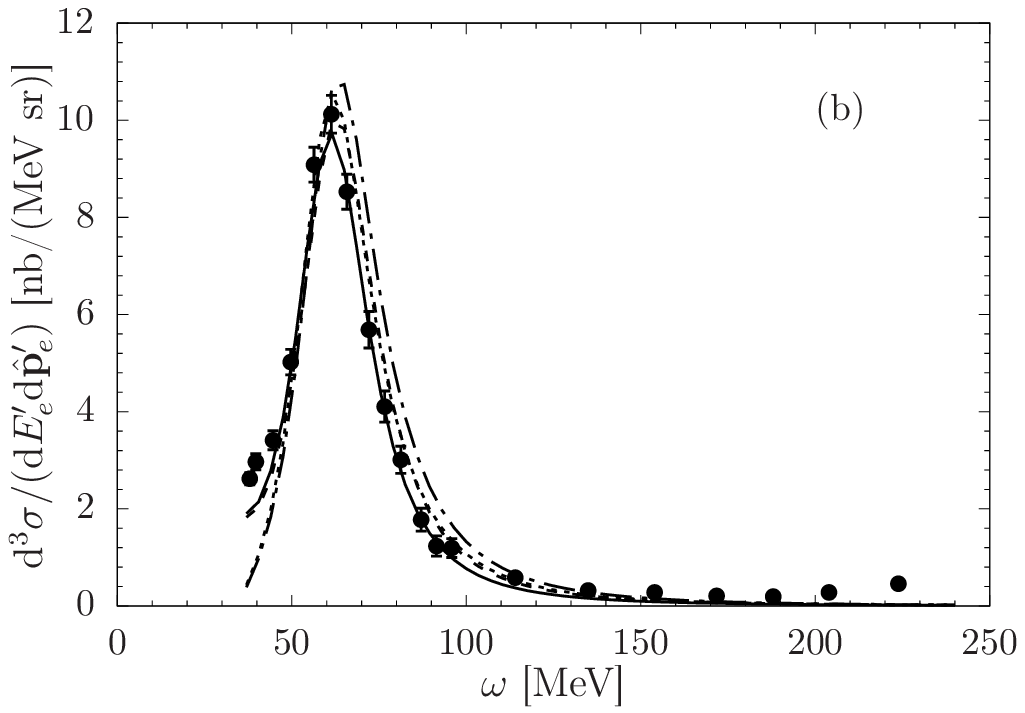}
\includegraphics[width=.45\textwidth,clip=true]{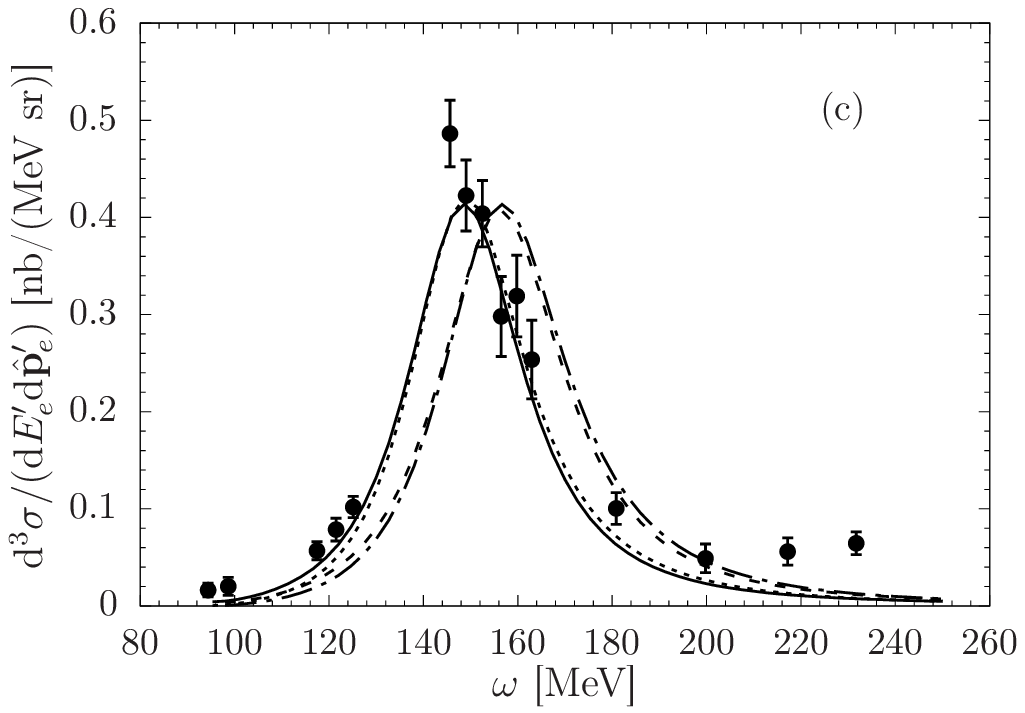}
\includegraphics[width=.45\textwidth,clip=true]{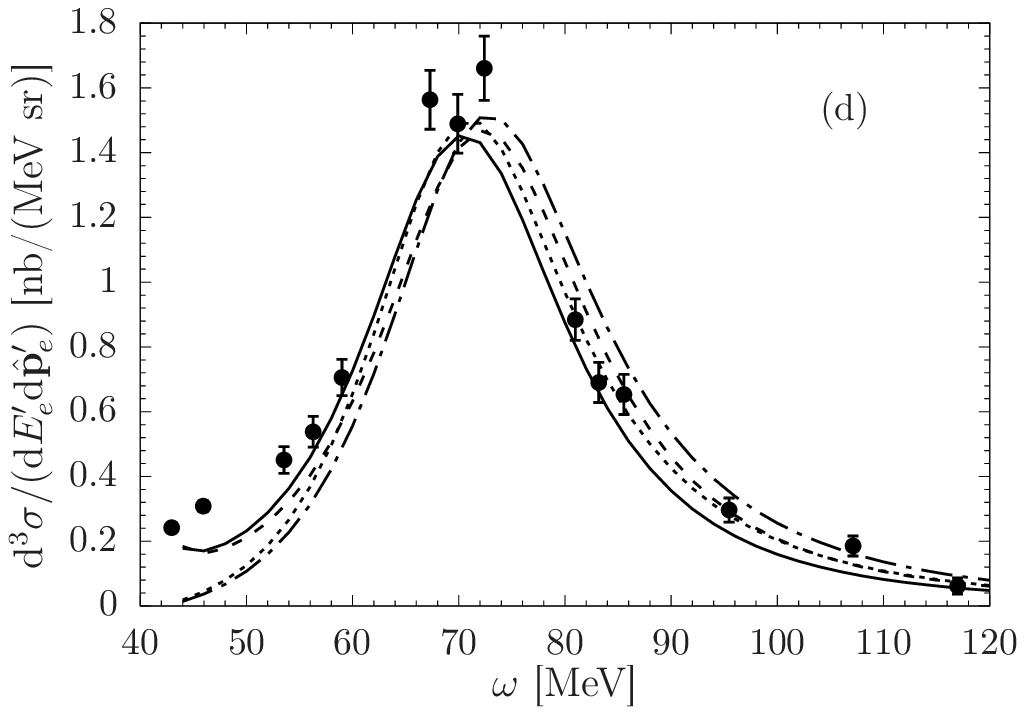}
\caption{The same as in Fig.~\ref{PRC37_1609_FIG08} 
but for the configurations from Fig.~10 in Ref.~\cite{PRC37_1609}. 
}
\label{PRC37_1609_FIG10}
\end{center}
\end{figure}

In Ref.~\cite{PRC37_1609} a rich data set for the ${\de^3 \sigma}/\left( {\de \energyel{e}{\mathbf{p}_e}{'} \de{\versor{p}_e'}} \right)$ 
differential cross sections 
(and the derived response functions) is compared with Arenh\"ovel and Leidemann's predictions 
as well as with the results obtained by Laget. Here, in Figs.~\ref{PRC37_1609_FIG08}-\ref{PRC37_1609_FIG10} we show our results 
corresponding to Figs. 8-10 in \cite{PRC37_1609}, restricting ourselves to the cross sections, although we could 
also calculate the nuclear response functions $R_L$ and $R_T$.

As before all of our calculations are performed neglecting two-nucleon contributions to the nuclear current operator.
The figures show mainly the quasi-elastic peak where these contributions should remain small. 
There is only a slight but visible shift between the positions
of the relativistic and nonrelativistic quasi-elastic peak. 
We see also the characteristic enhancement of the cross sections 
shown by {\em full} calculations 
(including two-nucleon final state interactions) close to the threshold (left slopes), 
where the 
internal two-nucleon energy is very small. This feature is definitely supported by the data
and absent in the plane wave based ({\em PW}) results.
On right slopes and for large energy transfers, where the internal two-nucleon energy exceeds the pion mass,
new channels (pion production, isobar $\Delta$ excitation) 
are open, which cannot be described by our theory.

Despite the incompleteness of our approach, the obtained data description is very good. 
That is observed especially for the energy transfers corresponding to the internal two-nucleon energies 
smaller than the pion mass. In all the eleven cases the relativistic predictions describe the experimental data better than the nonrelativistic ones.

\subsection{An excursion into polarization observables}

\noindent

So far we have dealt mainly with the unpolarized cross sections.
The investigations in this subsection are inspired 
by Sabine Jeschonnek and J. W. Van Orden's paper~\cite{PRC80.054001}.
The authors investigated four polarization observables,
$A_d^V$, $A_d^T$, $A_{ed}^V$, and $A_{ed}^T$.
The experimental data for these quantities were presented 
for example in Ref.~\cite{PRC95.024005}.
The cross section for the $\overrightarrow{^2{\rm  H}}({\vec e}$,e'p)n
reaction (polarized beam and target) can be expressed as~\cite{PRC80.054001}
\begin{equation}
	\sigma = \sigma_0 \left[
	1 + 
	   \sqrt{\frac32} p_z \left( A_d^V + h A_{ed}^V   \right)
	  +  \sqrt{\frac12} p_{zz} \left( A_d^T + h A_{ed}^T   \right)
	\right] \, ,
\label{eq1}
\end{equation}
where $\sigma_0$ is the unpolarized cross section,
$p_z$ and $p_{zz}$ are the vector and tensor polarizations of the deuteron target. 
The same can be written in terms of the tensor polarization coefficients 
$t_{10}$ and $t_{20}$:
\begin{equation}
	\sigma = \sigma_0 \left[
	1 + 
	   t_{10} \left( A_d^V + h A_{ed}^V   \right)
	  +  t_{20} \left( A_d^T + h A_{ed}^T   \right)
	\right] \, .
\label{eq2}
\end{equation}
Equations (\ref{eq1}) and (\ref{eq2}) are derived 
from the general density matrix  $ \rho = \rho_e \otimes \rho_d $,
which is a tensor product of the electron matrix $\rho_e$  
($h$ is the electron helicity)
\begin{equation}
\rho_e = \frac12 \left( 
        \begin{array}{cc}
		1+h & 0 \\
		0   & 1-h \\
	\end{array}
	\right)
\end{equation}
and the deuteron matrix $\rho_d$ 
\begin{equation}
\rho_d = 
\left(
\begin{array}{ccc}
 \frac{1}{3} \left(\sqrt{\frac{3}{2}}
	t_{10}+\frac{{t_{20}}}{\sqrt{2}}+1\right) &
	-\frac{{t_{11}}^*+{t_{21}}^*}{\sqrt{6}} &
	\frac{{t_{22}}^*}{\sqrt{3}} \\
	-\frac{{t_{11}}+{t_{21}}}{\sqrt{6}} & \frac{1}{3} \left(1-\sqrt{2}
	{t_{20}}\right) & -\frac{{t_{11}}^*-{t_{21}}^*}{\sqrt{6}} \\
	\frac{{t_{22}}}{\sqrt{3}} & -\frac{{t_{11}}-{t_{21}}}{\sqrt{6}} &
   \frac{1}{3} \left(-\sqrt{\frac{3}{2}}
	{t_{10}}+\frac{{t_{20}}}{\sqrt{2}}+1\right) \\
\end{array}
\right) \, ,
\end{equation}
assuming axially symmetric deuteron target polarization -- the most typical 
experimental situation.
(In Eqs. (\ref{eq1})-(\ref{eq2}) the axis of symmetry is the $z$-axis.)

The electron parameters used in Ref.~\cite{PRC80.054001} lead to $E_{c.m.}$ energies,
which exceed the pion mass and do not allow us to perform our {\em full} calculations. 
For example those of Figs.~2 and~8 ($E_e$= 5.5 GeV, $Q^2$= 2 GeV$^2$, $x \equiv Q^2/(2 m \omega)$= 1) 
yield $E_{c.m.} \approx$ 0.47 GeV.
We compared only our {\em plane wave} results with the PWIA predictions shown in Figs.~2, 3 and~8 in Ref.~\cite{PRC80.054001}
and obtained very similar results (not shown). 

Instead we decided to make our {\em full} calculations
for a kinematics close to the settings used 
in NIKHEF experiments~\cite{PhysRevLett.88.102302}: 
$\energyel{e}{\mathbf{p}_e}{}$= 565 MeV, $\theta_e$= 35 degree and $\energyel{e}{\mathbf{p}_e}{'}$= 376 MeV.
This choice leads to $E_{c.m.} \approx $ 159 MeV, thus somewhat above the pion 
production threshold
but the magnitude of the three-momentum transfer is rather small (approx. 335.5 MeV).
Once the electron arm is fixed, we deal with two-body kinematics and can label 
exclusive events by specifying additionally 
the polar and azimuthal angles $\theta_{pq}$ and $\phi_{pq}$
of the outgoing proton momentum. These kinematical variables are defined in the two-nucleon c.m. frame
(see Fig.~\ref{fig:kinematics_diagram}).

\begin{figure}[hb!]
\includegraphics[width=7cm]{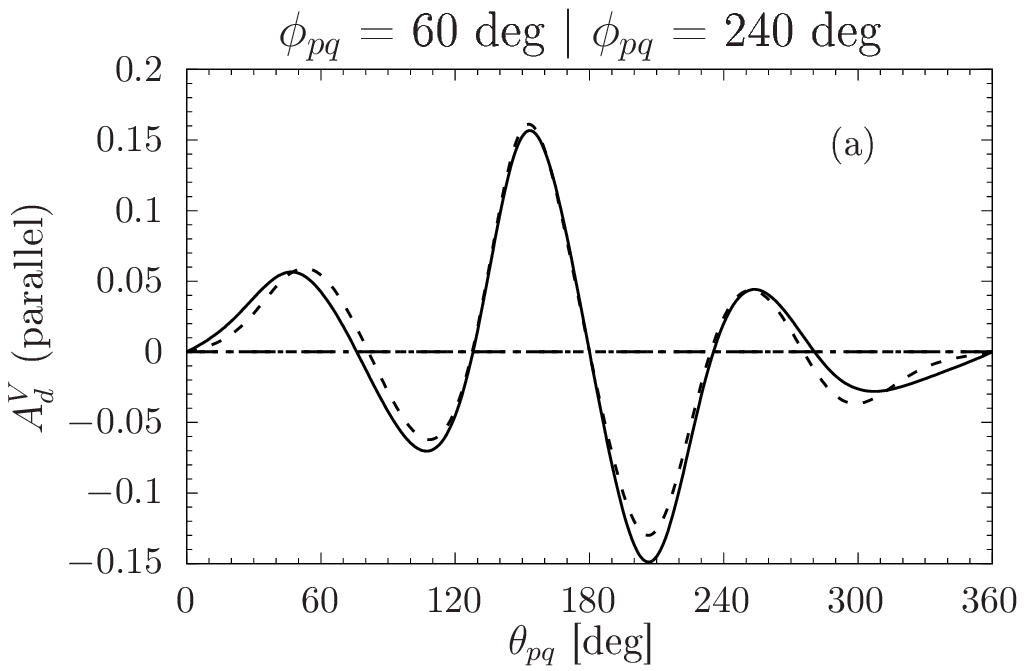}
\includegraphics[width=7cm]{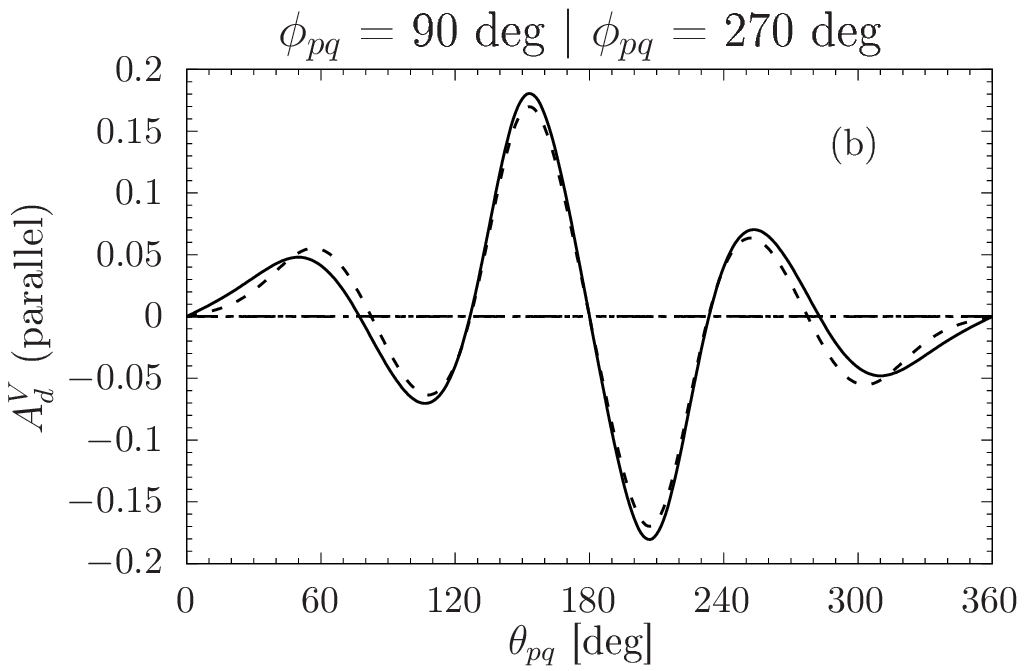}
\includegraphics[width=7cm]{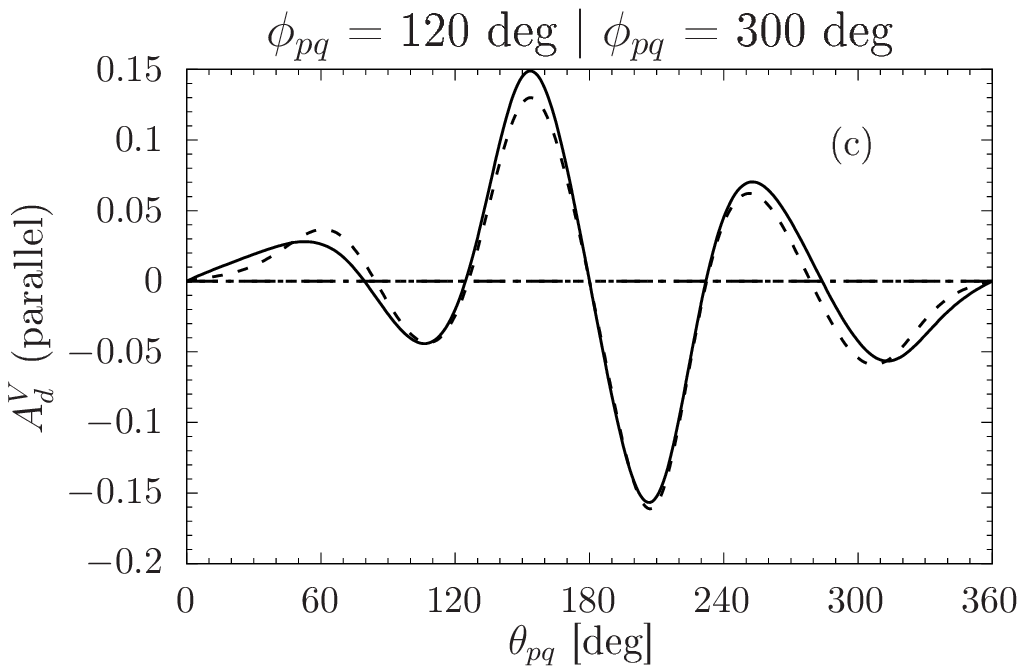}
\includegraphics[width=7cm]{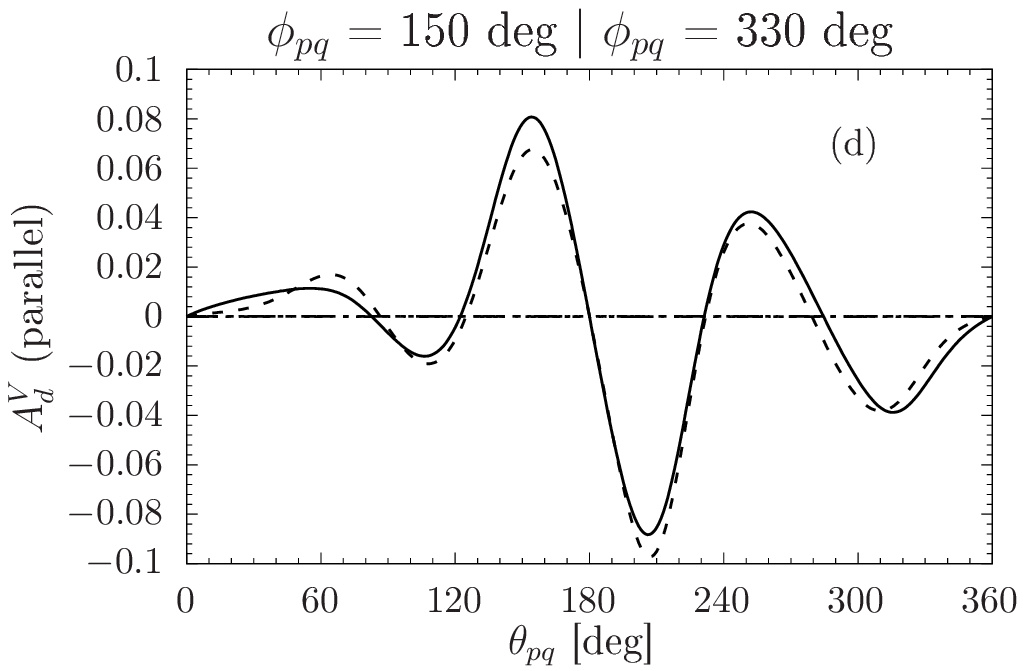}
\caption{
Predictions for $A_d^V$ at the ``NIKHEF'' kinematics for selected values 
of the azimuthal angle $\phi_{pq}$ as functions of the 
polar angle $\theta_{pq}$ for the deuteron polarization axis parallel to the three-momentum transfer.
The solid (dotted) line represents the results with (without) the contribution 
of the rescattering term calculated relativistically.
The dashed  (dash-dotted) line represents the results with (without) the contribution of the rescattering term and calculated within the non-relativistic treatment. For this observable both {\em plane wave} results are identically zero.
}
\label{fig:A1}
\end{figure}

\begin{figure}[hb!]
\includegraphics[width=7cm]{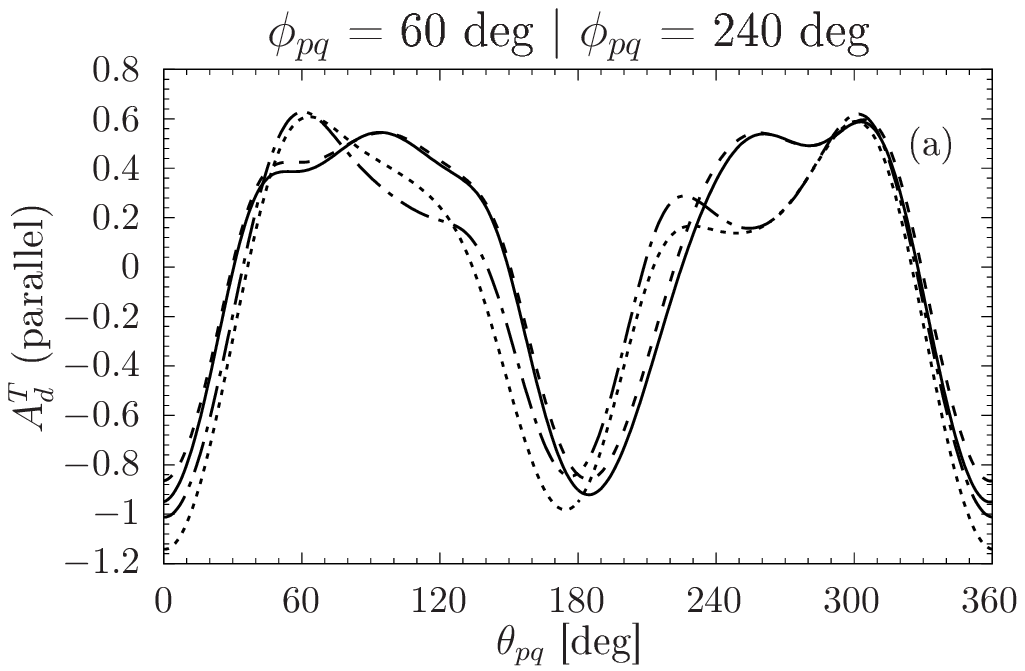}
\includegraphics[width=7cm]{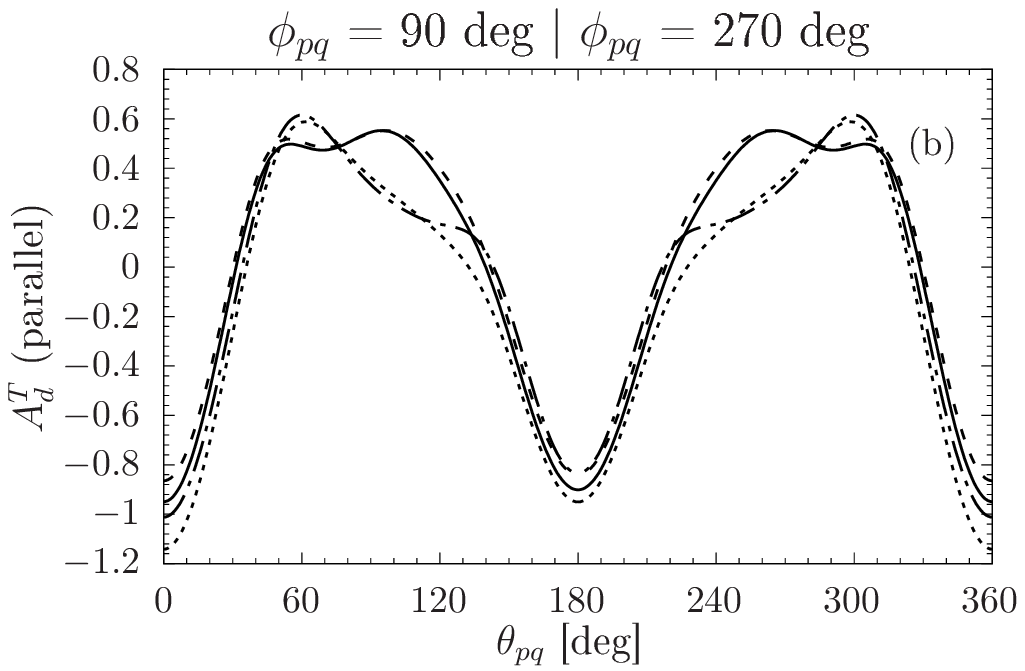}
\includegraphics[width=7cm]{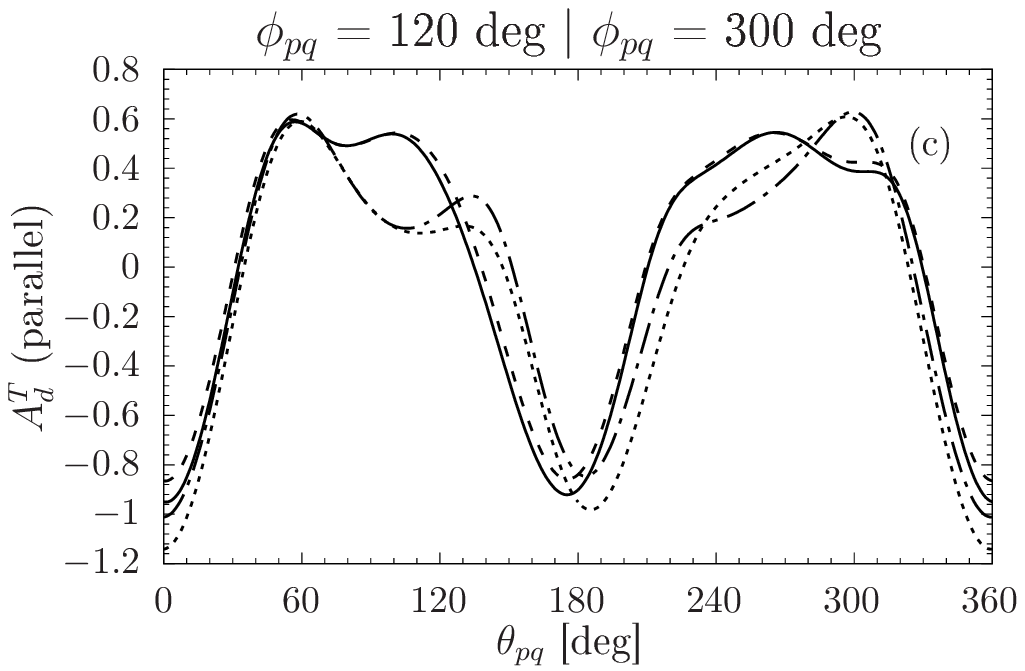}
\includegraphics[width=7cm]{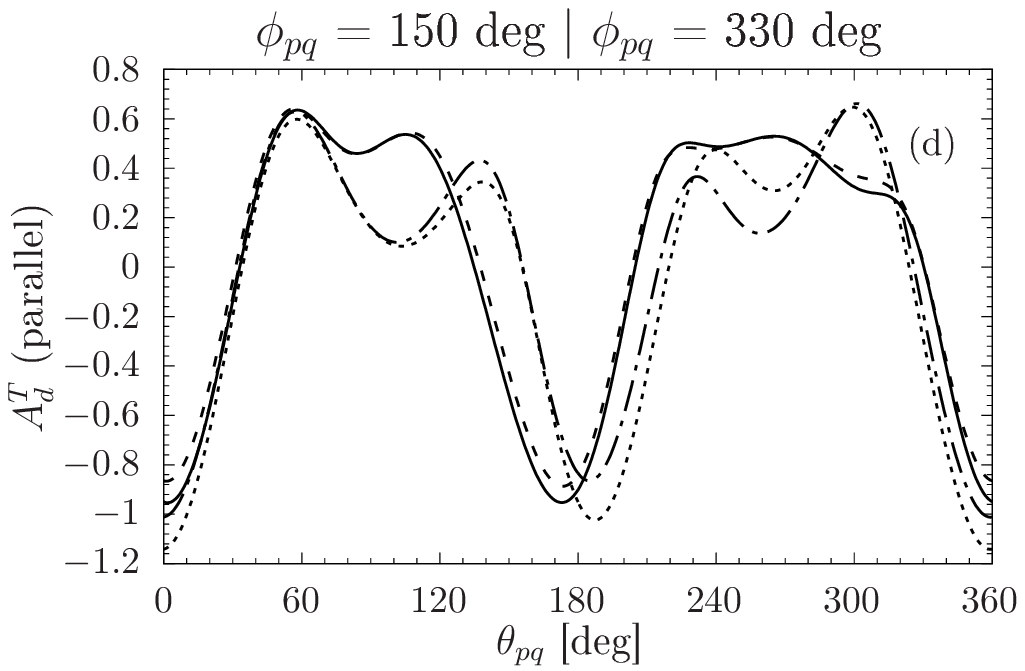}
\caption{
Same as in Fig.~\ref{fig:A1} for $A_d^T$.
}
\label{fig:A3}
\end{figure}

\begin{figure}[hb!]
\includegraphics[width=7cm]{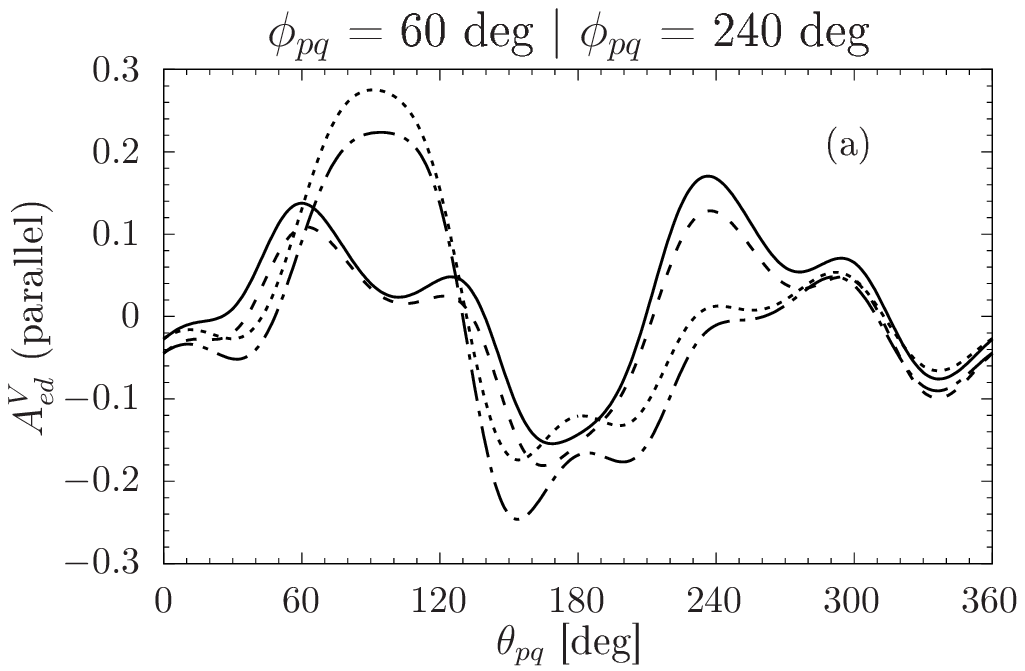}
\includegraphics[width=7cm]{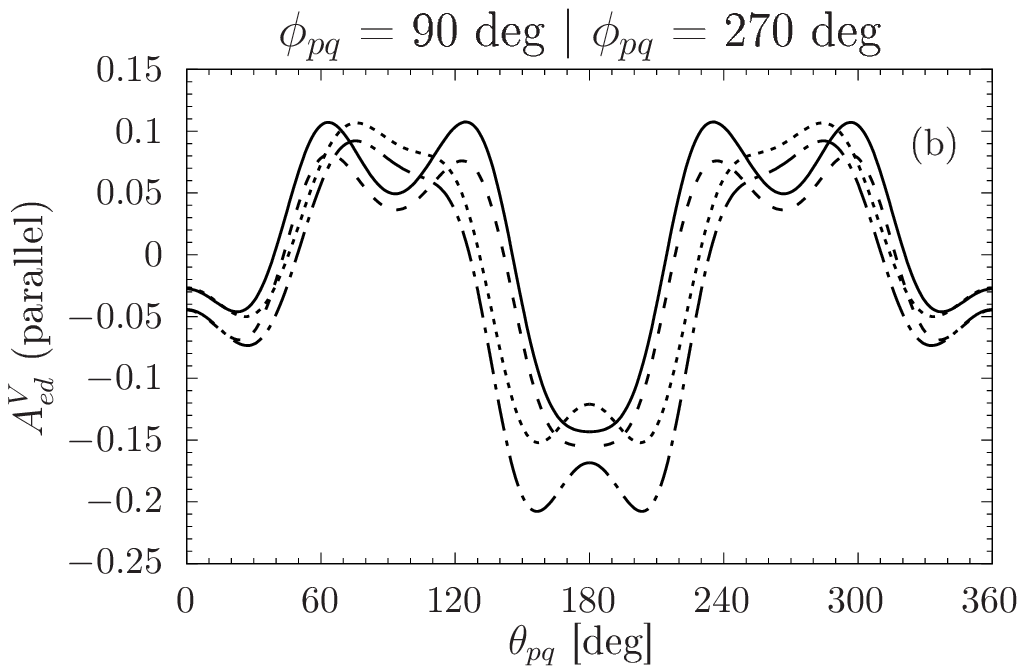}
\includegraphics[width=7cm]{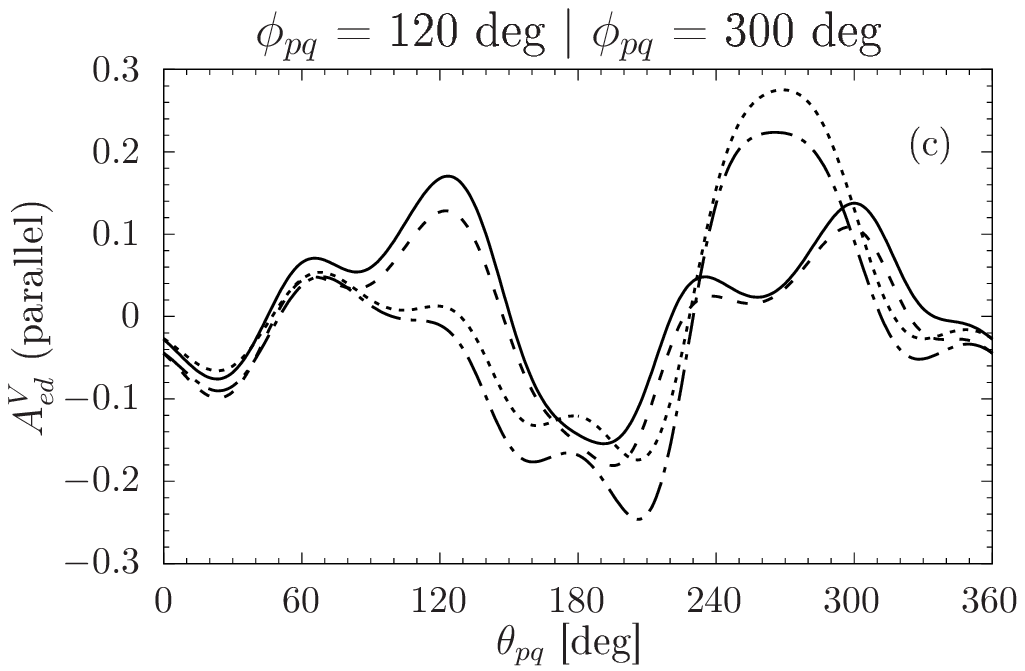}
\includegraphics[width=7cm]{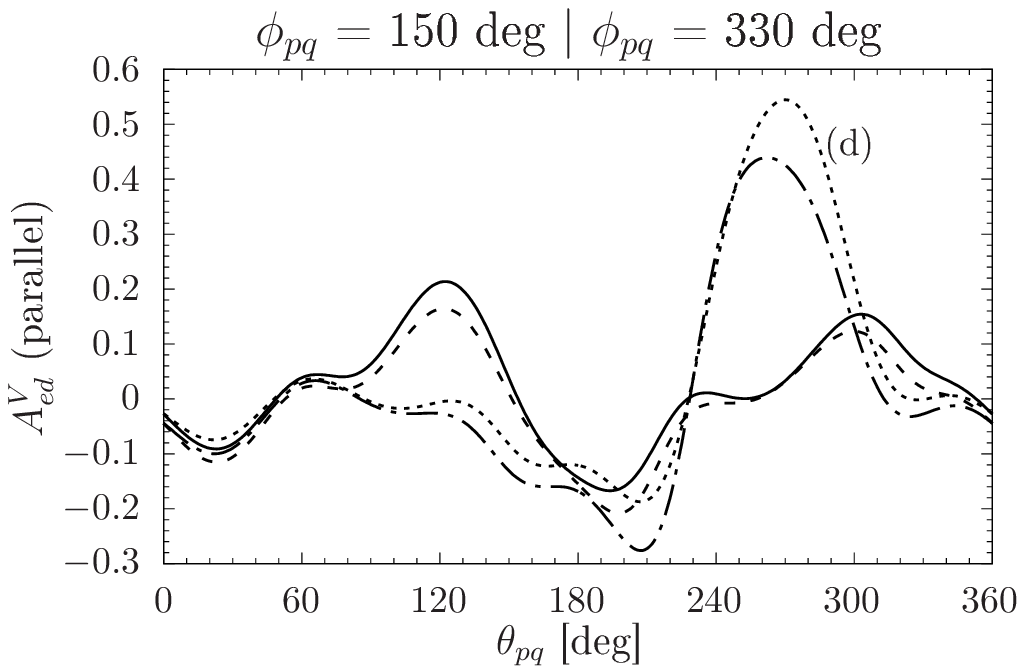}
\caption{
Same as in Fig.~\ref{fig:A1} for $A_{ed}^V$.
}
\label{fig:A5}
\end{figure}

\begin{figure}[hb!]
\includegraphics[width=7cm]{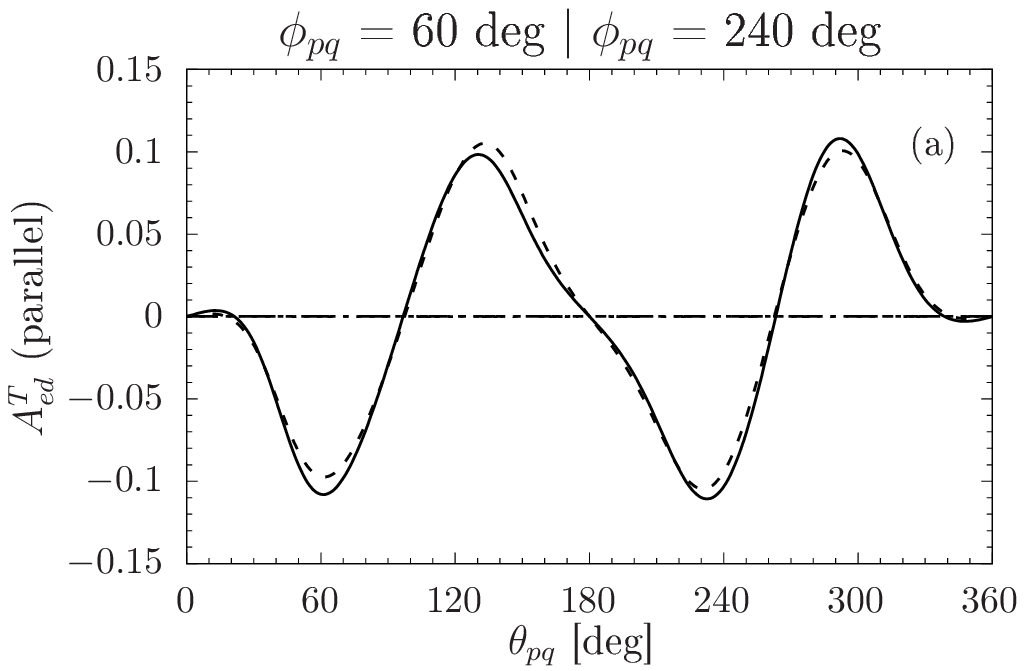}
\includegraphics[width=7cm]{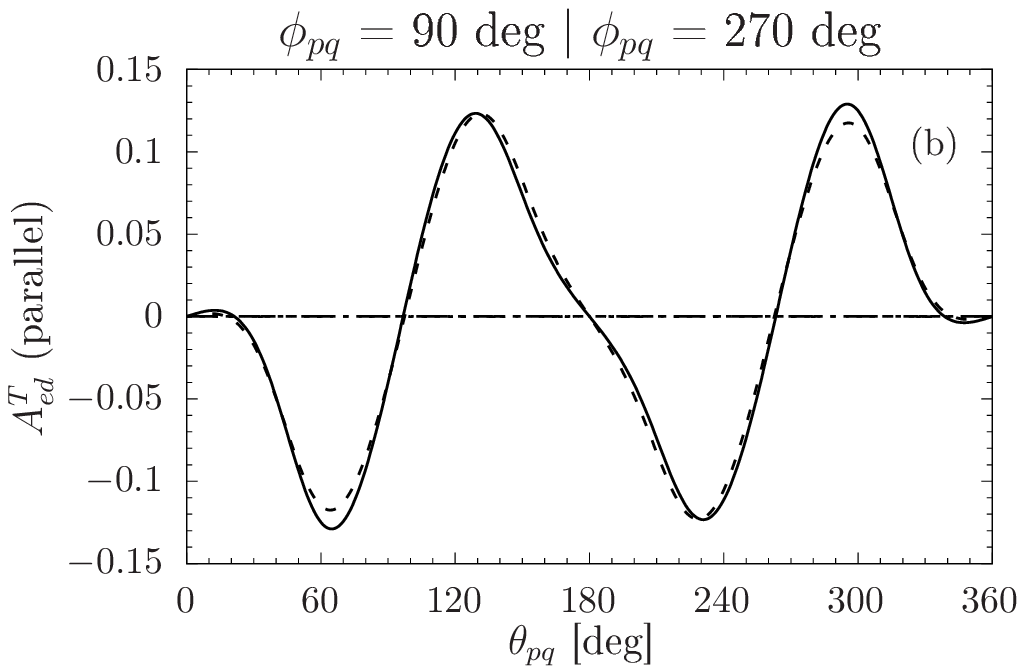}
\includegraphics[width=7cm]{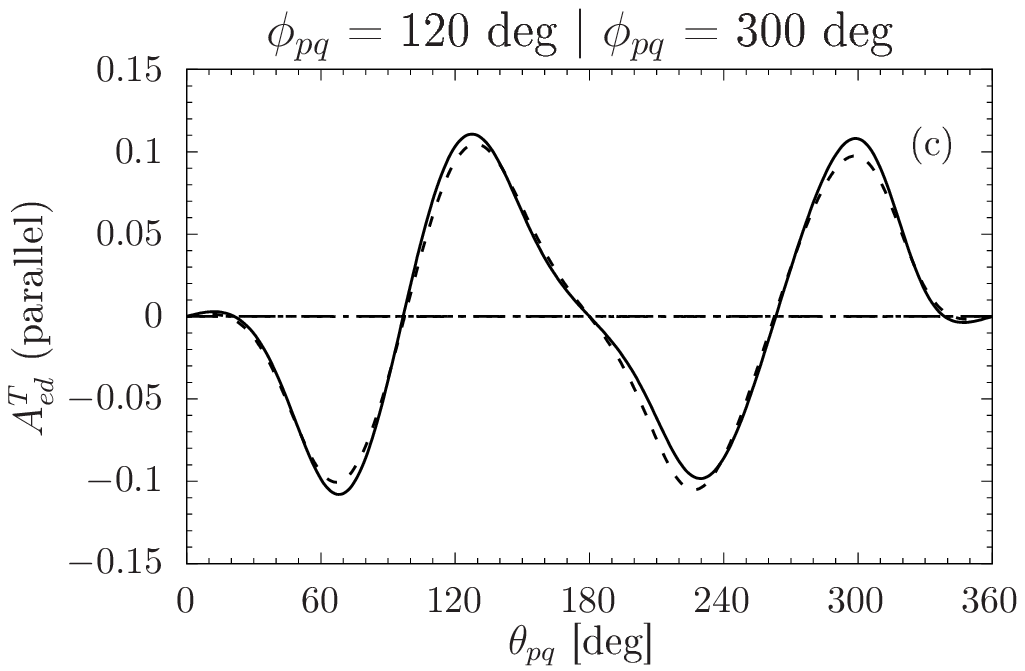}
\includegraphics[width=7cm]{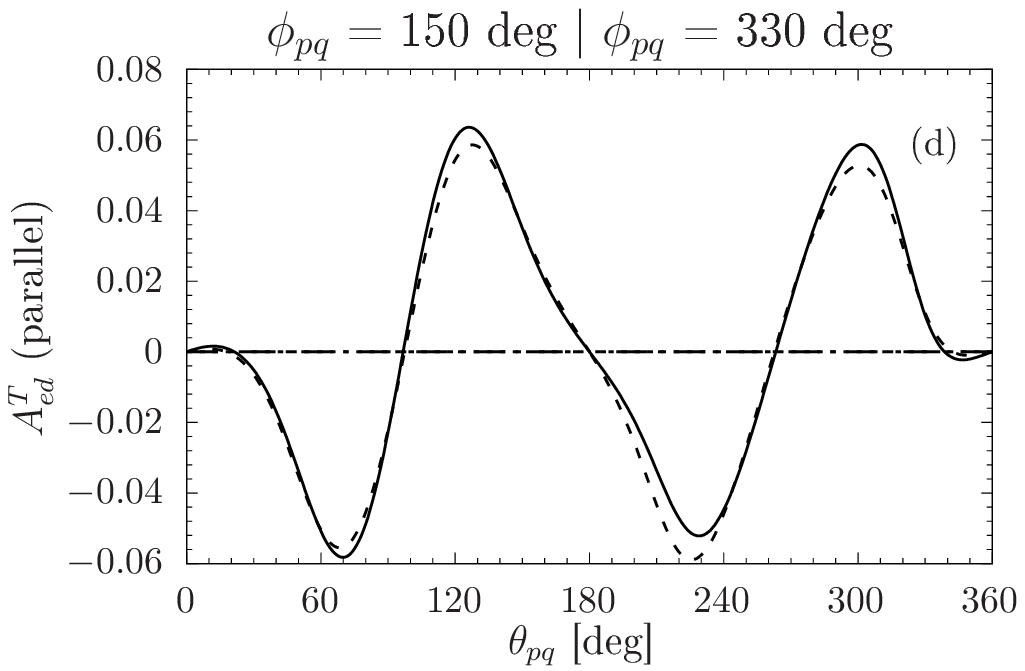}
\caption{
Same as in Fig.~\ref{fig:A1} for $A_{ed}^T$.
}
\label{fig:A7}
\end{figure}

In Figs.~\ref{fig:A1}-\ref{fig:A7} we show our predictions for 
$A_d^V$, $A_d^T$, $A_{ed}^V$, and $A_{ed}^T$ as functions of 
$\theta_{pq}$ for several values of $\phi_{pq}$. Actually we put together 
results for $\phi_{pq} = x$ and $\phi_{pq} = x + 180 \, {\rm deg}$, so $\theta_{pq} > 180 \, {\rm deg}$ at 
$\phi_{pq} = x$ is to be understood as  $360 \, {\rm deg} - \theta_{pq}$ at
$\phi_{pq} = x+ 180 \, {\rm deg}$.
The deuteron polarization axis is chosen parallel
to the three-momentum transfer ${\ve q}$.

Note that the polarization observables show additionally some symmetry properties.
In particular $A_d^V$ and $A_{ed}^T$ are equal zero for $\phi_{pq}$= 0 or 180 deg (not shown).
The {\em plane wave} predictions for $A_d^V$ and $A_{ed}^T$ are identically zero, regardless of $\phi_{pq}$.
Also the non-vanishing {\em plane wave} results for $A_d^T$ and $A_{ed}^V$ are quite different
from the corresponding {\em full} predictions. Actually the difference between the nonrelativistic and relativistic 
results is more pronounced for the {\em plane wave} calculations.

\begin{figure}[hb!]
\includegraphics[width=7cm]{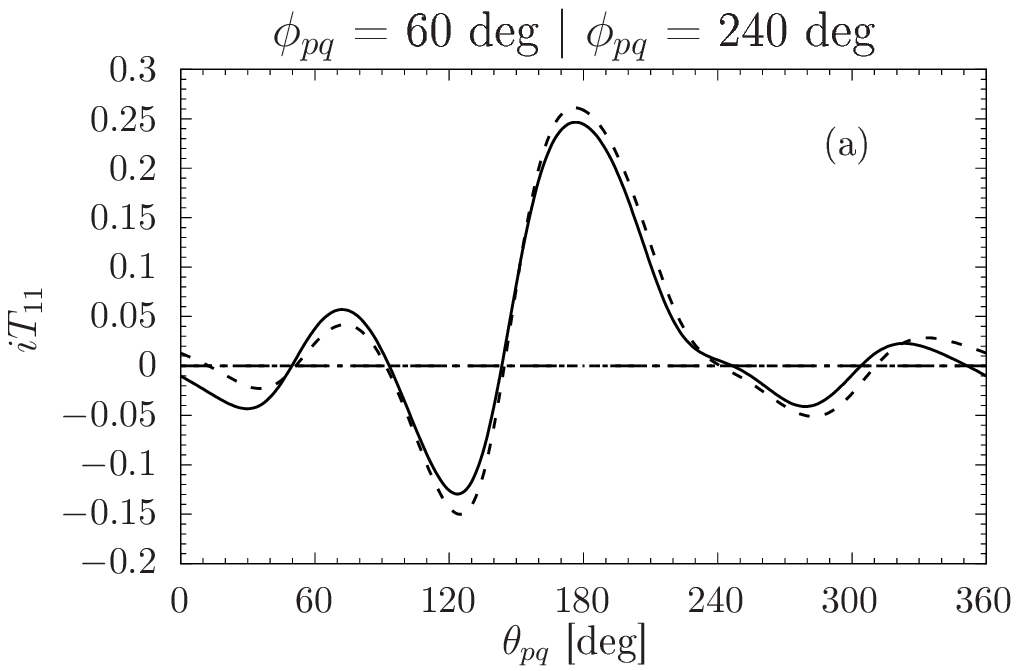}
\includegraphics[width=7cm]{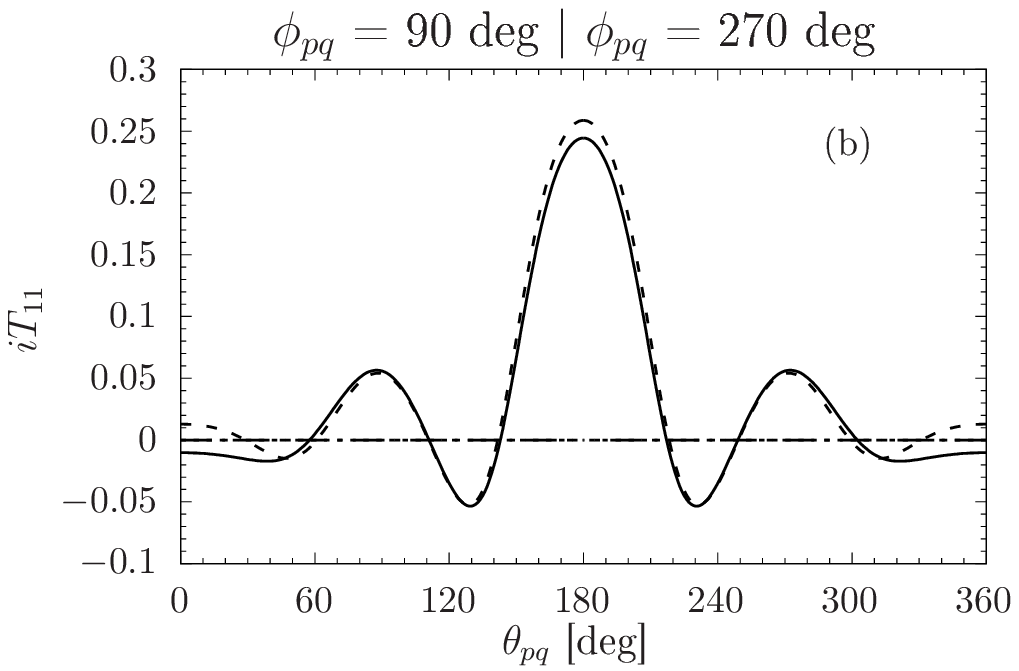}
\includegraphics[width=7cm]{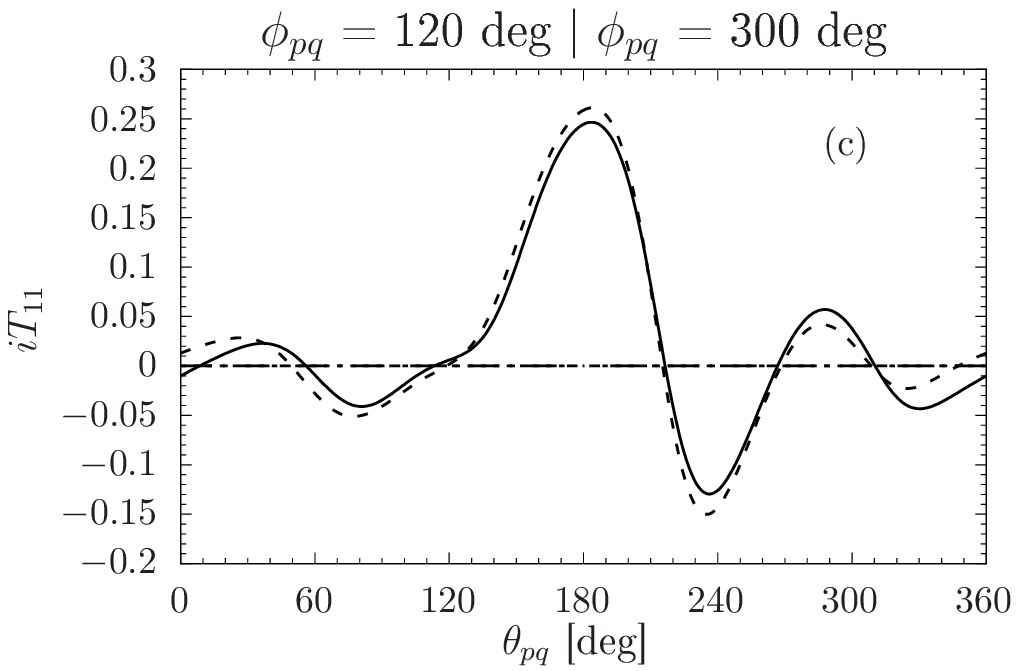}
\includegraphics[width=7cm]{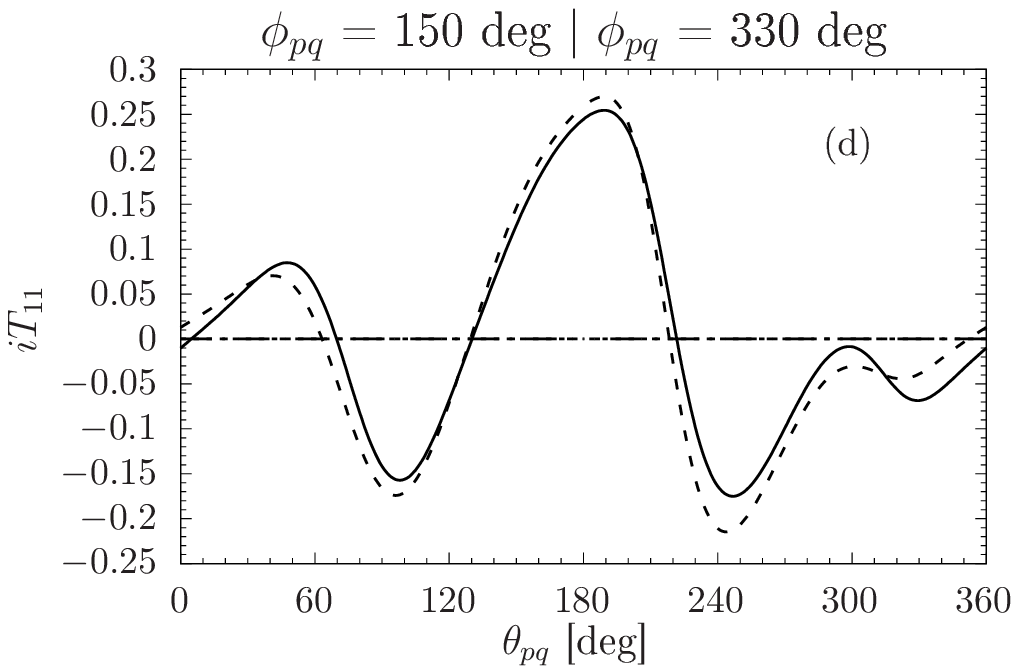}
\caption{
Predictions for the deuteron vector analyzing power $i T_{11}$ at the ``NIKHEF'' kinematics for selected values 
	of the azimuthal angle $\phi_{pq}$ as functions of the polar angle $\theta_{pq}$.
Lines are the same as in Fig.~\ref{fig:A1}.
}
\label{fig:iT11}
\end{figure}

\begin{figure}[hb!]
\includegraphics[width=7cm]{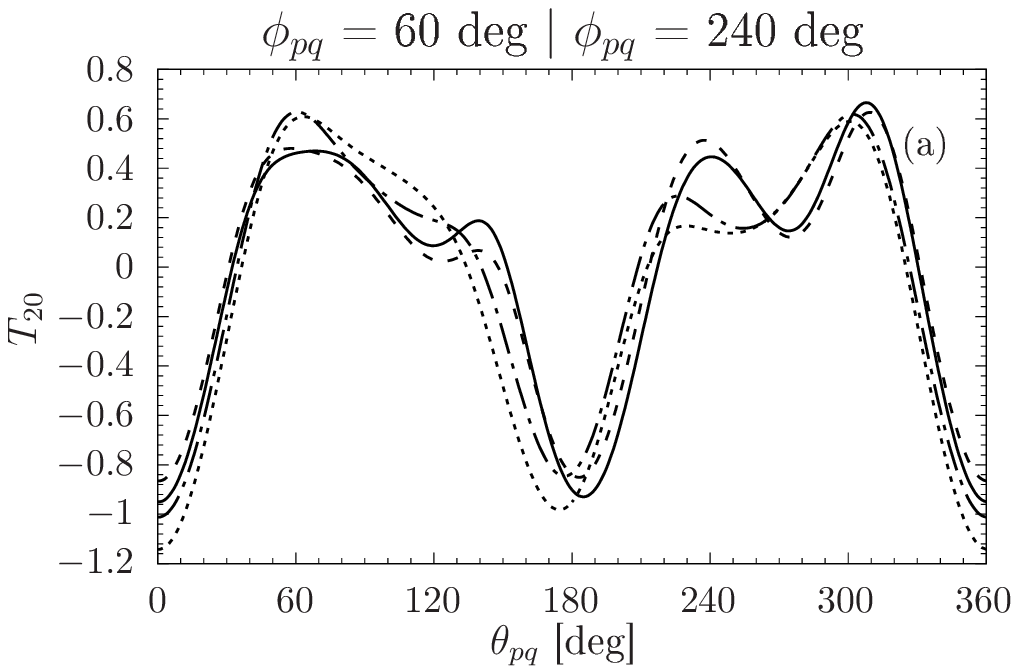}
\includegraphics[width=7cm]{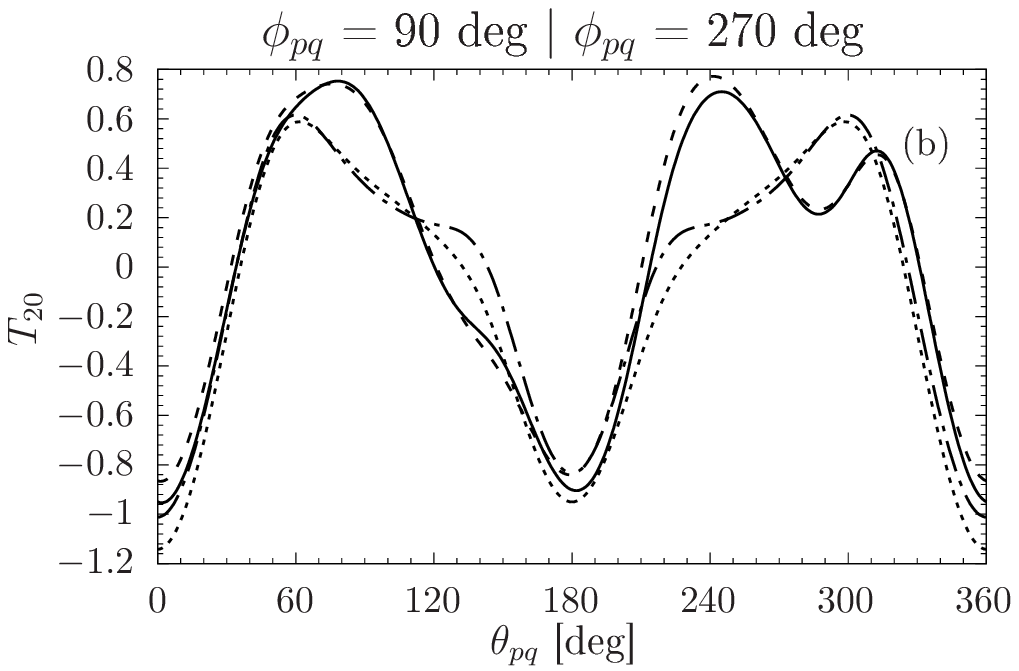}
\includegraphics[width=7cm]{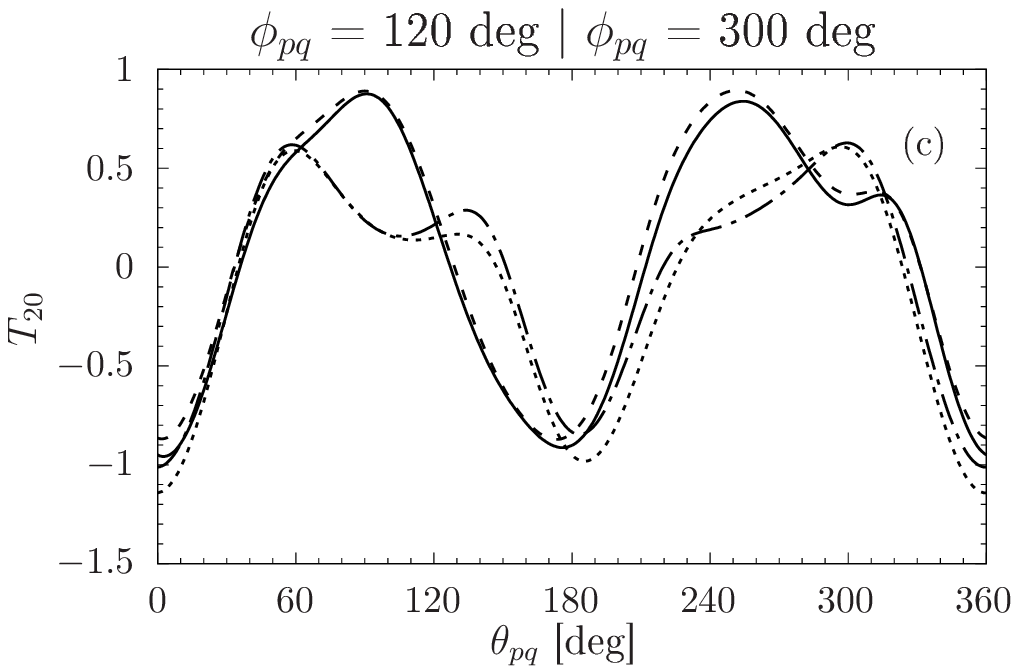}
\includegraphics[width=7cm]{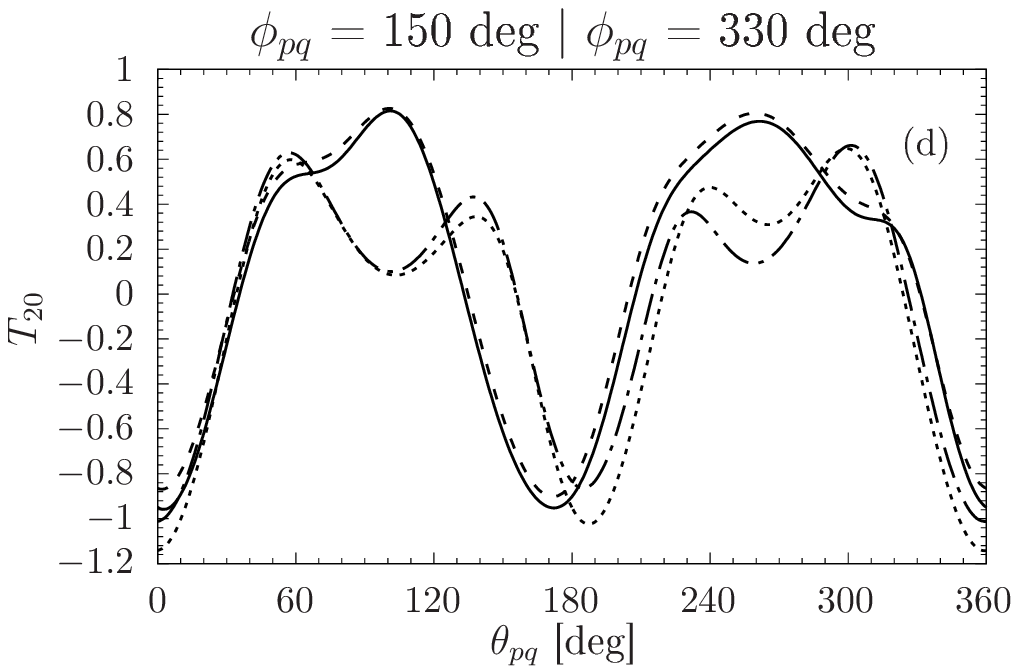}
\caption{
The same as in Fig.~\ref{fig:iT11} 
for the deuteron tensor analyzing power $T_{20}$.
}
\label{fig:T20}
\end{figure}

\begin{figure}[hb!]
\includegraphics[width=7cm]{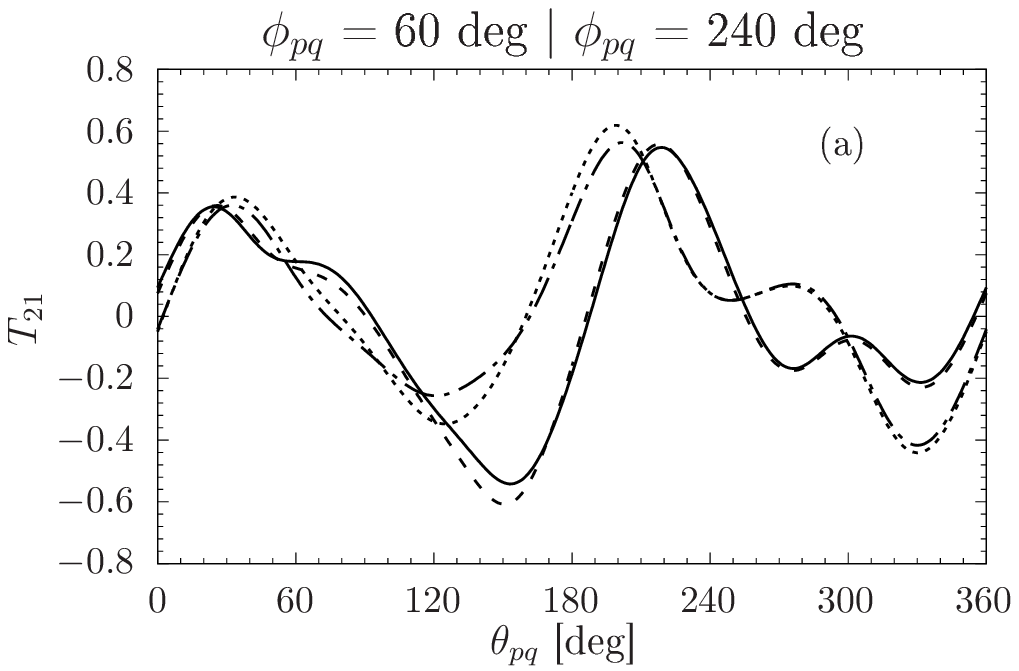}
\includegraphics[width=7cm]{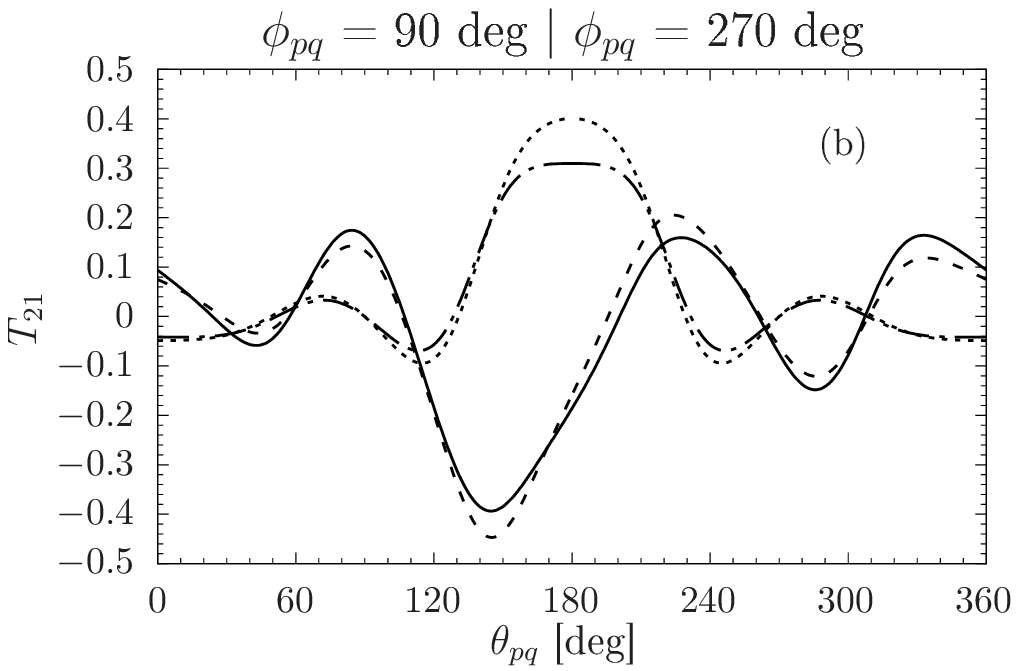}
\includegraphics[width=7cm]{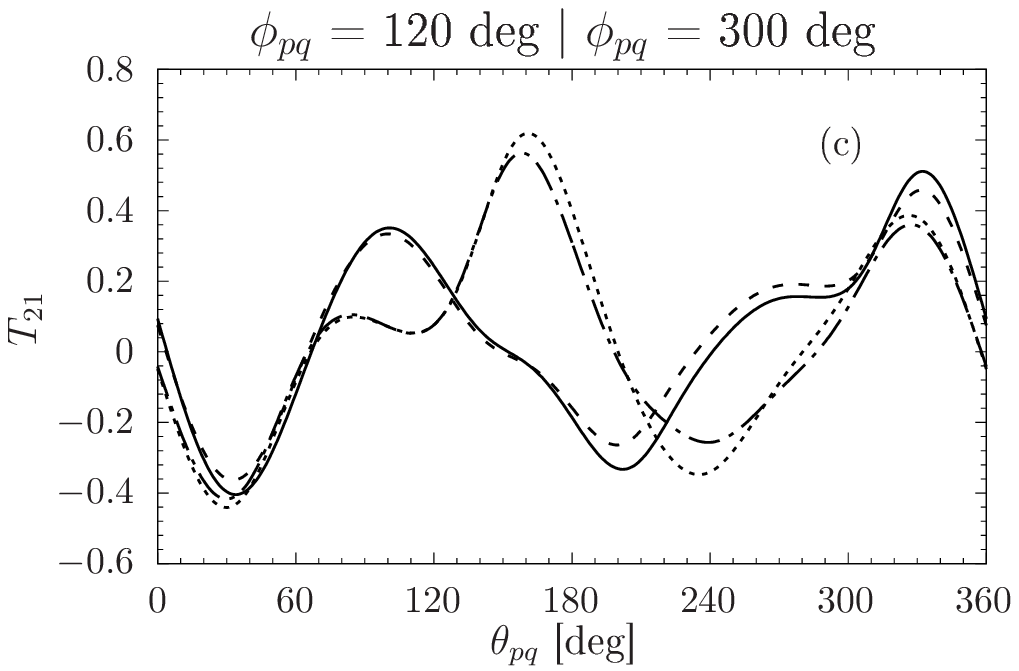}
\includegraphics[width=7cm]{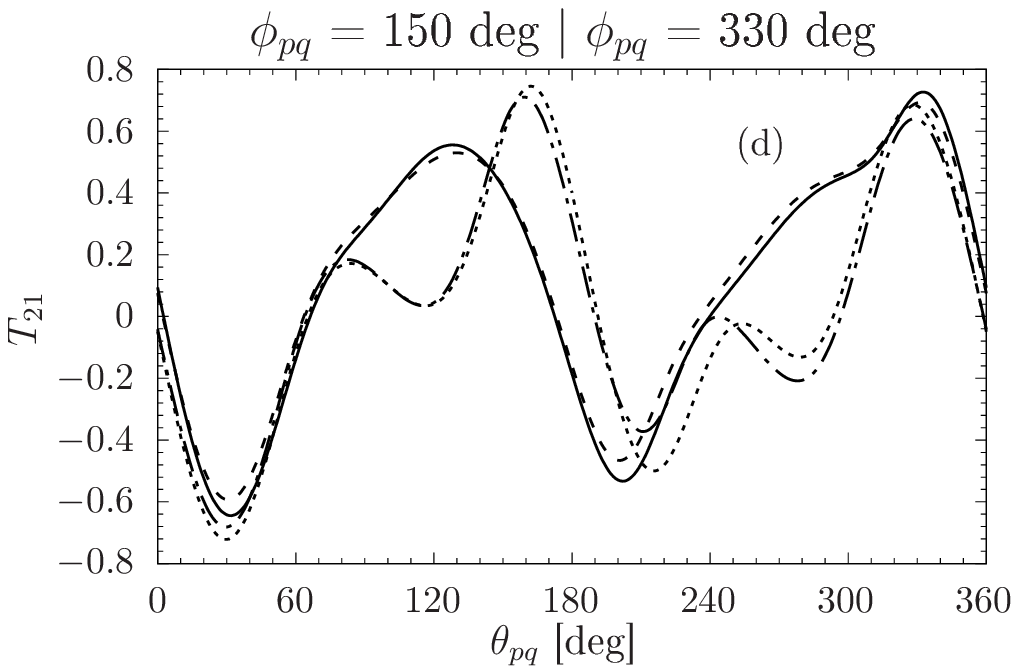}
\caption{
The same as in Fig.~\ref{fig:iT11} 
for the deuteron tensor analyzing power $T_{21}$.
}
\label{fig:T21}
\end{figure}

\begin{figure}[hb!]
\includegraphics[width=7cm]{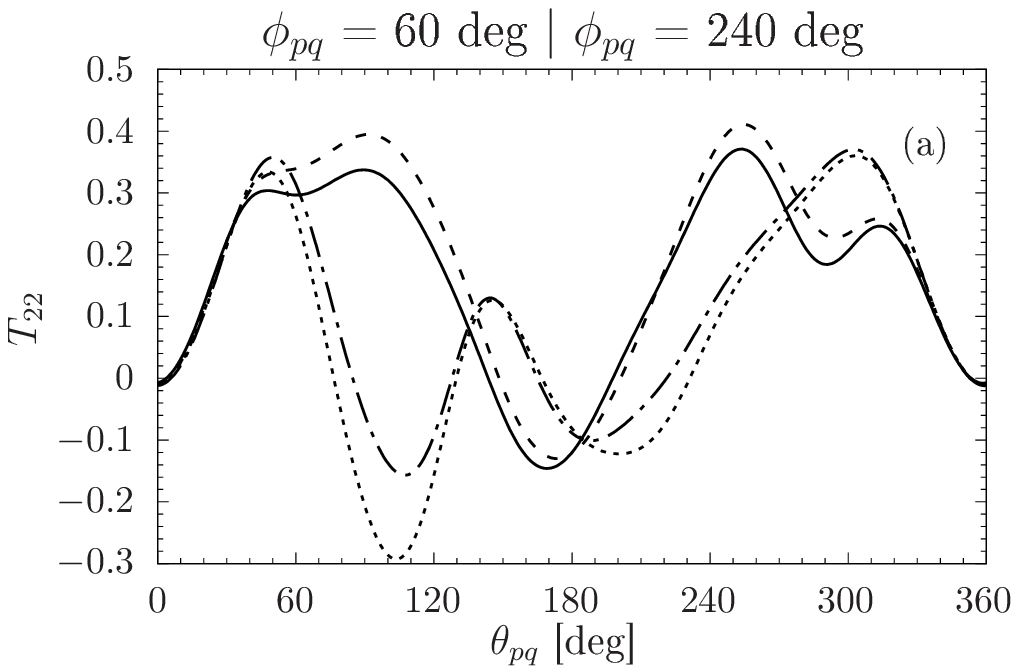}
\includegraphics[width=7cm]{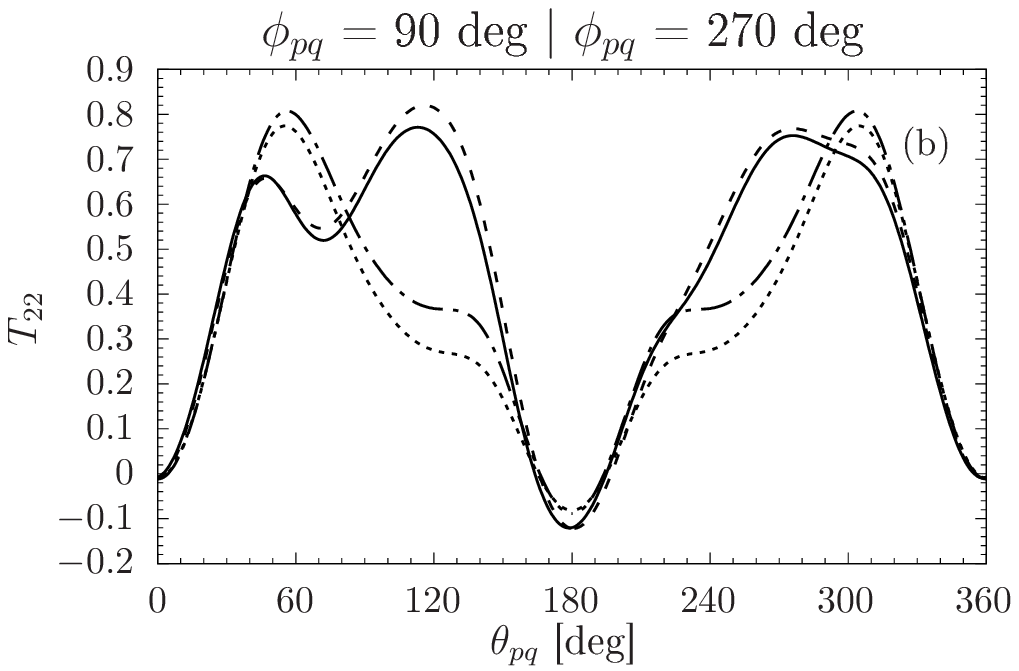}
\includegraphics[width=7cm]{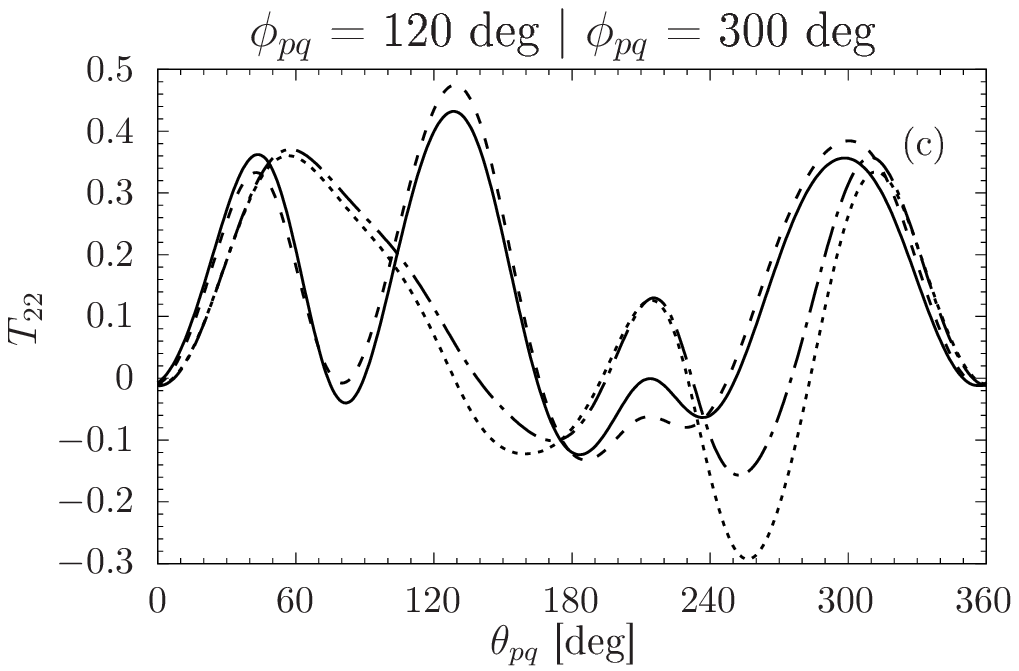}
\includegraphics[width=7cm]{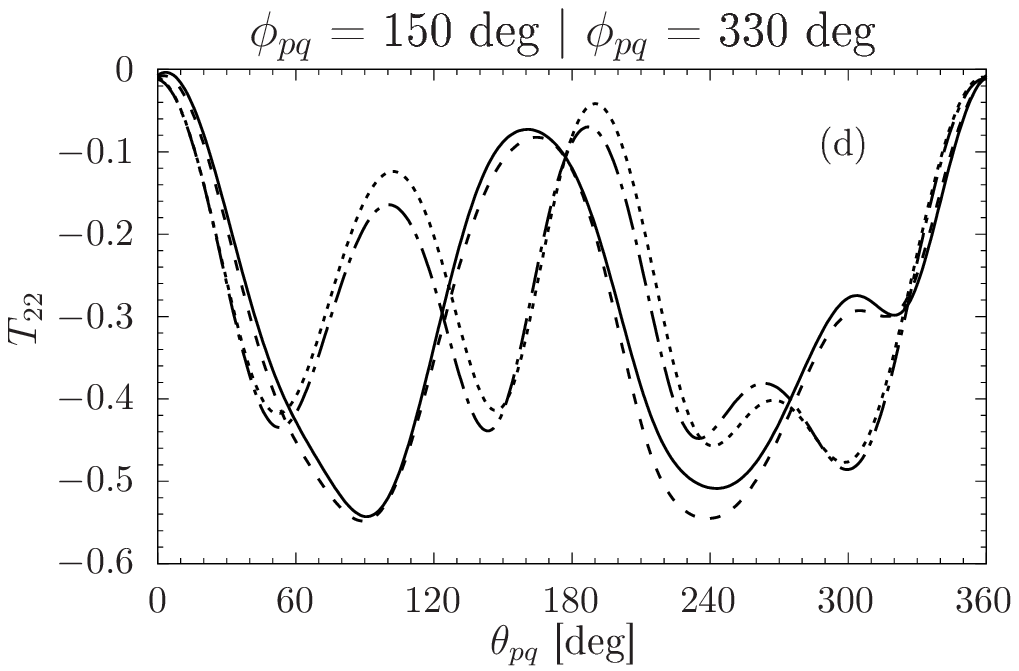}
\caption{
The same as in Fig.~\ref{fig:iT11} 
for the deuteron tensor analyzing power $T_{22}$.
}
\label{fig:T22}
\end{figure}

For the same ``NIKHEF'' electron kinematics we calculated also the more often considered deuteron
analyzing powers $i T_{11}$, $T_{20}$,  $T_{21}$ and $T_{22}$. They are displayed in Figs.~\ref{fig:iT11}-\ref{fig:T22}, respectively.
We see in Fig.~\ref{fig:iT11} that, as expected, $i T_{11}$ is zero in the {\em plane wave} approximation. For the chosen electron kinematics the differences between the relativistic and nonrelativistic predictions are visible but not very strong both 
for the {\em plane wave} and {\em full} results. Generally, the spread between the {\em plane wave} and {\em full} results is much more pronounced, although for some of the analyzing powers there are angular regions, where all the fours curves nearly overlap. 

\subsection{Dealing with ``$Q^2-p_{miss}$'' kinematics in electron induced breakup of $^2$H}

\subsubsection{Unpolarized cross sections}
In order to make transition from
$ {\de^5 \sigma } /\left( { \de\versor{p}_e' \de\energyel{e}{\mathbf{p}_e}{'} \de\versor{k} } \, \right) $
 to the often experimentally regarded 
$ {\de^2 \sigma }/ \left( { \de{Q^2} \, \de p_{\rm miss} } \, \right) $
we employ the relation
\[
 Q^2 \approx 4 \energyel{e}{\mathbf{p}_e}{} \, \energyel{e}{\mathbf{p}_e}{'} \, 
 \sin^2 \frac{\theta_e}{2} \, ,
\]
from which it is clear that one $Q^2$ value can be obtained taking 
various $( \energyel{e}{\mathbf{p}_e}{'} \, , \theta_e )$ pairs. 
Next, we take finite bins 
in $Q^2$ and $p_{\rm miss}$, scanning the whole four dimensional 
parameter space to see, which combinations of 
$( \theta_e , \energyel{e}{\mathbf{p}_e}{'} , \theta_k , \phi_k \, )$ 
lead to required $Q^2$ and $p_{\rm miss}$ bins.

\begin{figure}
\begin{center}
\includegraphics[width=.45\textwidth,clip=true]{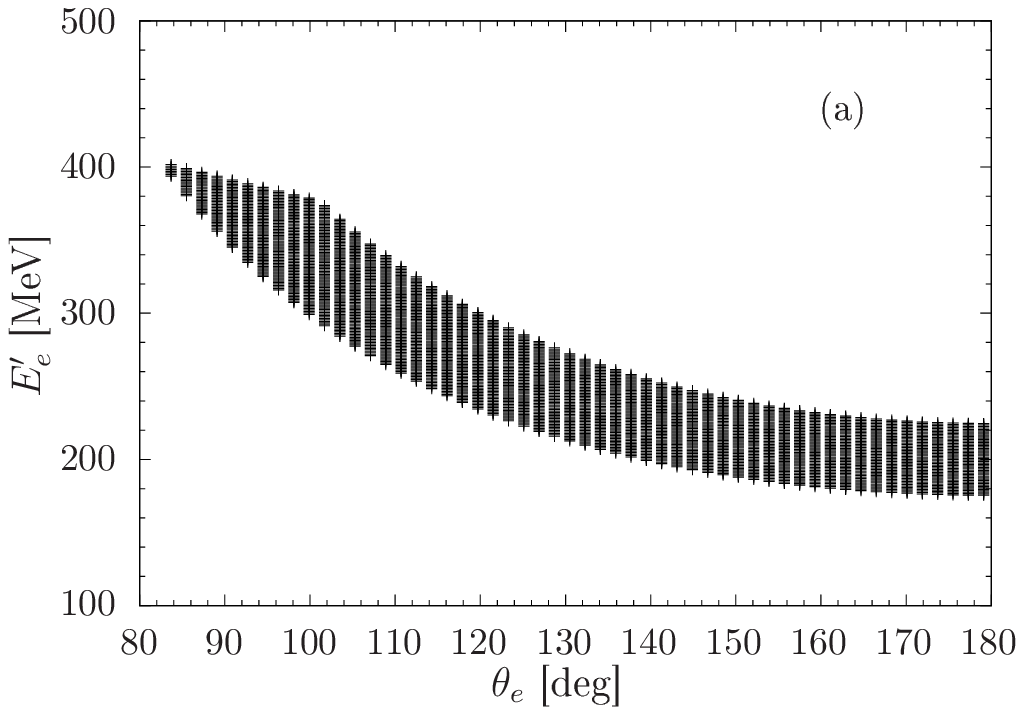}
\includegraphics[width=.45\textwidth,clip=true]{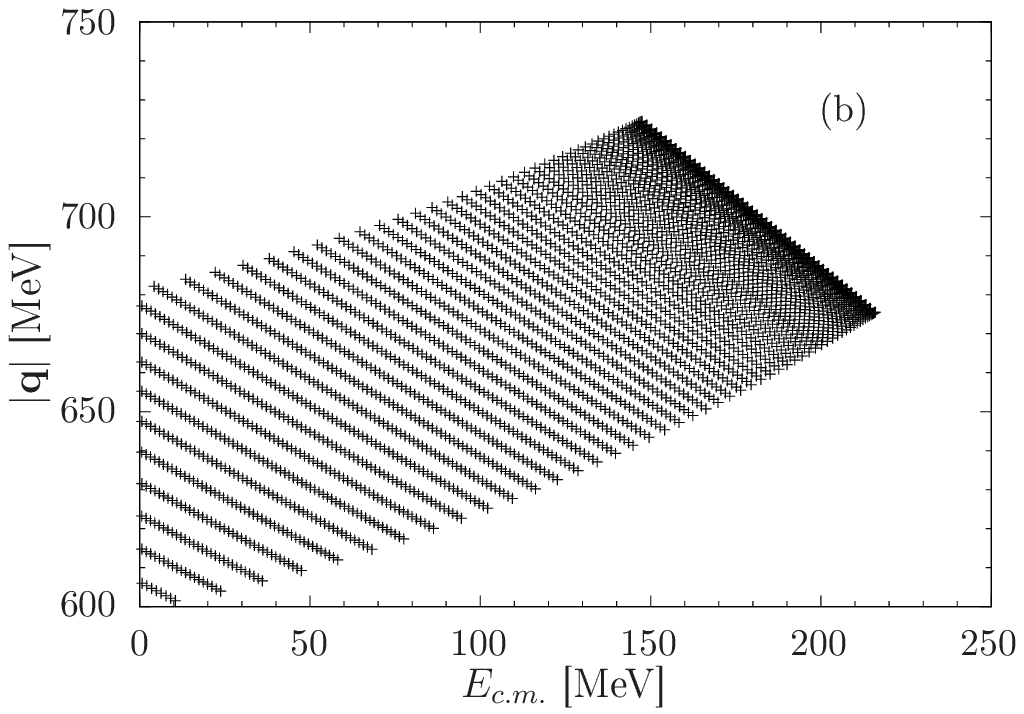}
\includegraphics[width=.45\textwidth,clip=true]{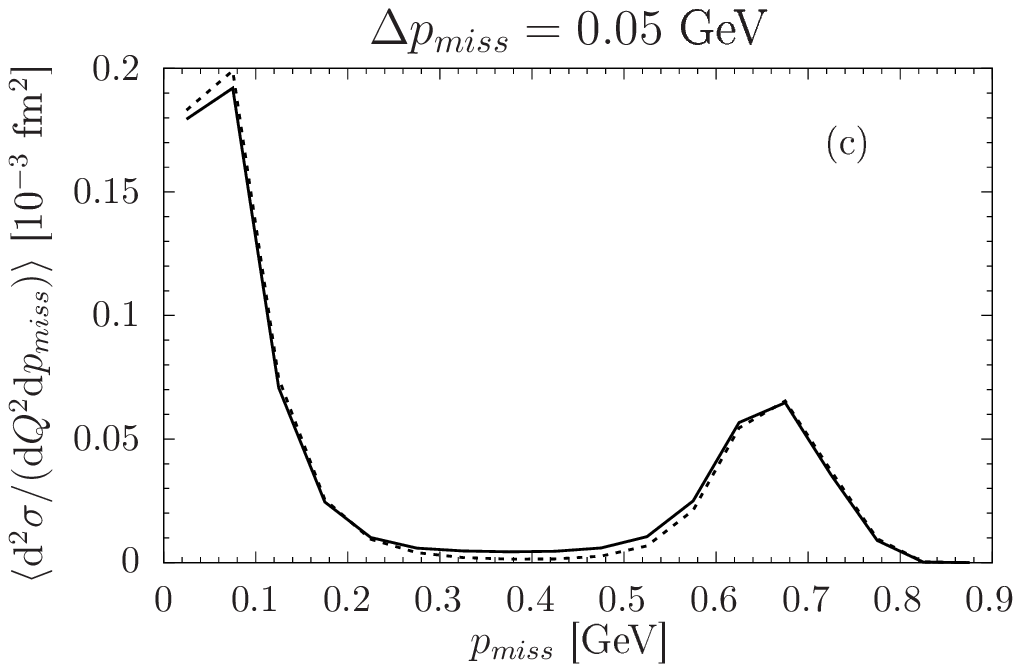}
\includegraphics[width=.45\textwidth,clip=true]{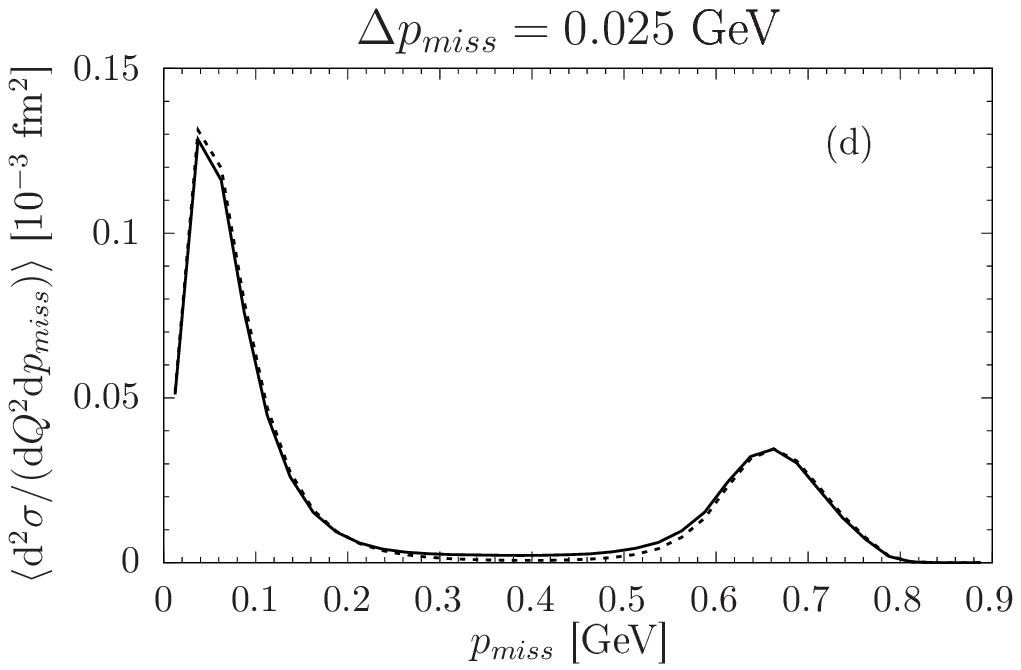}
\caption{
Kinematical regions in the $(\theta_e, E'_e)$ (a) 
and $(E_{c.m.} , |\ve{q}|) $ (b) planes 
for $ 0.35 \, {\rm GeV}^2 < Q^2 < 0.45 \, {\rm GeV}^2 $
and $\energyel{e}{\ve{p}_e}{}$=500~MeV. The lower panels show relativistic cross sections summed over the phase-space regions defined additionally by a finite $p_{miss}$ bin and labelled by the bin position for two bin widths: 0.05 GeV (a) and 0.025 GeV (b)   
obtained with the plane wave impulse approximation (dotted lines) 
and performing the {\em full} calculations (solid lines).
	}
\label{FIG(3)}
\end{center}
\end{figure}

\begin{figure}
\begin{center}
\includegraphics[width=.45\textwidth,clip=true]{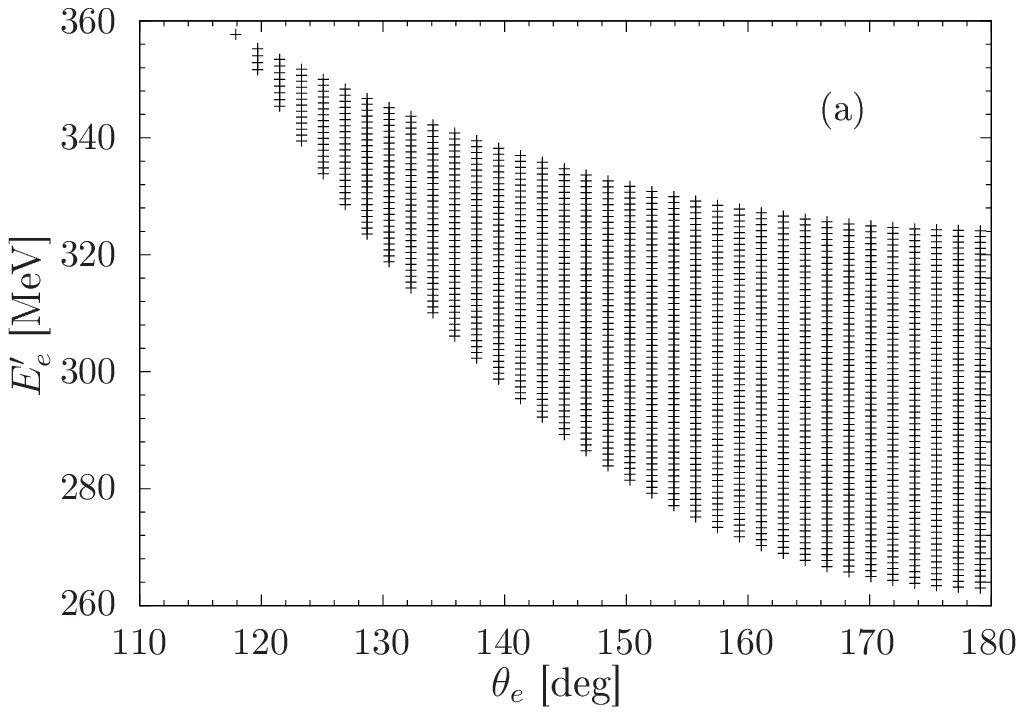}
\includegraphics[width=.45\textwidth,clip=true]{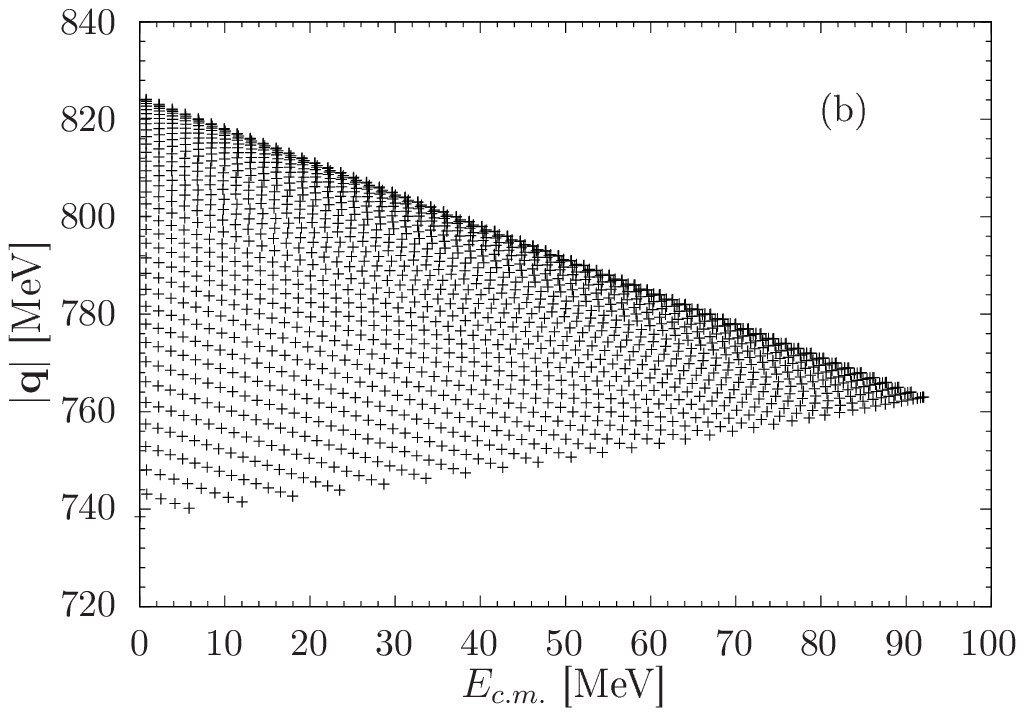}
\includegraphics[width=.45\textwidth,clip=true]{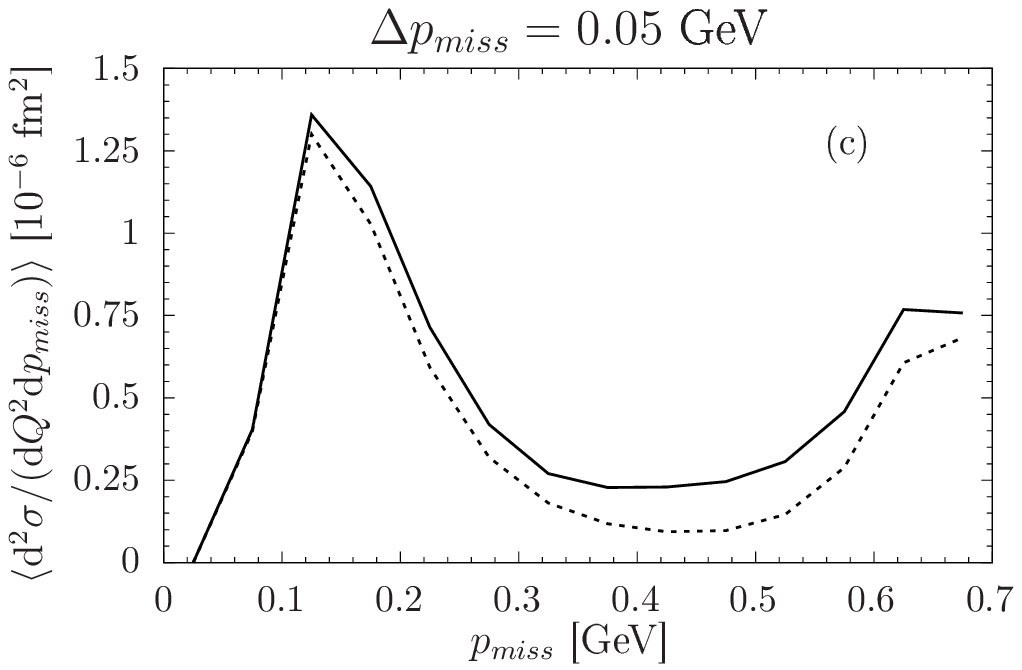}
\includegraphics[width=.45\textwidth,clip=true]{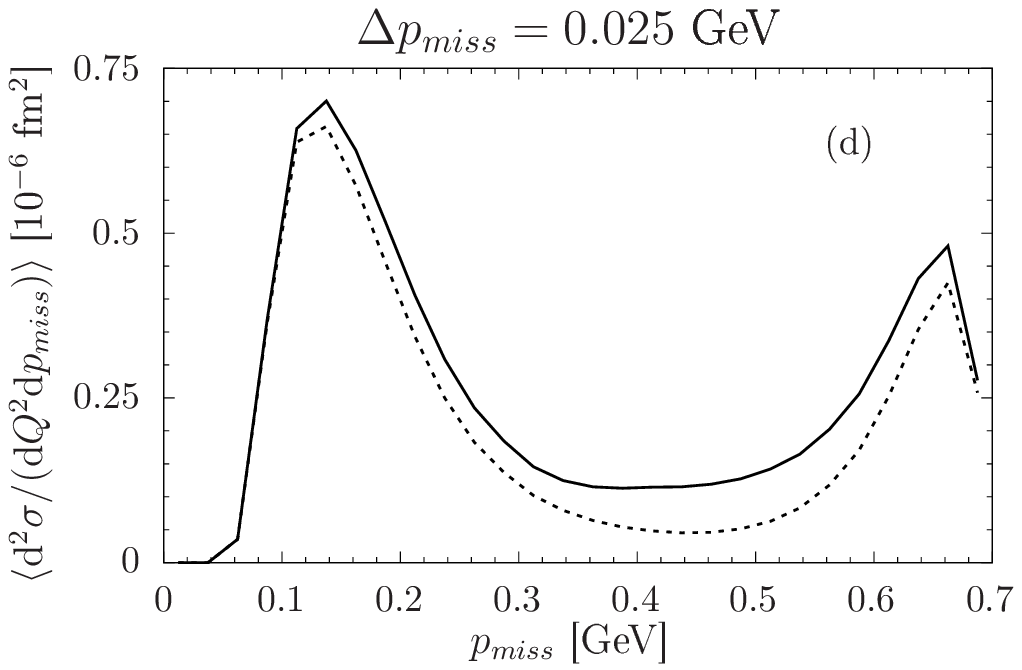}
\caption{The same as in Fig.~\ref{FIG(3)} for $ 0.525 \, {\rm GeV}^2 < Q^2 < 0.675 \, {\rm GeV}^2 $.}
\label{FIG(4)}
\end{center}
\end{figure}

A similar procedure is often used in experiments
(see for example Ref.~\cite{PRL98.262502}), where additionally 
requirements given by the experimental set-up need to be taken into account.
The experimental electron kinematics from Ref.~\cite{PRL98.262502} cannot
be used in our {\em full} calculations 
so we prepared an example for the initial electron energy 
$\energyel{e}{\mathbf{p}_e}{}$= 500 MeV and
chose four $Q^2$ intervals: 
$ \left( 0.0875, 0.1125 \right) {\rm GeV}^2 $, 
$ \left( 0.175, 0.225 \right) {\rm GeV}^2 $, 
$ \left( 0.35, 0.45 \right) {\rm GeV}^2 $, 
and $ \left( 0.525, 0.675 \right) {\rm GeV}^2 $.
For each $Q^2$ interval we proceeded in the following way:
we took 100 uniformly distributed $\theta_e$ points such that $ 0 < \theta_e (i) < \pi$.
For each $\theta_e (i)$ we calculated the kinematically allowed range
of the outgoing electron energies such that $Q^2$ fell into the desired interval.
The number of the  $ \energyel{e}{\mathbf{p}_e}{'} $ points depended on the length 
of this interval and varied from $1$ to $100$. Then for each 
$ \left( \energyel{e}{\mathbf{p}_e}{} , \theta_e (i) , \energyel{e}{\mathbf{p}_e}{'} (j) \right) $ set 
we ran a double loop over 72 $\theta_k (l) $ and 36 $\phi_k (n) $ values (again uniformly distributed
from $0$ to $\pi$ and from $0$ to $2\pi$, respectively)
and generated the differential cross section 
$ {\de^5 \sigma }/ \left( { \de\versor{p}_e' \de\energyel{e}{\mathbf{p}_e}{'} \de\versor{k} } \, \right) $.
The values of this cross section (calculated just with the plane wave approximation
or including also final state interactions) were written to a file 
together with the complete integral weight, which was
\[
	2 \pi \; \Delta\theta_e(i) \; \sin\left( \theta_e(i) \right) \;
	\Delta\energyel{e}{\mathbf{p}_e}{'}(j) \;
	\Delta\theta_k(l) \; \sin \left( \theta_k(l) \right) \;
	\Delta\phi_k(n) \;
\]
and with the value of the missing momentum (the magnitude of the neutron momentum).
During computations for each $Q^2$ interval we created a file with several millions lines
but that allowed us to sort these ''events'' according to the $p_{\rm miss}$ value
and sum up all contributions belonging to a desired interval of the $\Delta p_{\rm miss}$ length.
In this manner we obtained 
\[
\left\langle 	\frac{\de^2 \sigma}{\de{Q^2} \de{p_{\rm miss}}}  \right\rangle \equiv 
	\int\limits_{Q^2_{min}}^{Q^2_{max}} \de{Q^2} 
	\int\limits_{{\bar p}_{\rm miss} - \frac12\Delta p_{\rm miss}}^{{\bar p}_{\rm miss} + \frac12\Delta p_{\rm miss}} 
	\de{p_{\rm miss}} \, 
	\frac{\de^2 \sigma}{\de{Q^2} \de{p_{\rm miss}}} \, ,
\]
which will be then presented as a function of ${\bar p}_{\rm miss}$.
It turned out that $E_{c.m.}$ values for the
$ Q^2 \in \left( 0.0875, 0.1125 \right) {\rm GeV}^2 $
and
$ Q^2 \in \left( 0.175, 0.225 \right) {\rm GeV}^2 $ intervals
exceed by far the pion mass so we restrict ourselves 
to the two other cases 
$ Q^2 \in \left( 0.35, 0.45 \right) {\rm GeV}^2 $
and
$ Q^2 \in  \left( 0.525, 0.675 \right) {\rm GeV}^2 $,
see Figs.~\ref{FIG(3)} and \ref{FIG(4)}.
For these two $Q^2$ intervals we investigated 
$ \left\langle    \frac{\de^2 \sigma}{\de{Q^2} \de{p_{\rm miss}}}  \right\rangle $
under some kinematical dependencies
 calculated 
with two different $\Delta p_{\rm miss}$: 0.05 and 0.025 GeV.
Since the kinematics at these $Q^2$ values is definitely relativistic,
we show only our relativistic predictions.  
For $ Q^2 \in \left( 0.35, 0.45 \right) {\rm GeV}^2 $ the {\em full} results are very close to the {\em plane wave} predictions but for $ Q^2 \in \left( 0.525, 0.675 \right) {\rm GeV}^2 $
the two types of calculations yield different results for $p_{miss} \gtrsim $ 0.1 GeV.

\subsubsection{Polarization observables in the ``$Q^2-p_{miss}$'' kinematics}

Also polarization observables discussed in Refs.~\cite{PRC80.054001,PRC95.024005}
can be studied in the ``$Q^2-p_{miss}$'' kinematics or be summed over any part of the available phase space.
However, if we want to use exactly the same kinematics as for the cross section, where we integrate over the azimuthal 
angle $\Phi_p$, then $ \langle A_d^V \rangle = \langle A_{ed}^T \rangle = 0 $. 
It means that some other kinematics has to be chosen to study these two observables, which 
additionally vanish under the plane wave impulse approximation. 

\begin{figure}
\begin{center}
\includegraphics[width=.45\textwidth,clip=true]{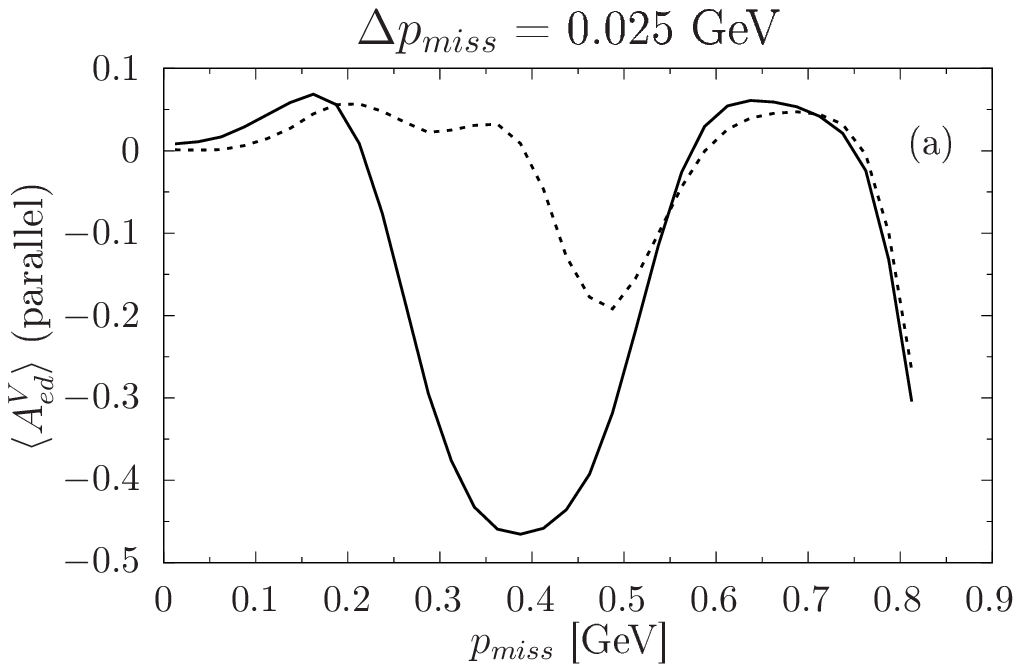}
\includegraphics[width=.45\textwidth,clip=true]{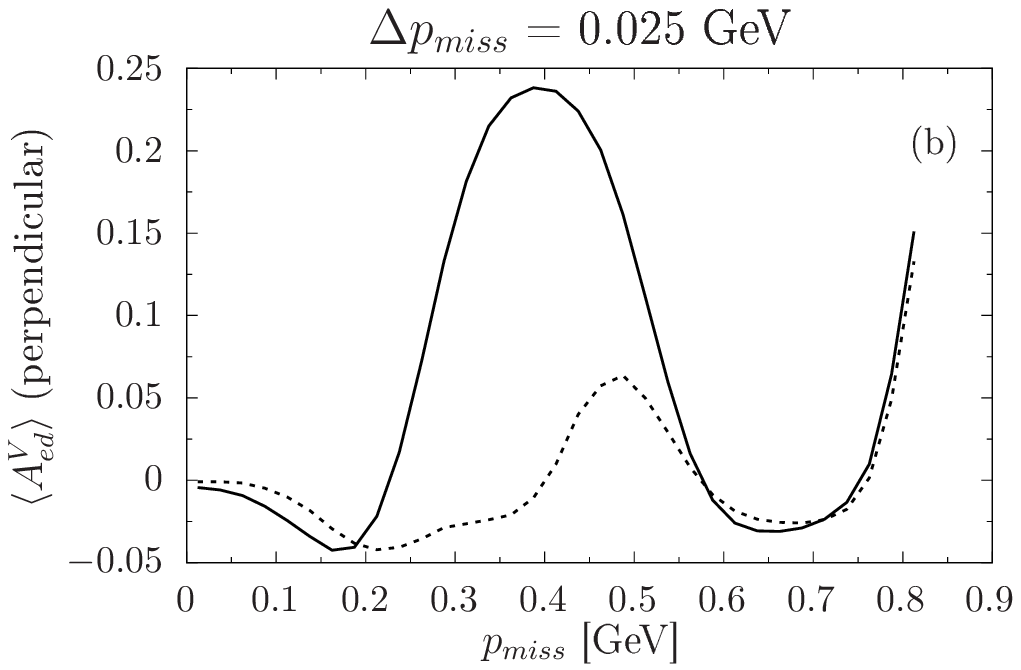}
\includegraphics[width=.45\textwidth,clip=true]{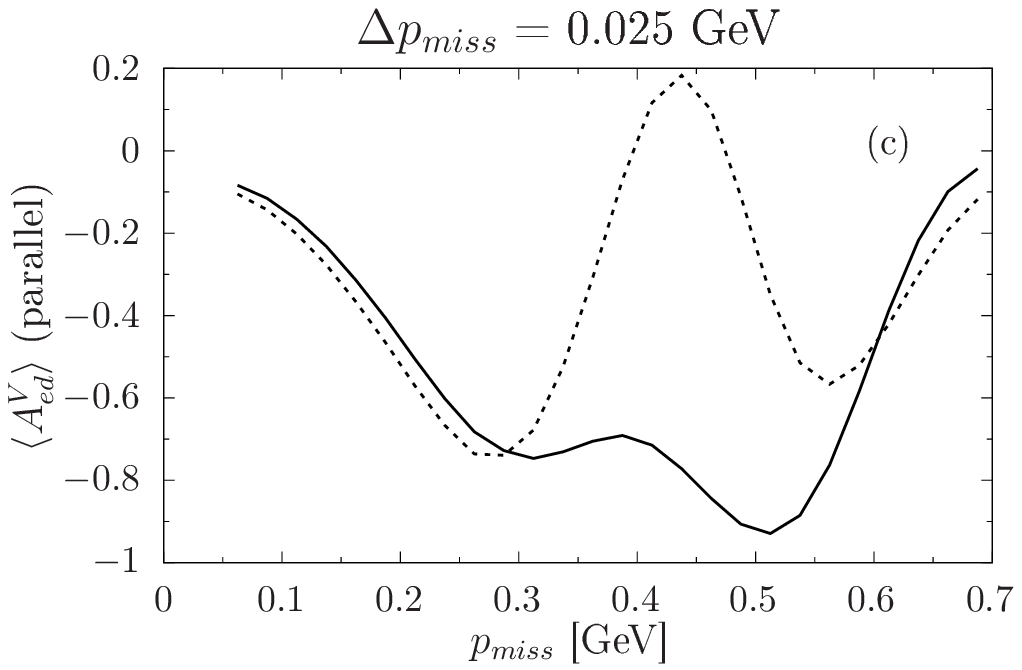}
\includegraphics[width=.45\textwidth,clip=true]{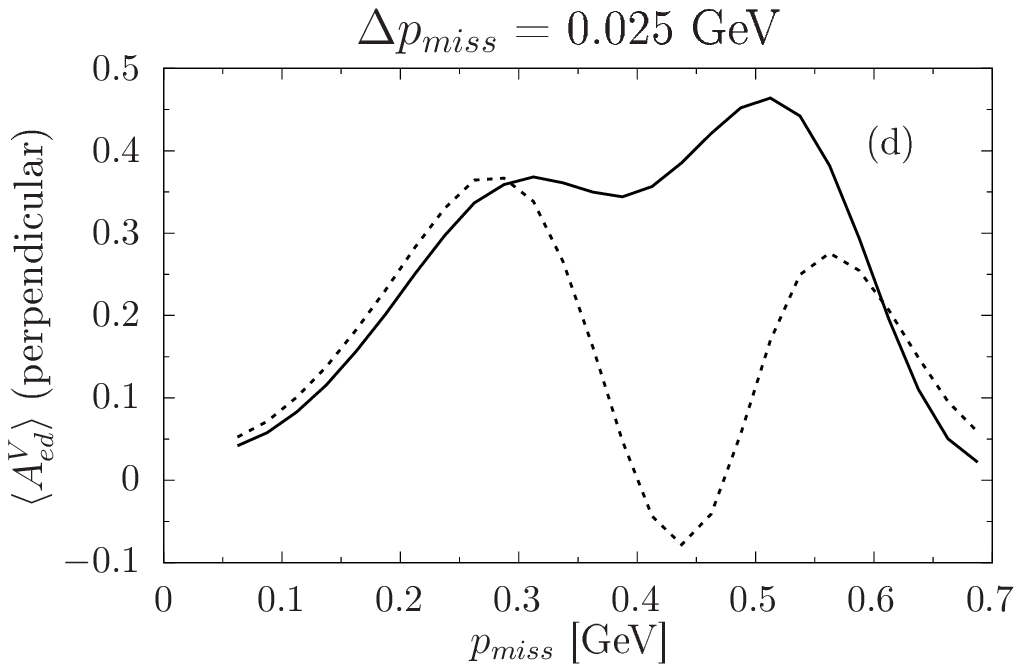}
\caption{
Averaged theoretical predictions for $A_d^T$ at
	$ 0.35 \, {\rm GeV}^2 < Q^2 < 0.45 \, {\rm GeV}^2 $ ((a) and (b))
	and
	$ 0.525 \, {\rm GeV}^2 < Q^2 < 0.675 \, {\rm GeV}^2 $ ((c) and (d)) 
	are shown as a function of the magnitude of the missing momentum $p_{miss}$ 
	for the deuteron polarization axis parallel ((a) and (c)) 
	and perpendicular ((b) and (d))
to the three-momentum transfer ${\ve{q}}$.
The {\em plane wave} ({\em full}) calculations are represented 
	by the dotted (solid) line.
}
\label{FIG(7)}
\end{center}
\end{figure}

\begin{figure}
\begin{center}
\includegraphics[width=.45\textwidth,clip=true]{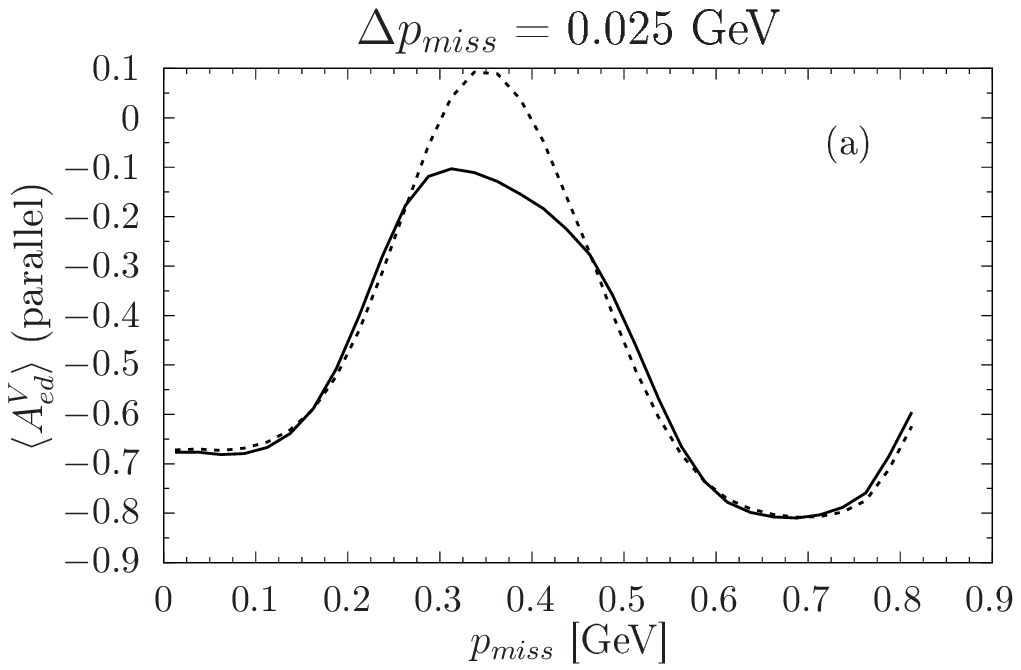}
\includegraphics[width=.45\textwidth,clip=true]{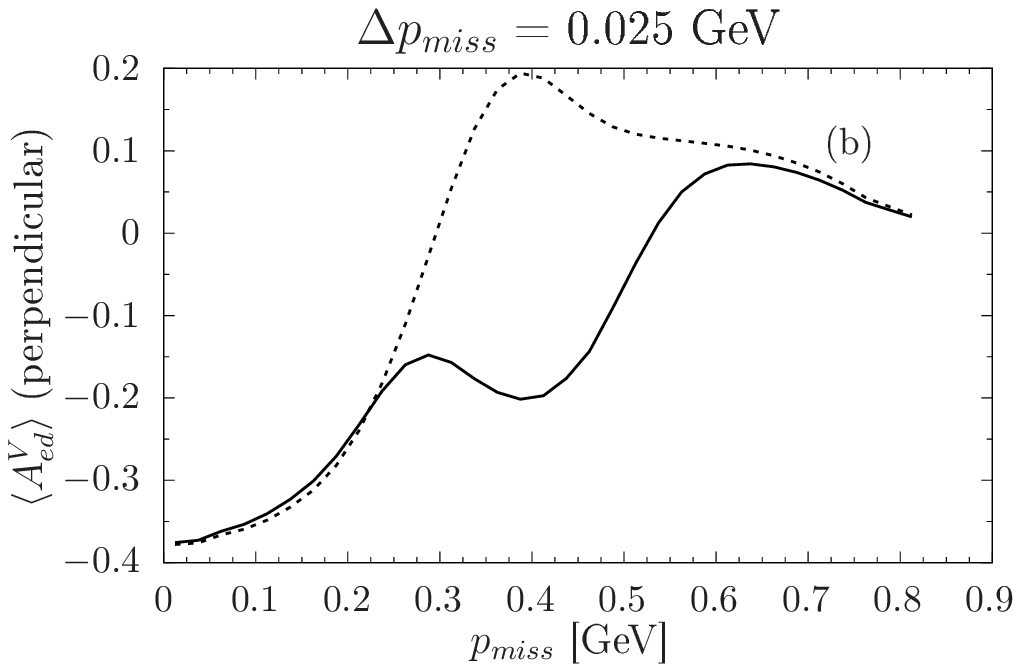}
\includegraphics[width=.45\textwidth,clip=true]{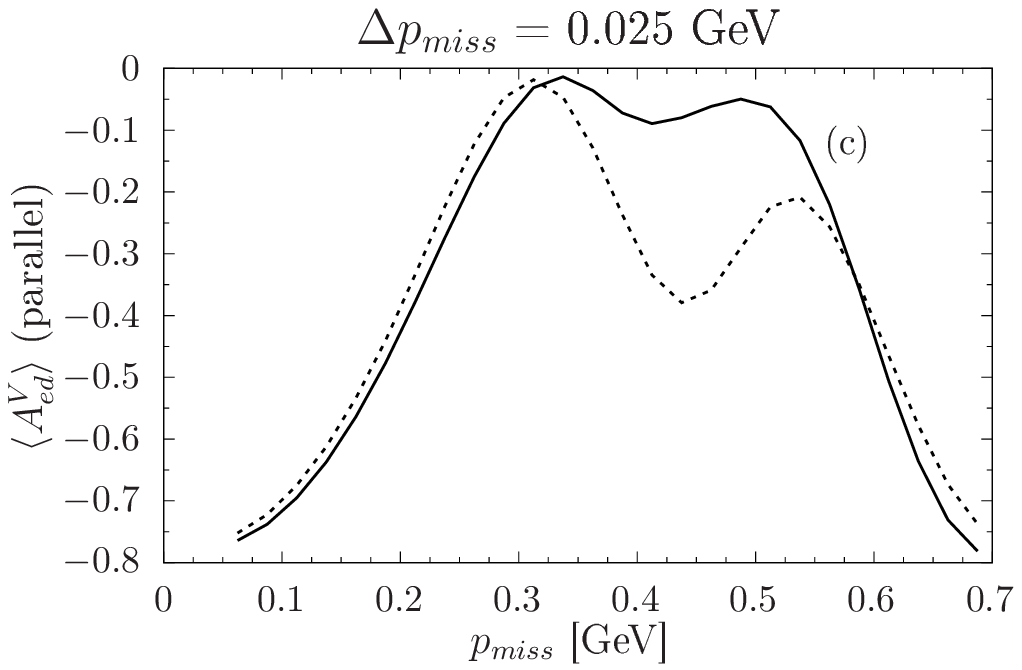}
\includegraphics[width=.45\textwidth,clip=true]{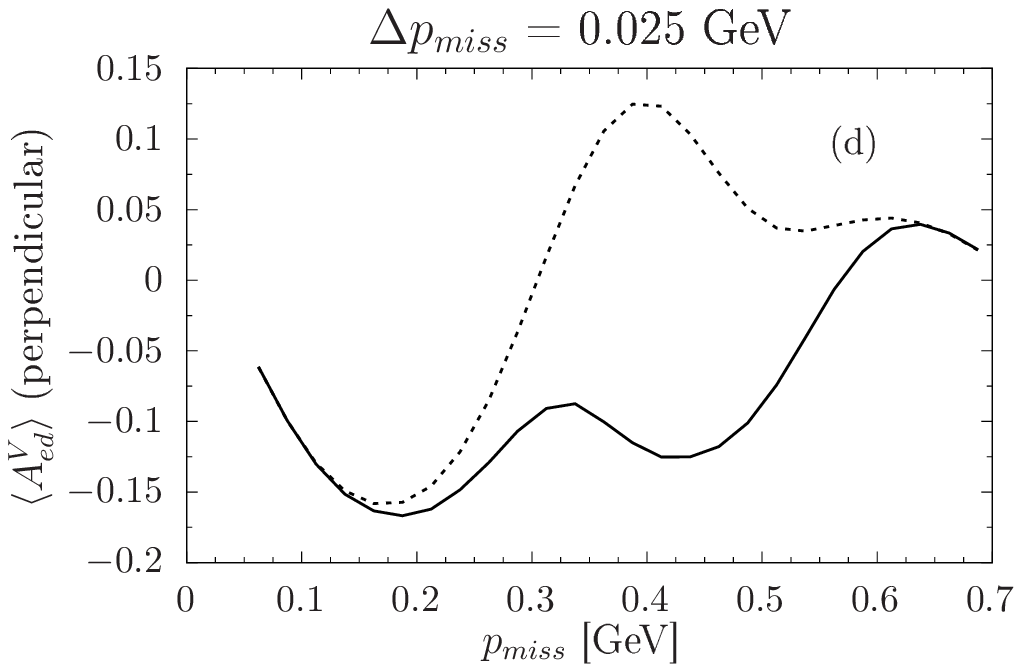}
	\caption{The same as in Fig.~\ref{FIG(7)} for $A_{ed}^V$.}
\label{FIG(12)}
\end{center}
\end{figure}

Thus we restrict ourselves to the remaining two: $A_d^T$ and $A_{ed}^V$. 
They are not just summed as it was the case for the cross sections but weighted with the (unpolarized) cross section calculated at the same points of the considered phase-space domain.  
Since these observables are defined for a given deuteron polarization axis,
we choose this time the deuteron polarization axis parallel and perpendicular 
to the three-momentum transfer ${\ve{q}}$. 
Since the results do not change significantly with ${\Delta}p_{miss}$,
we display them in Figs.~\ref{FIG(7)} and \ref{FIG(12)}  only for ${\Delta}p_{miss}$ = 0.025 GeV. 
Our {\em plane wave} and {\em full} predictions are quite different in the middle of the $p_{miss}$ intervals but come close together otherwise. Thus these two observables can provide more information about the final state interaction effects than the unpolarized cross sections.  

\subsection{Neutrino reactions with the deuteron}

\begin{figure}
\begin{center}
\includegraphics[width=0.45\textwidth,clip=true]{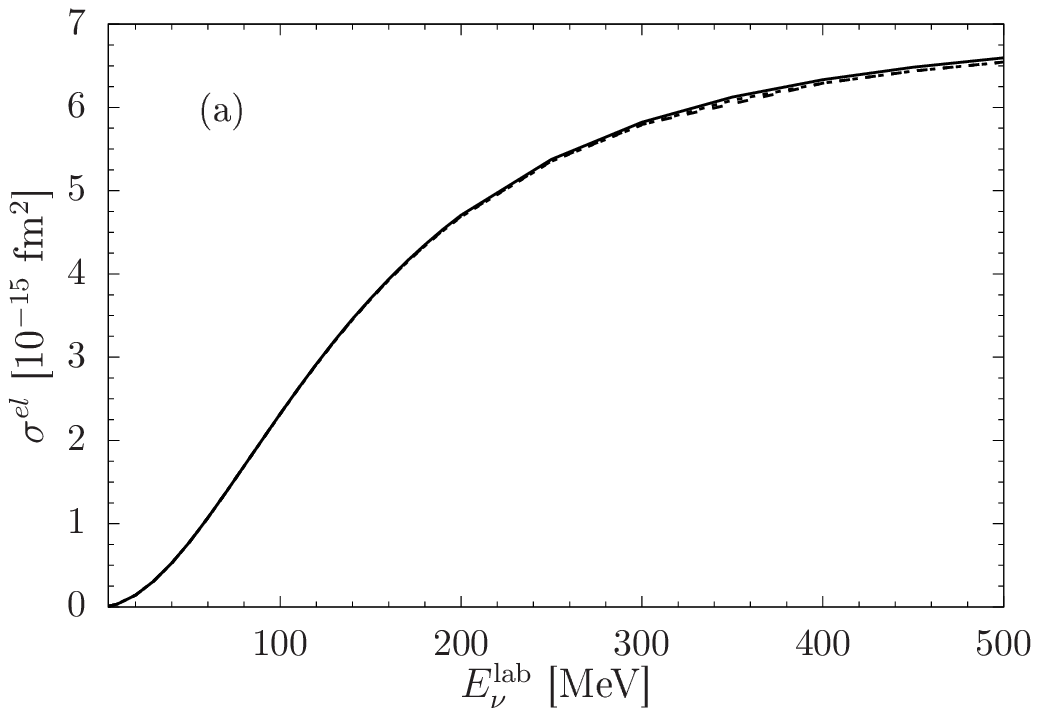}
\includegraphics[width=0.45\textwidth,clip=true]{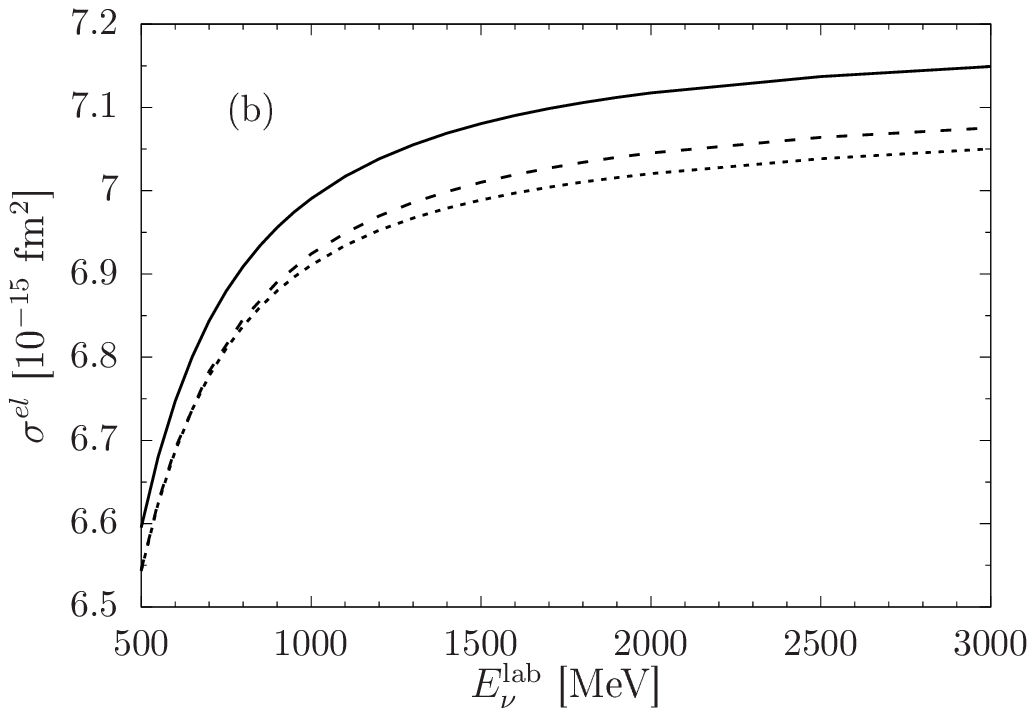}
\caption{Predictions for the total elastic cross section 
$\sigma^{\rm el}$ in neutrino-deuteron scattering
shown as a function of the laboratory neutrino energy for neutrino 
energies $E_{\nu}^{lab}$ (a) smaller and (b) bigger than 500 MeV. 
The cross was evaluated relativistically in the laboratory system (solid line) 
and in the center of mass system (dotted line), as well as non-relativistically 
in the laboratory system (dashed line).
}
\label{ncel_nud_consist}
\end{center}
\end{figure}

There also many reactions induced by neutrinos or antineutrinos.
We start with predictions for the elastic NC $ \nu + d \rightarrow \nu + d$ 
total cross section. They are obtained both in the laboratory 
and in the CM frame.
We expect some deviations 
between these two predictions due to the incomplete current operator.
This problem is illustrated
in Fig.~\ref{ncel_nud_consist}, where we show three predictions
for the total cross section in elastic NC driven neutrino-deuteron scattering.
The cross section is calculated nonrelativistically in the laboratory frame and 
relativistically both, in the laboratory frame and in the total momentum zero frame.
Up to approximately $E_\nu^{lab} \le $ 500 MeV all the calculations yield very similar results 
but the situation changes for higher neutrino energies.
Actually the deviation between the two relativistic predictions is larger
than the difference between the nonrelativistic and relativistic results obtained 
in the laboratory frame.
It is, however, evident that this observable hides all the differences bound 
in particular with the difference between the relativistic and nonrelativistic 
kinematics.

We calculated also breakup cross sections for two selected reactions,
$ \nu + d \rightarrow \nu + p + n$
as well as
$ {\bar\nu} + d \rightarrow e^{+} + n + n$.

	The formulas for the neutrino induced deuteron breakup reactions 
	are very similar to those for electron scattering. Adjusting Eq.~(\ref{pepmax}), 
	the formula for the total breakup cross section is
	\begin{eqnarray}
	\sigma^{br} = 
	\int \de{\versor{p}_\nu'} 
	\int_0^{|\ve{p}_\nu'|^{max}} \de|\ve{p}_\nu'| \,
	\frac{ \de^3\sigma}{\de \mathbf{\hat p}_\nu' \de|\ve{p}_\nu'| } \nonumber \\
	= 2\pi \, \int_0^{\pi} \de\theta_{p_\nu'} \sin\theta_{p_\nu'}  \, 
	\int_0^{|\ve{p}_\nu'|^{max}} \de|\ve{p}_\nu'| \,
	\left. \frac{ \de^3\sigma}{\de \mathbf{\hat p}_\nu' \de|\ve{p}_\nu'| } \right|_{\phi_{p_\nu'}=0} \, .
	\end{eqnarray}
	
	We also mention that for the 
	$ \bar{\nu}_e + d \rightarrow e^{+} + n + n $ charged-current driven reaction 
	the single-nucleon weak current operator assumes a well-known form
	\begin{eqnarray}
		\matrixelement{\ve{p}',\mu',\tau'}{J_{k, {\rm WCC}}^\mu(0)}{\ve{p},\mu , \tau} \nonumber \\
	= 
	\delta_{ \tau' -\oneHalf } \,
	\delta_{ \tau \oneHalf } \,
		{\bar u}(\ve{p}',\mu')
	\left(
	F_{1}^C(Q^2) \gamma^\mu
	+ \frac{i}{2m}\sigma^{\mu\nu}q_\nu F_{2}^C(Q^2) \right. \nonumber \\
	\left. 
	+F_{A}^C(Q^2) \gamma^\mu \gamma_5 
	+  \frac{q^\mu}{m}\gamma_5F_{P}^C(Q^2)
	\right)
	u(\ve{p},\mu)\, ,
	\label{j1wcc}
	\end{eqnarray}
	with the weak charged-current nucleon form factors $F_{i}^C$. For recent
	parametrizations of these quantities see for example \cite{PRC86.035503,BBA03}.
	The cross section formulas for this reaction have to be additionally multiplied
	by $\cos^2\theta_C$, where $\theta_C \approx 13^\circ$ is the Cabibbo angle 
	and the correction factor due to two identical particles in the final state
	has to be introduced. 

\begin{figure}
\begin{center}
\includegraphics[width=0.45\textwidth,clip=true]{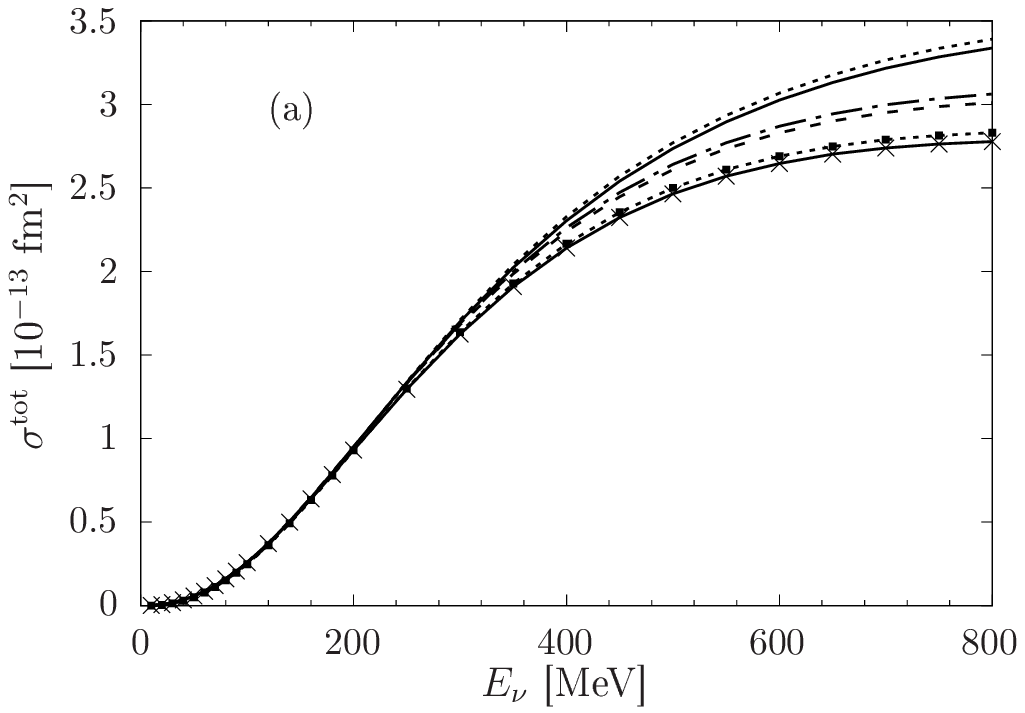}
\includegraphics[width=0.45\textwidth,clip=true]{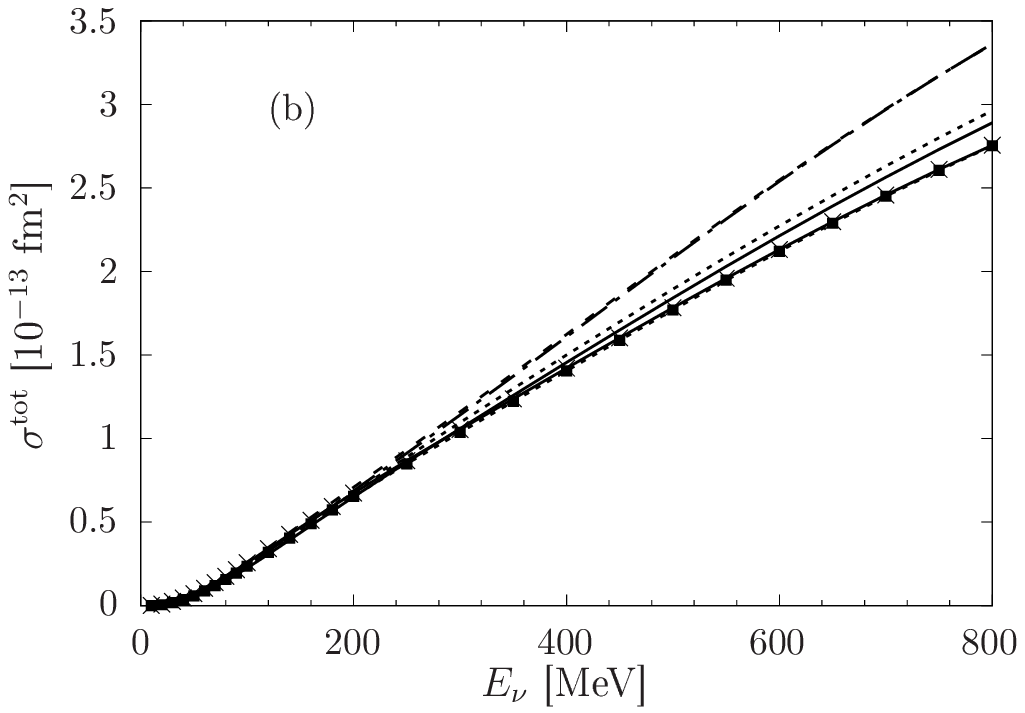}
	\caption{The {\em plane wave} and {\em full} predictions 
	for the total breakup cross section ${\sigma}^{\rm tot}$
	for the (a) $\nu + d \rightarrow \nu + p + n$ 
            and (b) $\bar{\nu}+d\rightarrow e^++n+n$
	reaction as a function 
	of the initial (anti)neutrino laboratory energy $E_{\nu}$. 
	The solid (dotted) line represents the differential cross section with (without) the contribution
        of the rescattering term calculated relativistically.
        The dashed  (dash-dotted) line represents the differential cross section with (without) the contribution of the rescattering term and calculated within the non-relativistic treatment.
	The solid (dashed) line with white (black) squares represents the result for the non-relativistic calculations 
	with (without) the contribution of the rescattering term and with the $(p/m)^2$   corrections to the nonrelativistic single-nucleon current operator.
}
\label{FIG9}
\end{center}
\end{figure}

In Fig.~\ref{FIG9} we display
relativistic and nonrelativistic predictions with ({\em full}) and without ({\em plane wave}) 
the rescattering contribution in the nuclear matrix elements 
and additionally two nonrelativistic predictions, where the nonrelativistic single-nucleon 
current operator is augmented by relativistic $(p/m)^2$ corrections. Strictly spoken,
our way of introducing rescattering effects is valid only for the center of mass energies
smaller than the pion mass, so the {\em full} results should be treated with great care.
We see, however, that these contributions in all the three cases are small.
Adding the corrections to the nonrelativistic current operator makes the difference 
between the nonrelativistic and relativistic predictions bigger for the NC induced reaction but 
brings the ``corrected'' nonrelativistic results closer to the relativistic predictions 
for the CC driven process.

\begin{figure}
\begin{center}
\includegraphics[width=0.45\textwidth,clip=true]{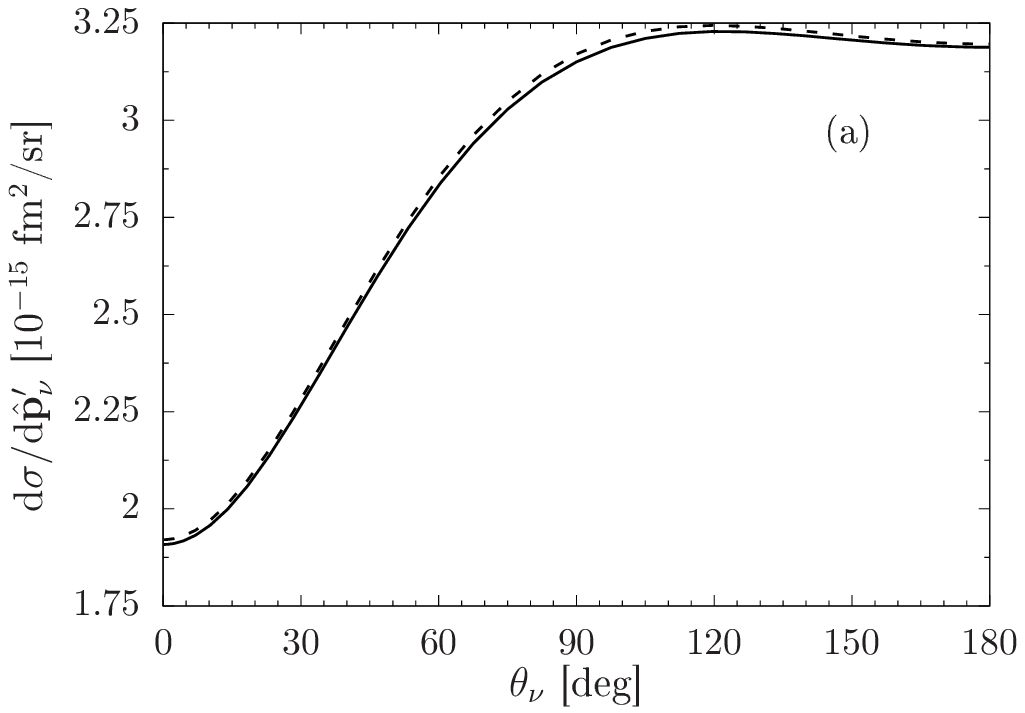}
\includegraphics[width=0.45\textwidth,clip=true]{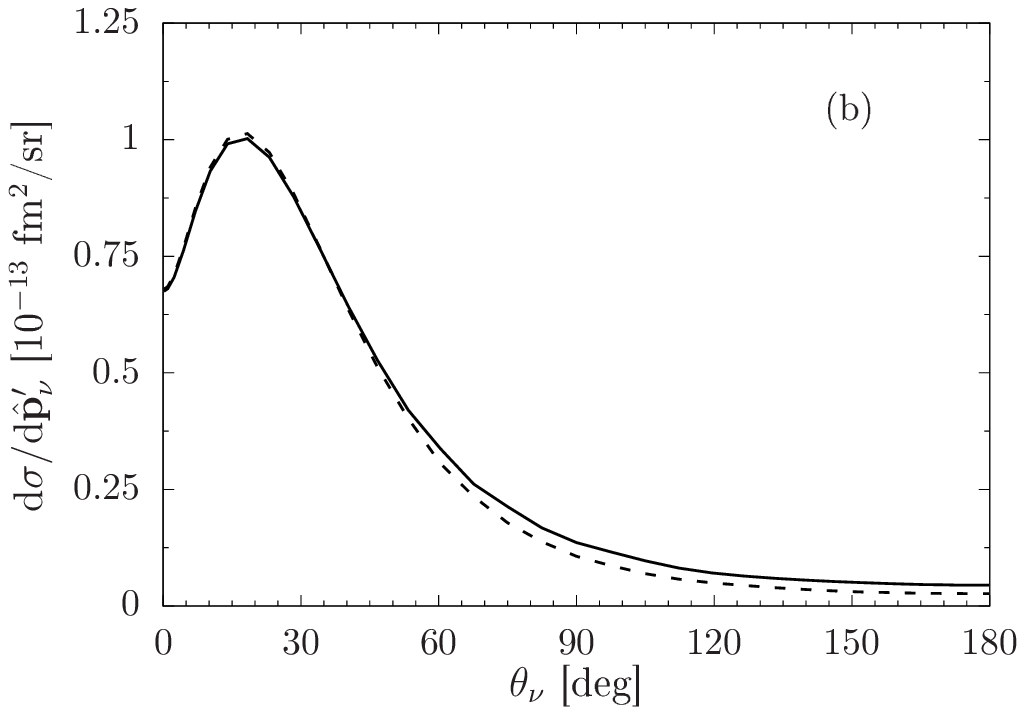}
	\caption{The differential breakup cross section $\de{\sigma}/\de\versor{p}_\nu'$
	for the $\nu + d \rightarrow \nu + p + n$ process
	calculated in the laboratory frame 
	for the initial neutrino energies $E_\nu$= 120 MeV (left panel) 
	and 700 MeV (right panel).
	The lines are the same as in Fig.~\ref{ncel_nud_consist}.
}
\label{FIG10}
\end{center}
\end{figure}

\begin{figure}
\begin{center}
\includegraphics[width=0.45\textwidth,clip=true]{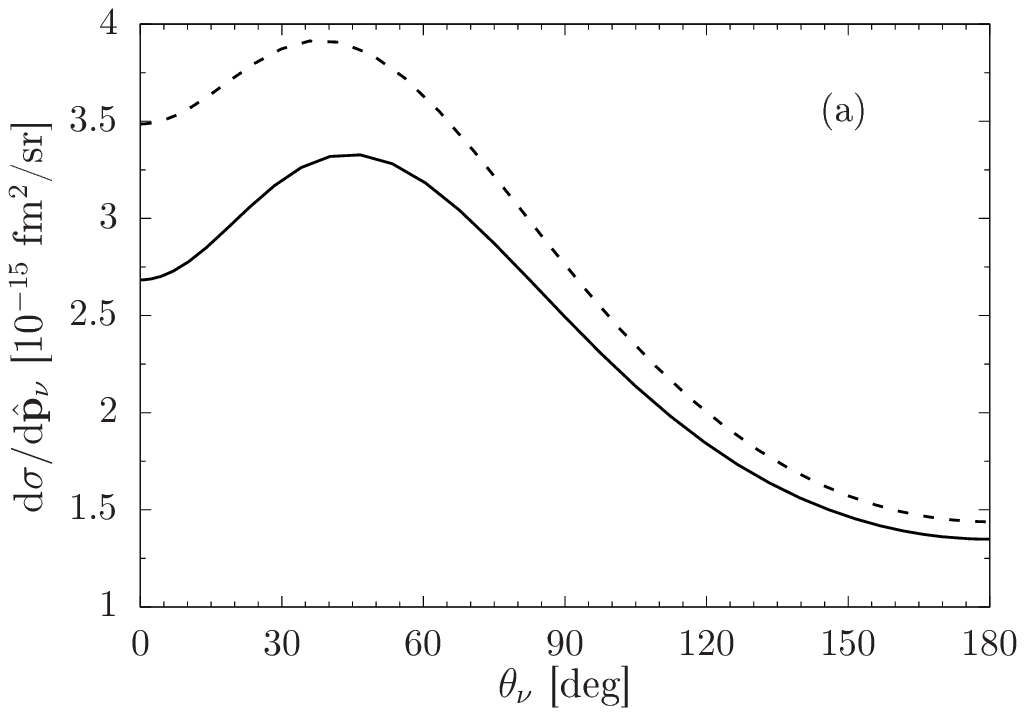}
\includegraphics[width=0.45\textwidth,clip=true]{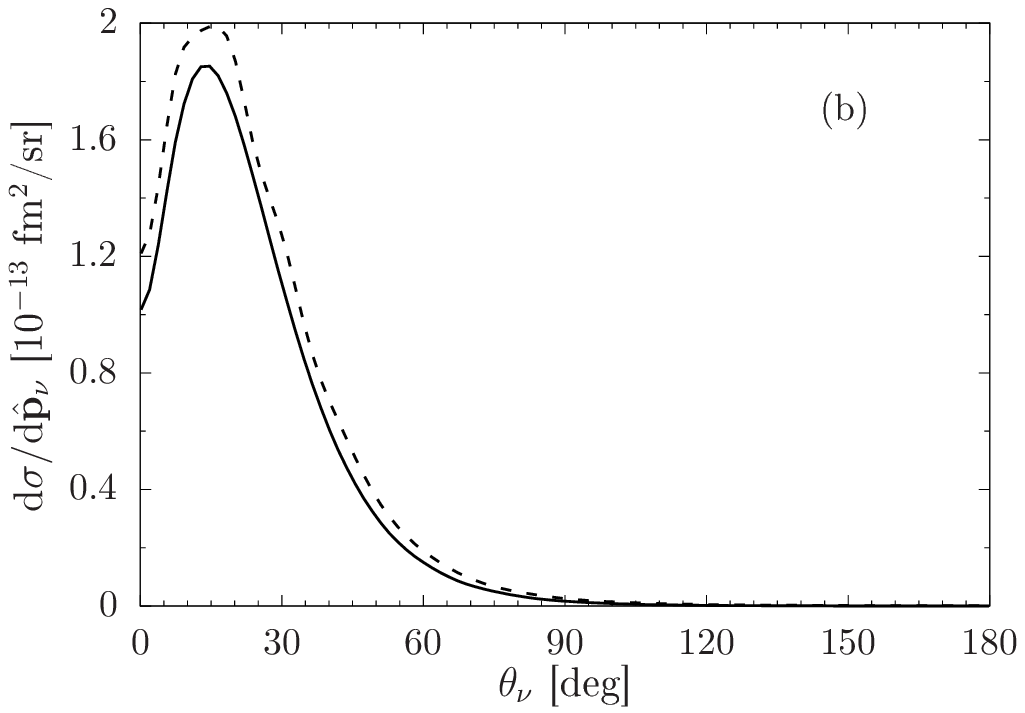}
	\caption{Same as Fig.~\ref{FIG10}, but for the ${\bar\nu} + d \rightarrow e^{+} + n + n$ process.
}
\label{FIG12}
\end{center}
\end{figure}

The total breakup cross section is obtained as a result of angular integration
over the whole solid angle corresponding to the final lepton momentum. 
In Fig.~\ref{FIG10} we show the differential breakup cross sections $\de{\sigma}/\de\versor{p}_\nu'$
at two different laboratory neutrino energies $E_\nu$= 120 MeV and 700 MeV
for the $\nu + d \rightarrow \nu + p + n$ process.
The angular distributions calculated relativistically and nonrelativistically show only small 
but visible differences and clearly change with the initial neutrino energy.
Also the shape of the angular distribution strongly depends on the initial (anti)neutrino energy.
The corresponding results shown in Fig.~\ref{FIG12} for the $\bar{\nu}+d\rightarrow e^++n+n$ reaction
are different, especially for the lower energy, where for the NC reaction no peak at forward angles 
is observed. The peak visible for both reactions at the higher energy is broader in the case of the NC induced reaction.
The relativistic features are more pronounced in the case of the CC driven process.

\section{Summary and outlook}

We give a complete relativistic formalism and construct tools to perform calculations of exclusive,
semi-exclusive and inclusive unpolarized cross sections and various
polarization observables in electron and neutrino
scattering experiments with deuteron targets.
In the present work the strong interaction dynamics is defined by an explicit dynamical unitary
representation of the Poincar\'e group\cite{Wigner1939}.
In the chosen framework representations of space translations and
rotations in the interacting and non-interacting
representations are identical \cite{Bakamjian:1953kh}.
The Argonne V18 potential \cite{Wiringa:1994wb} is the starting point
for building the relativistic nucleon-nucleon interaction reproducing
the experimental deuteron binding energy and
nucleon-nucleon scattering observables \cite{KAMADA2007119}.

Our formalism does not take into account the pion production channel
and neglects two-body contributions in the electromagnetic as well as in the weak
nuclear current operator. These limitations require additional studies
and will be addressed in subsequent investigations. Presently the
description of the deuteron form factor $B(Q^2)$ and the deuteron tensor analyzing power $T_{20}$
suggest the need for two-body contributions to the current.
The current model
is best applicable to kinematics, where the internal two-nucleon energy
remains below the pion production threshold but the magnitude of the three-momentum
transfer extends at least to several GeV. Here the
final-state nucleon-nucleon interactions can be included exactly.
In particular we demonstrate fair
agreement with the experimental data for deuteron electrodisintegration
in the region of quasi-elastic peak, where the dynamics is governed predominantly
by the single-nucleon current operator.
Our predictions for the total cross sections in the neutral-current
and charged-current induced reactions are also quite reliable, since
the two-nucleon contributions in these reactions were shown not to be strong \cite{PRC86.035503}.
We demonstrate that the use of the relativistic kinematics is mandatory for the magnitudes
of the three-momentum transfer comparable and higher than the nucleon mass.
Relativistic  $(p/m)^2$ corrections to the nonrelativistic single-nucleon current are
to be used with great caution. The fact that purely nonrelativistic and relativistic results
for the total elastic neutral-current driven neutrino-deuteron scattering cross section
agree very well does not justify the use of the nonrelativistic framework
in the relativistic domain, where already the nonrelativistic kinematics is wrong.
The frame dependence of our calculations of the total elastic cross section points to a need for 
corrections due to two-body currents.
The predictions obtained in the
laboratory frame and in the total momentum zero frame differ at 3 GeV by about 1.5 \%,
which can be traced back to the inadequacy of our weak nuclear current operator.
The kinematics of the electron and neutrino induced deuteron breakup is relatively simple
and allows one to easily consider {\em any} kinematical conditions. We show examples
in the so-called ``$Q^2-p_{miss}$'' kinematics, making predictions for the unpolarized
cross sections and selected polarization observables.
We are ready to analyze experimental data and plan to improve the present framework
by augmenting it with two-nucleon current contributions. 
Last not least, the Argonne V18 nucleon-nucleon potential can be replaced by the recently developed accurate chiral interaction~\cite{Reinert18}.

\appendix 

\section{Nuclear current matrix elements}
	
	Bearing in mind that we will also need nuclear matrix elements for the deuteron disintegration  
	reactions we actually calculate a chain of matrix elements, starting from 
	\begin{eqnarray}
	\langle \ve{p}_1', \mu_1', \tau_1' , \ve{p}_2', \mu_2' , \tau_2' \vert J_{nuc}^{\mu} (0)  \vert 
	\ve{p}_D,\mu_D ,D \rangle \, ,  
	\label{B}
	\end{eqnarray}
	then we insert the completeness relations to obtain
	\begin{eqnarray}
	\langle \ve{p}_1', \mu_1', \tau_1' , \ve{p}_2', \mu_2' , \tau_2' \vert J_{nuc}^{\mu} (0)  \vert 
	\ve{p}_D,\mu_D ,D \rangle 
	\nonumber \\
	= \sum_{\mu_1\tau_1}\sum_{\mu_2\tau_2}
	\int \de{\ve{p}_1}\de{\ve{p}_2}
	\matrixelement{\ve{p}_1',\mu_1',\tau_1',\ve{p}_2',\mu_2',\tau_2'}{J_{nuc}^{\mu}(0)}{\ve{p}_1,\mu_1,\tau_1,\ve{p}_2,\mu_2,\tau_2}
	\nonumber \\
	\times \,
	\sum_{jlst\mt\mu}\int \de{\ve{p}}\,\de{k}\,k^2\,
	\braket{\ve{p}_1,\mu_1,\tau_1,\ve{p}_2,\mu_2,\tau_2}{(j,k)\ve{p},\mu; l s t\mt}\,
	\nonumber \\
	\times \,
	\braket{(j,k)\ve{p},\mu; lst\mt}{\ve{p}_D,\mu_D,D}
	\end{eqnarray}
	and, using Eqs.~(\ref{b.4}) and (\ref{D_PWD}), we arrive at
	\begin{eqnarray}
	\langle \ve{p}_1', \mu_1', \tau_1' , \ve{p}_2', \mu_2' , \tau_2' \vert J_{nuc}^{\mu} (0)  \vert 
	\ve{p}_D,\mu_D ,D \rangle  \nonumber \\
		= {\cal N}^{-1}(\ve{p}_1,\ve{p}_2') 
	\sum_{\mu_1\tau_1}
	\bra{\ve{p}_1',\mu_1',\tau_1'}J^\mu_1(0)\ket{\ve{p}_1,\mu_1,\tau_1}
	\CG{\oneHalf}{\tau_1}{\oneHalf}{\tau_2'}{0}{0} \nonumber \newlineEq
	\sum_{l=0,2} \deutL{l}(k)\, 
	\sum_{\mu_l \mu_s } \Y{l \mu_l}{\versor{k}}\,\CG{l}{\mu_l}{ 1}{\mu_s}{1}{\mu_D} \nonumber
	\newlineEq
	\sum_{\mu_1''\mu_2''}
	\CG{\oneHalf}{\mu_1''}{\oneHalf}{\mu_2''}{1 }{\mu_s}
		\Dmatrix{\oneHalf}{\mu_1 \mu_1''}{R_w\left( \ve{p}_D/m_{120} , \ve{k} \right)}
		\Dmatrix{\oneHalf}{\mu_2'\mu_2''}{R_w\left( \ve{p}_D/m_{120} ,-\ve{k} \right)}\,,
	\label{termB}
	\end{eqnarray}
	where $\ve{k}=\ve{k}(\ve{p}_1,\ve{p}_2')$, $k=|\ve{k}|$, $\ve{p}_1=\ve{p}_D-\ve{p}_2$
	and ${\cal N} (\ve{p}_1,\ve{p}_2) $ is given in Eq.~(\ref{b.6}).
	
	For the semi-exclusive observables in the deuteron breakup process,
	where we integrate over all the nuclear states for the fixed final 
	lepton scattering angle and energy, it is convenient 
	to prepare matrix elements 
	\begin{eqnarray}
	\langle \ve{k}', \ve{p}', \mu_1', \tau_1' , \mu_2' , \tau_2' \vert J_{nuc}^{\mu} (0)  \vert \ve{p}_D,\mu_D ,D \rangle \, .
	\label{Bp}
	\end{eqnarray}
	They take the following form
	\begin{eqnarray}
	\langle \ve{k}', \ve{p}', \mu_1', \tau_1' , \mu_2' , \tau_2' \vert J_{nuc}^{\mu} (0)  \vert \ve{p}_D,\mu_D ,D \rangle 
	\nonumber \\
	= \sum_{\mu_1''\mu_2''}\int \de{\ve{p}_1''}\,\de{\ve{p}_2''}\,
	\braket{\ve{k}',\ve{p}',\mu_1',\mu_2'}{\ve{p}_1'',\mu_1'',\ve{p}_2'',\mu_2''}
	\nonumber \\
	\times \, 
		\matrixelement{\ve{p}_1'',\mu_1'',\tau_1',\ve{p}_2'',\mu_2'',\tau_2'}{J_{nuc}^\mu(0)}{\ve{p}_D,\mu_D,D}\,,
	\end{eqnarray}
	where
	\begin{eqnarray}
	\braket{\ve{k}',\ve{p}',\mu_1',\mu_2'}{ \ve{p}_1'',\mu_1'',\ve{p}_2'',\mu_2''}
	\nonumber \\
		= {\cal N}(\ve{p}_1'',\ve{p}_2'')\,
	\delta(\ve{p}_1''-\ve{p}_1(\ve{k}',\ve{p}'))\,
	\delta(\ve{p}_2''-\ve{p}_2(\ve{k}',\ve{p}')) \nonumber \\
	\times
	\,
		\Dmatrix{\oneHalf \, \ast}{\mu_1'\mu_1''}{R_w\left( \ve{p}'/m_{120}' , \ve{k}' \right)}\,
		\Dmatrix{\oneHalf \, \ast}{\mu_2'\mu_2''}{R_w\left( \ve{p}'/m_{120}' ,-\ve{k}' \right)}
	\, .
	\label{eq:overlapP1P2PK}
	\end{eqnarray}
	
	The rescattering contributions to the breakup matrix elements are calculated 
	using partial wave states. That requires that also matrix elements 
	\begin{eqnarray}
	\langle (j',k' ) \ve{p}' ,\mu' ; l' ,s' ; t' \mt' \vert J_{nuc}^{\mu} (0) \vert \ve{p}_D,\mu_D ,D \rangle
	\label{Bpp}
	\end{eqnarray}
	are evaluated. 
	These matrix elements are
	\begin{eqnarray}
	\langle (j',k' ) \ve{p}' ,\mu' ; l' ,s' ; t' \mt' \vert J_{nuc}^{\mu} (0) \vert \ve{p}_D,\mu_D ,D \rangle 
	\nonumber \\
	=\sum_{\mu_1''\mu_2''}\sum_{\tau_1''\tau_2''}
	\int \de{\ve{k}}\,\de{\ve{p}}\,
	\braket{ (j',k' ) \ve{p}' ,\mu' ; l' ,s' ; t' \mt' }{\ve{k},\ve{p},\mu_1'',\tau_1'',\mu_2'',\tau_2''}\,
	\nonumber \\
	\times \,
	\matrixelement{\ve{k},\ve{p},\mu_1'',\tau_1'',\mu_2'',\tau_2''}{J^\mu_1(0)}{\ve{p}_D,\mu_D,D}\,,
	\end{eqnarray}
	where
	\begin{eqnarray}
	\braket{ (j',k' ) \ve{p}' ,\mu' ; l' ,s' ; t' \mt'}{\ve{k},\ve{p},\mu_1'',\tau_1'',\mu_2'',\tau_2''}
	\nonumber \\
	=  \delta(\ve{p'}-\ve{p})\,
	\frac{\delta(k-k')}{k^2} \CG{\oneHalf}{\tau_1''}{\oneHalf}{\tau_2''}{t'}{\mt'}
	\nonumber \\
	\times \, 
	\sum_{\mu_l' \mu_s' } 
	\CG{l'}{\mu_l'}{s'}{\mu_s'}{j'}{\mu'}\,
	\CG{\oneHalf}{\mu_1''}{\oneHalf}{\mu_2''}{s'}{\mu_s'}\,
	\Yconj{l'\mu_l'}{\versor{k}'}
	\, .
	\label{eq:overlapKPtopartial}
	\end{eqnarray}
	The result for matrix elements (\ref{Bpp}) can be also used to easily calculate 
	nuclear matrix elements for the elastic scattering reactions, since
	\begin{eqnarray}
	\langle \ve{p}_D', \mu_D' ,D \vert J_{nuc}^{\mu} (0)  \vert \ve{p}_D,\mu_D ,D \rangle \nonumber \\
	= \, \int \de  \ve{p}' \, \sum\limits_{l'=0,2} \int \de k' {k'}^2 \, 
	\langle \ve{p}_D', \mu_D' ,D \vert (1,k' ) \ve{p}' ,\mu' ; l' ,1;  0 0 \rangle \nonumber \\
	\times \, 
	\langle (1,k' ) \ve{p}' ,\mu' ; l' ,1 ; 0 0 
	\vert  J_{nuc}^{\mu} (0)  \vert \ve{p}_D,\mu_D ,D \rangle \nonumber \\
	= \sum\limits_{l'=0,2} \int \de k' {k'}^2 \, \deutL{l'} (k') \, 
	\langle (1,k' ) \ve{p}_D' ,\mu_D' ; l' ,1 ; 0 0 
	\vert  J_{nuc}^{\mu} (0)  \vert \ve{p}_D,\mu_D ,D \rangle \, .
	\label{A}
	\end{eqnarray}
	
The rescattering part (\ref{rescatt}) 
of the ${{}^{(-)}\langle} \ve{p}_1', \mu_1' , \tau_1' , \ve{p}_2', \mu_2' , \tau_2' \vert J_{nuc,EM}^{\mu} (0) \vert \ve{p}_D,\mu_D ,D \rangle$
matrix element is calculated in two steps. We calculate first 
	\begin{eqnarray}
	\matrixelement{(j',k')\ve{p}',\mu';l's't'\mt'}
	{
		t( E + i \epsilon) 
		G_0 ( E + i \epsilon) \,
		J^\mu_{nuc}(0)}{\ve{p}_D\mu_D D} \nonumber \\
	=
	\sum_{j\mu l s t\mt}
	\int \de{\ve{p}}\,
	\int \de{k}\,k^2\,
	\matrixelement{(j',k')\ve{p}',\mu';l's't'\mt'}
	{
		t( E + i \epsilon) 
		G_0 ( E + i \epsilon) \,
	}{(j,k)\ve{p},\mu;lst\mt} \nonumber \\
	\times \, 
	\matrixelement{(j,k)\ve{p},\mu;lst\mt}{J^\mu_{nuc}(0)}{\ve{p}_D,\mu_D,D} \, ,
	\end{eqnarray}
	where 
	\begin{eqnarray}
	\matrixelement{(j',k')\ve{p}',\mu';l's't'\mt'}
	{
		t( E + i \epsilon) 
		G_0 ( E + i \epsilon) \,
	}{(j,k)\ve{p},\mu;lst\mt} \nonumber \\
	=
	\delta(\ve{p}'-\ve{p})
	\delta_{jj'}
	\delta_{\mu\mu'}
	\delta_{ss'}
	\delta_{tt'}
	\delta_{\mt\mt'} \nonumber \\
	\times \, 
	\frac{
		\matrixelement{(jk')\mu';l's't'\mt'}{t(E+i\epsilon,p')}{(jk)\mu;lst\mt}
	}{E+i\epsilon -\sqrt{4(m^2+k^2)+|\ve{p}'|^2}}
	\end{eqnarray}
	and obtain
	\begin{eqnarray}
	\matrixelement{(j',k')\ve{p}',\mu';l's't'\mt'}
	{
		t( E + i \epsilon) 
		G_0 ( E + i \epsilon) \,
		J^\mu_{nuc}(0)}{\ve{p}_D,\mu_D,D} \nonumber \\
	=
	\sum_{l }
	\int \,\de{k}\,k^2\,
	\frac{\matrixelement{(j',k'),\mu';l's't'\mt'}{t(E+i\epsilon,p')}{(j',k),\mu';ls't'\mt'}}{E+i\epsilon -\sqrt{4(m^2+k^2)+|\ve{p}'|^2}}
	\nonumber \\ \times \,
	\matrixelement{(j',k)\ve{p}',\mu';ls't'\mt'}{J^\mu_{nuc}(0)}{\ve{p}_D,\mu_D,D}\,.
	\label{tg0j1}
	\end{eqnarray}
	The pole in (\ref{tg0j1}) is treated by dividing and multiplying the integrand
	by 
	\[
	h(E,\left|\ve{p}'\right|,k) \equiv
	\frac{1}{4}(E+\sqrt{4(m^2+k^2)+\ve{p}'^2}) \,,
	\]
	which leads to 
	\begin{eqnarray}
	\matrixelement{(j',k')\ve{p}',\mu';l's't'\mt'}
	{
		t( E + i \epsilon) 
		G_0 ( E + i \epsilon) \,
		J^\mu_{nuc}(0)}{\ve{p}_D,\mu_D,D} \nonumber \\
	=\,m \sum_{l } \int \,\de{k}\,k^2\, h(E,\left|\ve{p}'\right|,k) \nonumber \\
	\times \, 
	\frac{\matrixelement{(j',k'),\mu';l's't'\mt'}{t(E+i\epsilon,p')}{(j',k),\mu';ls't'\mt'}}{k_0^2-k^2+i\epsilon}
	\nonumber \\
	\times \, 
	\matrixelement{(j',k)\ve{p}',\mu';ls't'\mt'}{J^\mu_{nuc}(0)}{\ve{p}_D,\mu_D,D}\,,
	\label{tg0j1.2}
	\end{eqnarray}
	where
	\begin{equation}
	k_0=  \frac12\sqrt{E^2 - |\ve{p}'|^2 - 4 m^2}\,.
	\end{equation}
	The resulting integral is calculated using standard subtraction techniques.
	
	The result prepared in Eq.~(\ref{tg0j1.2}) for
	\[
	\matrixelement{(j',k')\ve{p}',\mu';l's't'\mt'}
	{
		t( E + i \epsilon) 
		G_0 ( E + i \epsilon) \,
		J^\mu_{nuc}(0)}{\ve{p}_D,\mu_D,D} 
	\]
	together with the two overlaps given
	in Eqs.~(\ref{b.4}) and (\ref{eq:overlapKPtopartial})
	can be used to calculate the final nuclear matrix elements 
	\begin{equation*}
	\matrixelement{\ve{p}_1',\mu_1',\tau_1',\ve{p}_2',\mu_2',\tau_2'}
	{
		t( E + i \epsilon) 
		G_0 ( E + i \epsilon) \,
		J^\mu_{nuc}(0)}{\ve{p}_D,\mu_D,D}
	\end{equation*}
	and
	\begin{equation*}
	\matrixelement{\ve{k}',\ve{p}',\mu_1',\tau_1',\mu_2',\tau_2'}
	{
		t( E + i \epsilon) 
		G_0 ( E + i \epsilon) \,
		J^\mu_{nuc}(0)}{\ve{p}_D,\mu_D,D} \, ,
	\end{equation*}
	which are needed for the exclusive and
	semi-exclusive (and eventually total for neutrino induced reactions)
	cross sections.

\acknowledgments
One of the authors (J.G.) gratefully acknowledges the financial support of the JSPS International Fellowships for Research in Japan (ID=S19149). 
One of the authors (W.P.) gratefully acknowledges
support of this research by the US Department of Energy, Office of Science,
grant number DE-SC0016457.
The numerical calculations were partly performed on the supercomputers of the JSC, J\"ulich, Germany.

	
\bibliographystyle{h-physrev}
\bibliography{biblio}
	
\end{document}